\documentclass{jpp}

\usepackage{graphicx}
\usepackage{natbib}

\ifCUPmtlplainloaded \else
  \checkfont{eurm10}
  \iffontfound
    \IfFileExists{upmath.sty}
      {\typeout{^^JFound AMS Euler Roman fonts on the system,
                   using the 'upmath' package.^^J}%
       \usepackage{upmath}}
      {\typeout{^^JFound AMS Euler Roman fonts on the system, but you
                   dont seem to have the}%
       \typeout{'upmath' package installed. JPP.cls can take advantage
                 of these fonts, if you use 'upmath' package.^^J}%
       \providecommand\upi{\pi}%
      }
  \else
    \providecommand\upi{\pi}%
  \fi
\fi

\ifCUPmtlplainloaded \else
  \checkfont{msam10}
  \iffontfound
    \IfFileExists{amssymb.sty}
      {\typeout{^^JFound AMS Symbol fonts on the system, using the
                'amssymb' package.^^J}%
       \usepackage{amssymb}%
         
       \let\ge=\geqslant  \let\geq=\geqslant
      }{}
  \fi
\fi

\ifCUPmtlplainloaded \else
  \IfFileExists{amsbsy.sty}
    {\typeout{^^JFound the 'amsbsy' package on the system, using it.^^J}%
     \usepackage{amsbsy}}
    {\providecommand\boldsymbol[1]{\mbox{\boldmath $##1$}}}
\fi

\usepackage[hidelinks]{hyperref}
\usepackage[EM,jfm]{jpnotation}
\usepackage[nooneline,small,tight]{subfigure}
\usepackage{paralist}
\usepackage{amsmath}
\usepackage{amssymb}
\usepackage{color}

\graphicspath{{plots/}}

\newcommand\Real{\mbox{Re}} 
\newcommand\Imag{\mbox{Im}} 

\newsavebox{\astrutbox}
\sbox{\astrutbox}{\rule[-5pt]{0pt}{20pt}}

\title%
[Fourier--Hermite spectral representation for the Vlasov--Poisson system]%
{Fourier--Hermite spectral representation for the Vlasov--Poisson system in the weakly collisional limit}

\author[J. T. Parker and P. J. Dellar]%
{J.\ns T.\ns P\ls A\ls R\ls K\ls E\ls R$^1$%
  \thanks{Email address for correspondence: parkerj@maths.ox.ac.uk}\ls \and
P.\ns J.\ns D\ls E\ls L\ls L\ls A\ls R$^1$}

\affiliation{$^1$
Mathematical Institute, University of Oxford, Andrew Wiles Building, Radcliffe Observatory Quarter, Woodstock Road, Oxford, U.K.,
OX2 6GG} 

\pubyear{2014}
\volume{}
\pagerange{}
\date{?; revised ?; accepted ?. - To be entered by editorial office}
\begin{document}

\maketitle

\begin{abstract}
We study Landau damping in the 1+1D Vlasov--Poisson system using a Fourier--Hermite spectral representation. 
We describe the propagation of free energy in phase space using forwards and backwards propagating Hermite modes recently developed for gyrokinetics [Schekochihin et al. (2014)]. 
The change in the electric field corresponds  to the net Hermite flux via a free energy evolution equation. 
In linear Landau damping, decay in the electric field corresponds to forward propagating Hermite modes; 
in nonlinear damping, the initial decay is followed by a growth phase characterised by the generation of backwards propagating Hermite modes 
by the nonlinear term. 
The free energy content of the backwards propagating modes increases exponentially until balancing that of the forward propagating modes. 
Thereafter there is no systematic net Hermite flux, so the electric field cannot decay and the nonlinearity effectively suppresses Landau damping. 
These simulations are performed using the fully-spectral 5D gyrokinetics code \sgk\ [Parker et al. 2014],  modified to solve the 1+1D Vlasov--Poisson system. 
This captures Landau damping via an iterated L\'enard--Bernstein collision operator or via Hou--Li filtering in velocity space. 
Therefore the code is applicable even in regimes where phase-mixing and filamentation are dominant.
\end{abstract}

\begin{PACS}
Authors should not enter PACS codes directly on the manuscript, as these must be chosen during the online submission process and will then be added during the typesetting process (see http://www.aip.org/pacs/ for the full list of PACS codes)
\end{PACS}


\newcommand{\agks}{\textsc{SpectroGK}}
\newcommand{\alfven}{Alfv\'en}
\renewcommand{\sgk}{\textsc{SpectroGK}}
\newcommand{\bvp}{boundary value problem}
\renewcommand{\cvk}{Case--Van Kampen}
\newcommand{\collop}{collision operator}
\newcommand{\DFT}{discrete Fourier transform}
\newcommand{\dop}{Kirkwood}
\newcommand{\FHH}{Fourier--Hankel--Hermite}
\newcommand{\FH}{Fourier--Hermite}
\newcommand{\fp}{Fokker--Planck}
\newcommand{\gkm}{GK-M}
\newcommand{\gkeqn}{gyrokinetic equation}
\newcommand{\gkpot}{gyrokinetic potential}
\newcommand{\gstwo}{\textsc{GS2}}
\newcommand{\hh}{Hankel--Hermite}
\newcommand{\ibvp}{initial boundary value problem}
\newcommand{\ic}{initial condition}
\newcommand{\ilb}{iterated L\'enard--Bernstein}
\newcommand{\LB}{L\'enard--Bernstein}
\newcommand{\lbd}{Kirkwood}
\newcommand{\ps}{pseudospectral}
\newcommand{\psly}{pseudospectrally}
\newcommand{\tkmk}{tokamak}
\newcommand{\toab}{third-order Adams--Bashforth}
\newcommand{\vk}{Van Kampen}
\newcommand{\Vk}{Van Kampen}
\renewcommand{\VP}{Vlasov--Poisson}
\newcommand{\vp}{Vlasov--Poisson}
\newcommand{\VPS}{Vlasov--Poisson system}
\newcommand{\vps}{Vlasov--Poisson system}


\newcommand{\Fig}{Figure}
\newcommand{\fig}{Figure}
\newcommand{\Sec}{Section}
\renewcommand{\sec}{section}
\newcommand{\tbl}{Table}

%
\renewcommand{\e}{\textrm{e}}
\renewcommand{\i}{\textrm{i}}
\renewcommand{\O}{O}

\newcommand{\fulldist}{F}
\newcommand{\oneddist}{\tilde{f}}

\renewcommand{\a}{\mathsf{a}}
\newcommand{\at}{\tilde{a}}
\newcommand{\atpm}{\tilde{a}^{\pm}}
\newcommand{\atmp}{\tilde{a}^{\mp}}
\newcommand{\atp}{\tilde{a}^{+}}
\newcommand{\atm}{\tilde{a}^{-}}
\newcommand{\bbR}{\mathbb{R}}
\renewcommand{\E}{\hat{E}}
\newcommand{\Ev}{\boldsymbol{E}}
\renewcommand{\f}{\hat{f}}
\renewcommand{\F}{{\cal F}}
\renewcommand{\Finv}{{\cal F}^{-1}}
\newcommand{\Fmat}{{\mathsf{F}}}
\newcommand{\Fmatinv}{{\mathsf{F}}^{-1}}
\newcommand{\I}{{\cal I}}
\newcommand{\indjj}{\I_{jj'}}
\renewcommand{\k}{\boldsymbol{k}}
\newcommand{\kperpv}{\k_{\perp}}
\renewcommand{\N}{{\cal N}}
\renewcommand{\n}{\boldsymbol{n}}
\newcommand{\Nt}{N_{\vartheta}}
\newcommand{\Nmat}{{\mathsf{N}}}
\newcommand{\RE}{{\cal R}}
\newcommand{\sgn}{\textrm{sgn}}
\newcommand{\sumi}[1]{\sum_{#1=0}^{\infty}}
\newcommand{\sumii}[1]{\sum_{#1=-\infty}^{\infty}}
\newcommand{\sumo}[2]{\sum_{#1=0}^{#2}}
\newcommand{\tsop}{\mathsf{L}}
\renewcommand{\tsop}{{\cal L}}
\renewcommand{\tsop}{{\cal T}}
\renewcommand{\tsop}{{\cal A}}
\renewcommand{\vpara}{{v}}
\newcommand{\xperp}{{x}_{\perp}}


\newcommand{\vpstilde}{\eqref{eq:TildeKinetic}--\eqref{eq:TildePoisson}}
\newcommand{\ThreeDVP}{\eqref{eq:ThreeDKinetic}--\eqref{eq:ThreeDPoisson}}
\newcommand{\ctseqns}{\eqref{eq:CtsFourierHermite}--\eqref{eq:CtsFourierHermitePoisson}}
\newcommand{\dvp}{\eqref{eq:DiscreteHermiteMomentKineticEquation}--\eqref{eq:DiscretePoissonEquation}}


\newcommand{\Ex}[1]{(\ref{#1})}
\newcommand{\Exac}[1]{(\ref{#1}\textit{-c})}
\newcommand{\dd}[2]{\frac{\mathrm{d} #1}{\mathrm{d} #2}}

\section{Introduction}
\label{sec:Introduction}

Many phenomena in astrophysical and fusion plasmas require a kinetic
rather than a fluid description. The fundamental quantity is the
distribution function $\fulldist(\x,\v,t)$ that determines the number
density of particles at position $\x$ moving with velocity $\v$ at
time $t$. Numerical computations of the evolution of the distribution function in its six-dimensional
phase space are thus very demanding on resources. Even simulations using the reduced
five-dimensional gyrokinetic formulation \citep[e.g. ][]{Howes06,Krommes12} are restricted to modest resolutions in each dimension. For example, simulations by \citet{Highcock11} with
the gyrokinetic code \texttt{GS2} \citep{GS2} used
$64\times32\times14$ points in physical space, and the equivalent of
$24\times16$ points in velocity space. This has motivated the
development of our fully spectral gyrokinetic code \sgk\ \citep{SGK}.
A fully spectral representation of the distribution function may
be expected to make optimal use of the limited number of degrees of freedom
possible in each dimension. Moreover, we have established that the spectral
representation in \sgk\ correctly captures Landau damping in a reduced linear problem
for ion temperature gradient driven instabilities \citep{Hypercollisions}.

The Vlassov--Poisson and Vlassov--Poisson--Fokker--Planck systems are canonical mathematical models for kinetic
phenomena in plasmas \citep[\eg ][]{Glassey96}. They describe a single active species moving
in a fixed background charge distribution, which we take to be uniform with unit density for convenience:
\begin{subequations}
\begin{align}
  \label{eq:TildeKinetic3D}
  \partial_t F + \mathbf{v} \cdot \nabla_\mathbf{x} F - \mathbf{E} \cdot \nabla_\mathbf{v} F
 =   \nu C[F],
  \\
  \mathbf{E} = - \nabla \Phi,
  \\
  \label{eq:TildePoisson3D}
  -\nabla^2 \Phi = 1 - \intii\d \mathbf{v} ~ F.
\end{align}
\end{subequations}
Here $\nabla_\mathbf{x}$ and $\nabla_\mathbf{v}$ denote gradients with respect to $\mathbf{x}$ and $\mathbf{v}$,
$\mathbf{E}$ is the electric field derived from the electrostatic potential $\Phi$. The \rhs\ of the Poisson equation (\ref{eq:TildePoisson3D}) contains the uniform background charge, and the charge due to the particles described by $F$. Physically,
this system describes electron-scale Langmuir turbulence in which the much more massive ions remain immobile. The \rhs\ of (\ref{eq:TildePoisson3D}) represents particle collisions with rate $\nu$ using a Fokker--Planck operator $C[F]$ as described below.

The collisionless ($\nu=0$) linearised 1+1 dimensional form of the \vps\ is a canonical mathematical
model for the ``filamentation'' or ``phase-mixing'' that forms infinitesimally fine scale structures in velocity space
due to the shearing effect of the particle streaming term
$\mathbf{v} \cdot \nabla_\mathbf{x} F$. \cite{Landau46} showed that this system supports
solutions in which the potential $\Phi$ decays exponentially in time
\citep[see also][]{Balescu63,Lifshitz81}. Landau obtained this solution via a Laplace transform in time,
and a deformation of the contour in the integral \Ex{eq:TildePoisson3D} defining $\Phi$ to ensure its analytic continuity. The distribution function $F$ does not itself decay in time, but instead develops ever finer scales. The Landau-damping
solution is thus not an eigenfunction of the collisionless system. Instead, the system has a continuous spectrum of real eigenvalues
associated with non-decaying singular eigenfunctions called \cvk\ modes \citep{VanKampen55,Case59}. The \cvk\ modes are complete, so the Landau-damped solution may be expressed as an infinite superposition of them.

\cite{Lenard58} showed that the velocity-space diffusion due to a  Fokker--Planck collision term $\nu C[F]$
with any strictly positive frequency $\nu > 0$ creates a smooth eigenfunction whose frequency and damping rate approached those
of the electric field in Landau's solution as $\nu \to 0$. \cite{Ng99,Ng04} showed that the collisionally regularised system in fact
has a discrete spectrum of  smooth eigenfunctions that form a complete set. A subset of these eigenfunctions have eigenvalues that tend to solutions of Landau's dispersion relation (see \sec~\ref{sec:LinearLandauDamping}) in the limit of vanishing collisions \citep{Ng06}.
As $\nu \to 0$, the smooth eigenfunctions develop boundary layers with widths proportional to the decay rates $|\gamma|$ of the modes, 
in which $F$ oscillates with a wavelength proportional to $\nu^{-1/4}$ \citep{Ng06}.

Any strictly positive collision frequency $\nu > 0$ thus suffices to change the spectrum of the
integro-differential equation from continuous to discrete.
However, for numerical computations it is necessary to make the velocity space discrete, either through introducing a grid, or through
representing $F$ as a finite sum of orthogonal functions (see below). Either approach introduces a finest resolved scale
in the velocity space. The transition from collisional to collisionless behaviour then occurs at 
some finite collision frequency $\nu^*$, for which the oscillations in the eigenfunctions are just coarse enough
to be resolved \citep{Hypercollisions}. This critical frequency $\nu^*$ is resolution-dependent, 
and tends to zero in the limit of infinite resolution.
When $\nu<\nu^*$, the system's behaviour is ``collisionless'': all eigenfunctions decay more slowly than the Landau rate.
When $\nu=0$ the eigenfunctions are discrete \cvk\ modes with real eigenvalues, and hence no decay.
One cannot obtain a discrete analogue of the Landau-damped solution from a linear combination of this finite
set of eigenfunctions. 

Only when $\nu \ge \nu^*$ do we find a decaying eigenmode in the discretised system that is resolved, and whose
decay rate approximates the Landau rate. Since  $\nu^*\to0$ as resolution increases, the decay rate of the slowest
decaying resolved mode tends to the Landau rate with increasing velocity space resolution. We have found that 
very accurate approximations to the Landau rate can be achieved with a very modest number, around 10, degrees of freedom in velocity space by using an iterated \LB\ collision operator \citep{Hypercollisions}.


The above discussion applies to linearised 1+1 dimensional kinetic
theory. The question of how to capture Landau damping numerically also
arises in the much more complex nonlinear and multi-dimensional
simulations of astrophysical and fusion plasmas, for which the
canonical model is the five-dimensional ``gyrokinetic'' system
\citep[see reviews by][]{Howes06,Krommes12}. Charged particles
in magnetic fields spiral around the field lines. When the magnetic
field is sufficiently strong, the fast timescales and short lengthscales
of this  ``gyromotion'' may be eliminated by averaging over the particle
gyrations. This averaging also reduces the dimensionality of phase
space from 6 to 5, with velocity space components  parallel and perpendicular to the magnetic field.
 
A linearized 1+1 dimensional electrostatic version of gyrokinetics for motions parallel to the magnetic field is obtained by integrating out the velocity dependence perpendicular to the magnetic field (as before) and taking the limit of vanishing perpendicular wavenumber.  
The perturbation $f$ of the ion distribution function $\oneddist = f + f_0$ relative to a Maxwellian $f_0$ then evolves according to the gyrokinetic system
\begin{subequations}
\begin{align}
  \label{eq:GyrokineticEquation}
  \pd{f}{t} + v\pd{f}{z} = E\pd{f_0}{v} ,
  \\
  E = -\pd{\Phi}{z} , 
  \\
  \label{eq:Quasineutrality}
  \Phi = \intii \d v ~ f,
\end{align}
\end{subequations}
where $z$ and $v$ are the physical space and velocity space coordinates parallel to the magnetic field.
The previous Poisson equation \eqref{eq:TildePoisson3D} has been replaced by the quasineutrality condition
\Ex{eq:Quasineutrality}. This condition holds at lengthscales much larger than the Debye length, and on slow
ion timescales for which the electrons may be assumed to adopt an instantaneous Maxwell--Boltzmann distribution
proportional to $\exp(\Phi/(k_B T_e))$.
The system \Exac{eq:GyrokineticEquation} is otherwise identical to the perturbative form of the \vps\ derived in \sec~\ref{sec:VlasovPoissonSystem}.

The availability of high quality numerical solutions to the 1+1
dimensional \vps\ makes it a good benchmark for novel numerical
algorithms. Due to the high dimensionality, gyrokinetic simulations
can typically only afford relatively coarse resolution in each
dimension.  
Our
fully spectral gyrokinetic code \sgk\ 
uses 
a spectral representation in each dimension 
to make optimal use of a limited number of degrees of freedom.  
The 1+1
dimensional version of  \sgk\ reduces to a \FH\ representation with
spectral filtering, or hypercollisions, to provide dissipation at the
smallest resolved scales in $z$ and $v$.

\cite{Burnett35,Burnett36} expanded the distribution
function as a sum of spherical harmonics multiplying polynomials that are orthogonal \wrt\ the Gaussian weight
function that appears in the Maxwell--Boltzmann equilibrium distribution. The Hermite polynomials have this
orthogonality property in one dimension \citep{AbramowitzStegun} so \cite{Grad49Note,Grad49Kinetic,GradHandbuch}
introduced sets of tensor Hermite polynomials as a Cartesian alternative to Burnett's expansion.
Both expansions conveniently
convert an integro-differential kinetic equation into an  infinite hierarchy of partial differential equations
for the expansion coefficients. 


The same expansion in Hermite polynomials for velocity space, and in Fourier modes for physical space,
was used in early simulations of the 1+1 dimensional \vps, such as by \cite{Armstrong67}, \cite{Grant67} and \cite{Joyce71}, albeit
with different forms of dissipation and,  inevitably, much lower resolution than is currently feasible.

However, as higher dimensional models became computationally feasible, interest turned instead
to particle-in-cell (PIC) methods. These represent the distribution function using a set of
macro-particles located at discrete points ($\x_i, \v_i)$ in phase space, each of which represents many physical ions or electrons \citep{Dawson83,Hockney88,Birdsall04}.
The method exploits the structure of the \lhs\ of the kinetic equation \Ex{eq:TildeKinetic3D}
as a derivative along a characteristic in phase space. A PIC method evolves the solution by propagating macro-particles along their characteristics, analogous to the Lagrangian formulation of fluid dynamics.
The representation of the continuous function $\fulldist(\x,\v,t)$ by a discrete set of $n$ macro-particles creates
an $O(1/\sqrt{n})$ sampling error, sometimes called ``shot noise'', that creates particular difficulties in the tail
of the distribution where $\fulldist$ is much smaller than its maximum value.

More recent  multi-dimensional gyrokinetics codes have returned to Eulerian representations
of velocity space, using fixed grids either for   parallel velocities \citep{GENE,GKW} or for pitch angles \citep{GYRO,GS2}.
However Hermite polynomials have been used to develop reduced kinetic \citep{Zocco11} 
and gyrofluid models \citep{Hammett93,Parker95},
as well as the Hermite index being used as a quantity of interest in characterizing velocity space behaviour \citep{Schekochihin14,Kanekar14,Plunk14}.
This has reignited interest in using Hermite polynomials for computation in new reduced-dimension \citep{DNA,Viriato} gyrokinetics codes, and the fully five-dimensional \sgk.

In this paper we illustrate the solution of the \vps\ with the Fourier--Hermite method using a modified version of the \sgk\ gyrokinetics code. 
We demonstrate that both hypercollisionality \citep[\eg][]{Joyce71} and a velocity space form of the
\cite{Hou07} spectral filter suffice to 
prevent recurrence
and result in correct calculations even in regimes where filementation and Landau damping are dominant.



We derive the one-dimensional system in \sec\ \ref{sec:VlasovPoissonSystem} and the \FH\ spectral representation in \sec~\ref{sec:SpectralRepresentation}, 
before testing the implementation by studying nonlinear Landau damping and the two stream instability in \sec~\ref{sec:NumericalResults}.


\section{The 1+1-dimensional Vlasov--Poisson system}
\label{sec:VlasovPoissonSystem}
\newcommand{\vpseqns}{\eqref{eq:KineticEquation}--\eqref{eq:PoissonEquation}}

We derive a 1+1-dimensional form of the Vlasov--Poisson system (1.1a-c) by seeking solutions 
which have spatial dependence in the $z$ direction only, 
and integrating over the velocity components $\vperpv$ perpendicular to the $z$ direction.
The reduced distribution function $\oneddist(z,v,t) = \int \d^2\vperpv~\fulldist(\x,\v,t)$
obeys the system
\begin{align}
  \label{eq:TildeKinetic}
  \pd{\oneddist}{t} + v\pd{\oneddist}{z} - E\pd{\oneddist}{v} = 
  \nu C[\oneddist],
  \\
  E = - \pd{\Phi}{z},
  \\
  \label{eq:TildePoisson}
  -\pdd{\Phi}{z} = 1 - \intii\d v ~ \oneddist,
\end{align}
where $v$ and $E$ are the components of $\v$ and $\Ev$ in the $z$ direction.
We take the domain of the problem to be $(z,v)\in\Omega\times\bbR$
where $\Omega=[0,L]$,
and consider periodic boundary conditions for $\oneddist$ in space, while in velocity space $\oneddist(z,v,t)\to0$ as $|v|\to\infty$.
Overall charge neutrality requires that the integral of \eqref{eq:TildePoisson} over $\Omega$ vanishes,
so that $E=-\tpd{\Phi}{z}$ is periodic on $\Omega$.
A detailed discussion of other boundary conditions may be found in \cite{Heath12}.

It is useful to consider the decomposition $\oneddist=f_0+f$ where $f_0(v)$ is a stationary, spatially-uniform distribution function satisfying
\begin{align}
  1 = \int_{-\infty}^{\infty} \d v ~ f_0 .
  \label{eq:EquilibriumIntegral}
\end{align}
Equations \eqref{eq:TildeKinetic}--\eqref{eq:TildePoisson} then become
\begin{align}
  \label{eq:KineticEquation}
  \pd{f}{t} + v\pd{f}{z} - E \pd{f}{v} = E\pd{f_0}{v} ,
  \\
  E = -\pd{\Phi}{z} , 
  \\
  \label{eq:PoissonEquation}
  \pdd{\Phi}{z} = \intii \d v ~ f,
\end{align}
and the overall charge neutrality condition becomes 
\begin{align}
  \int_{\Omega}\d z ~ \intii \d v ~ f(z,v,t) = 0. \label{eq:OverallNeutrality}
\end{align}
Equation \eqref{eq:KineticEquation} implies that this condition holds for all subsequent times,
provided it holds initially. The decomposed system \vpseqns\ holds for any decomposition satisfying \eqref{eq:EquilibriumIntegral}, 
but it is particularly useful for small perturbations about an equilibrium,
for when $|f|\ll|f_0|$ and $|\tpd{f}{v}|\ll|\tpd{f_0}{v}|$, 
the linearized system may be readily obtained by neglecting the single nonlinear term $-E\tpd{f}{v}$ in \eqref{eq:KineticEquation}.

For comparison, the corresponding linear 1+1 dimensional form of the gyrokinetic equations
originally targetted by \sgk\ is
\begin{subequations}
\begin{align}
  \pd{f}{t} + v\pd{f}{z} = E\pd{f_0}{v} ,
  \\
  E = -\pd{\Phi}{z},
  \\
 \Phi = \intii \d v ~ f.
\end{align}
\end{subequations}
The nonlinear
term in \eqref{eq:KineticEquation} does not appear even in the nonlinear gyrokinetic system
under the standard ordering that assumes lengthscales in $z$, oriented along the background
magnetic field, are much larger than lengthscales in the perpendicular directions.

The dissipationless Vlasov--Poisson system \vpstilde\ with $\nu=0$ conserves the total momentum $P$ and energy $H$
given by
\begin{align}
 \label{eq:Momentum}
  P &= \int_{\Omega}\d z \intii \d v ~ v \oneddist(z,v,t),
  \\
  \label{eq:TotalEnergy}
  H &= \frac{1}{2}\int_{\Omega}\d z \int \d v ~ v^2 \oneddist(z,v,t)
  + \frac{1}{2}\int_{\Omega}\d z ~ |E(z,t)|^2.
\end{align}
It also conserves a family of Casimir invariants
\begin{align}
  C &= \int_{\Omega} \d z \intii \d v ~ {\cal C}(\oneddist),
  \label{eq:CasimirInvariants}
\end{align}
where $ {\cal C}(\oneddist)$ is any function of $\oneddist$ alone. For example,
taking  $ {\cal C}(\oneddist) = \oneddist$ shows that the system conserves the
total particle number
\begin{align}
  \label{eq:Particles}
  N &= \int_{\Omega}\d z \intii \d v ~ \oneddist(z,v,t).
\end{align}
This ensures that the time evolution preserves the overall neutrality condition \eqref{eq:OverallNeutrality}.
Another conserved quantity of this form is the the spatially-integrated Boltzmann entropy
\begin{align}
  {\cal H}[\oneddist] = \int_{\Omega}\d z ~ \intii \d v ~ \oneddist \log \oneddist ,
  \label{eq:BoltzmannEntropy}
\end{align}
for which ${\cal C}(\oneddist) = \oneddist \log  \oneddist$.

In conjuction with the decomposition $\oneddist=f_0+f$  it is useful to consider the
spatially-integrated relative entropy \cite[\eg][]{Bardos93,Pauli00}
\begin{align}
  \RE[\oneddist|f_0] = \int_{\Omega}\d z ~\intii \d v ~ \oneddist\log\lp\frac{\oneddist}{f_0}\rp - \oneddist + f_0.
  \label{eq:RelativeEntropyExact}
\end{align}
This quantity has been employed to establish rigorous hydrodynamic limits of the Boltzmann equation
\citep{Lions01,Golse04}, to establish the existence and long-time attractive properties of steady
solutions of the Vlasov--Poisson system \citep{Bouchut93,Dolbeault99} and for other plasma
applications \citep{Krommes94,Hallatschek04}.
Expanding \eqref{eq:RelativeEntropyExact} for small perturbations $f = \oneddist - f_0 \ll f_0$ gives
\begin{align}
  \begin{split}
  \RE[\oneddist|f_0] 
  &= \int_{\Omega}\d z ~\intii \d v ~  \frac{f^2}{2f_0}  +  \O\lp f^3 \rp  ,
  \end{split}
  \label{eq:RelativeEntropyLinearized}
\end{align}
so the relative entropy provides a sign-definite quadratic measure of small perturbations from a uniform state.

However, the relative entropy is not itself conserved in a plasma (unlike the Boltzmann equation for neutral
particles) because $f_0$ couples to the electric field through $\tfd{f_0}{t}=-E\tpd{f_0}{v}$. Evaluating the relative entropy for the Maxwell--Boltzmann distribution $f_0=\upi^{-1/2}\e^{-v^2}$ in standard dimensionless variables gives
\begin{align}
  \begin{split}
  \RE[\oneddist|f_0] 
  &=\int_{\Omega}\d z~ \intii \d v ~ \oneddist\log \oneddist  - \oneddist\log f_0  - \oneddist + f_0
  \\
  &= {\cal H}[\oneddist] 
  + \int_{\Omega}\d z~ \intii \d v ~  \lp \frac{1}{2} \log \upi - 1\rp \oneddist +  v^2 \oneddist   + f_0 ,
  \end{split}
\end{align}
so the free energy defined
\begin{align}
  \begin{split}
  W_\mathrm{exact}
  &= \RE
  + \int_{\Omega}\d z ~ |E|^2
  \\
  &= {\cal H}[\oneddist] + \lp \frac{1}{2}\log\upi -1\rp N + 2H + \int_{\Omega}\d z ~1 , 
  \end{split}
  \label{eq:FreeEnergy}
\end{align}
is a conserved quantity. Approximating the relative entropy by its quadratic form \eqref{eq:RelativeEntropyLinearized}
gives the quadratic expression
\begin{align}
  \begin{split}
  W
  &= W_f + W_E
  \\
  W_f &= \int_{\Omega}\d z ~\intii \d v ~  \frac{f^2}{2f_0} 
  \\
  W_E &= \int_{\Omega} \d z ~  |E|^2 ,
  \end{split}
  \label{eq:FreeEnergyLinearized}
\end{align}
which may be expressed neatly in terms of the Fourier--Hermite expansion coefficients of $f$ using Parseval's theorems 
(see \sec~3).

\section{\FH\ spectral representation}
\label{sec:SpectralRepresentation}

We solve the \vps\ \vpseqns\ using a \FH\ representation.
In space we represent the distribution function with a Fourier series, properties of which are well known.
In velocity space we expand the distribution function as the sum of Hermite functions.
For this we introduce the Hermite polynomials $H_m$ and re-normalized Hermite functions $\phi^m$ defined by
\begin{align}\label{eq:HermFunDef}
  H_m(\vpara) = (-1)^m \e^{\vpara^2}\fd{^m}{\vpara^m}\lp \e^{-\vpara^2}\rp,
  \hspace{1cm}
  \phi^m(\vpara) = \frac{H_m(\vpara)}{\sqrt{2^m m!}}, 
\end{align}
for $m=0,1,2,\ldots$.
The Hermite functions $\phi^m$ are orthonormal with respect to the Maxwellian weight $\e^{-v^2}/\sqrt{\upi}$,
so that introducing the dual Hermite functions $\phi_m(v)=\e^{-v^2}\phi^m(v)/\sqrt{\pi}$
we have the bi-orthonormality condition
\begin{align}\label{eq:orthogonality}
  \int_{-\infty}^{\infty} \phi_n(v)\phi^m(v)\,\d v =
   \delta_{nm}, ~~~\forall n \ge 0,m \ge 0.
\end{align}
Each $\phi_m$ satisfies the velocity space boundary condition $\phi_m(v)\to0$ as $v\to\pm\infty$,
and the set of dual Hermite functions is complete for functions that are analytic on a strip in the complex $v$ plane and satisfy the
decay condition $|f(v)| < c_1 \e^{-c_2v^2/2}$ for some constants $c_1 > 0$ and $c_2 > 1$ \citep{Boyd01}.
The Hermite functions oscillate with characteristic wavelength $\upi (2/m)^{1/2}$ so that higher-order functions represent finer velocity space scales.
Neighbouring modes are related by the recurrence relation
\begin{align}\label{eq:rr}
 \vpara\phi_m(\vpara) = \sqrt{\frac{m+1}{2}}\phi_{m+1}(\vpara) + \sqrt{\frac{m}{2}}\phi_{m-1}(\vpara),
\end{align}
and velocity derivatives are related to a single neighbouring mode
\begin{align}
  \pd{\phi_m}{v} = -\sqrt{2(m+1)}\phi_{m+1},
  \hspace{1cm}
  \pd{\phi^m}{v} = \sqrt{2m}\phi^{m-1}.
  \label{eq:HermiteDerivatives}
\end{align}

We expand the distribution function in a series of dual Hermite functions
and, to obtain a finite sum, truncate after the first $N_m$ (slowest-oscillating) Hermite modes. 
This truncation is equivalent to a velocity space discretization on the roots of the Hermite polynomial $H_{N_m}$. The spacing between these roots decreases like $1/\sqrt{N_m}$ as $N_m\to\infty$.
However as the \VPS\ is linear in velocity space, so there is no need to explicitly discretize in $v$.

\subsection{Discretized system}
\label{sec:DiscretizedSystem}

We solve \vpseqns\ using the 
\FH\ representation
\begin{align}
  \label{eq:SemiDiscreteFinal}
  f(z,v,t) = \sum_{m=0}^{N_m-1}\sum_{j=-\Nt}^{\Nt}  a_{jm}(t) \e^{\i k_jz} \phi_m(v),
\end{align}
with inverse
\begin{align}
  \label{eq:SemiDiscreteInverseFinal}
  a_{jm}(t) = \frac{1}{L}\intii\d v\int_0^L \d z ~ f(z,v,t) \e^{-\i k_jz} \phi^m(v),
\end{align}
where $k_j=2\pi j/L$.
Thus the continuous function $f(z,v,t)$ is defined by a discrete, finite set of coefficients $a_{jm}$, which are implicitly a function of time.
Putting \eqref{eq:SemiDiscreteFinal} into the \VPS\ \vpseqns\ and applying the operator
\begin{align}
  \frac{1}{L}\intii\d v\int_0^L \d z ~ \e^{-\i k_jz} \phi^m(v),
\end{align}
we derive the discrete system
\begin{align}
  \label{eq:DiscreteHermiteMomentKineticEquation}
  \fd{a_{jm}}{t} + \i k_j \lp \sqrt{\frac{m+1}{2}}a_{j,m+1} + \sqrt{\frac{m}{2}}a_{j,m-1}\rp 
  + \Nmat_{jm}
  = -\sqrt{2}\E_j\delta_{m1}  ,
  \\
  \label{eq:DiscreteElectricField}
  \hat{E}_j = -\i k_j\hat{\Phi}_j , 
  \\
  \label{eq:DiscretePoissonEquation}
  -k_j^2 \hat{\Phi}_j = a_{j0},
\end{align}
where the nonlinear term $\Nmat$ is the discrete Fourier convolution
\begin{align}
  \Nmat_{jm} 
  = \sqrt{2m}\sum_{j'=-\Nt}^{\Nt} \hat{E}_{j'} a_{j-j',m-1},
  \label{eq:ModifiedNonlinearTerm}
\end{align}
and $\hat{E}$ and $\hat{\Phi}$ are the Fourier coefficients of $E$ and $\Phi$.
The system \dvp\ is an infinite moment hierarchy in Hermite space, 
where mode coupling results from
particle streaming and velocity derivatives via the relations \eqref{eq:rr} and \eqref{eq:HermiteDerivatives},
and from the nonlinear term through the electric field.
Because $\tpd{f_0}{v}$ can be expressed in a Hermite series, the \rhs\ of \eqref{eq:DiscreteHermiteMomentKineticEquation} is a finite number of source terms appearing at fixed Hermite modes $m$,
and the system is closed but for the term $a_{j,N_m+1}$ which appears in the particle streaming in the highest moment equation.
The system \dvp\ is also exactly the system obtained using a continuous \FH\ representation on an infinite spatial domain,
but restricted to the discrete wavenumbers $k_j$ and Hermite modes $m<N_m$. 

Calculating the nonlinear term \eqref{eq:ModifiedNonlinearTerm} directly for each grid point requires $\O(N_mN_k^2)$ operations,
but this is reduced to $\O(N_mN_k\log N_k)$ operations if it is calculated \psly, \ie\ via a grid in $z$-space using \DFT s.
For this we require a discrete version of \eqref{eq:SemiDiscreteFinal} and \eqref{eq:SemiDiscreteInverseFinal}.
Specifically
\eqref{eq:SemiDiscreteFinal} must hold at every grid point $z_l=lL/N_k$,
\begin{align}
  \label{eq:SemiDiscreteFinalGridPoints}
  f(z_l,v) = \sum_{m=0}^{N_m-1}\sum_{j=-\Nt}^{\Nt}  a_{jm} \e^{\i k_jz_l} \phi_m(v),
\end{align}
and we must replace the inverse $z$-integral \eqref{eq:SemiDiscreteInverseFinal} 
with a finite sum of Fourier modes.
The choice of uniform spatial grid $z_l=lL/N_k$ is 
motivated by the resolution of the identity for Fourier modes
\begin{align}
  \delta_{jj'} 
  = \frac{1}{N_k}\sum_{l=0}^{N_k-1} \e^{2\upi\i (j-j')l/N_k}  
  = \frac{1}{N_k}\sum_{l=0}^{N_k-1} \e^{\i (k_j-k_j')lL/N_k} ,
\end{align}
so that multiplying \eqref{eq:SemiDiscreteFinalGridPoints} 
by $\e^{-\i k_{j'}z_l}$ and summing over $l$ we obtain
\begin{align}
  a_{jm} = \frac{1}{N_k}\sum_{l=0}^{N_k-1}\intii\d v~f(z_l,v) \e^{-\i k_jz_l}\phi^m(v) ,
\end{align}
with the nonlinear term calculated as
\begin{align}
  \label{eq:DiscreteNonlinearTerm}
  \Nmat_{jm} =  -\i\sqrt{2m}~\Fmat_{jl} \lb \Fmatinv_{ln}\lp k_n\hat{\Phi}_n\rp \Fmatinv_{ln'}\lp a_{n',m-1}\rp\rb
  ,
\end{align}
where $\Fmat$ is the \DFT\ operator
\begin{align}
  \label{eq:DFTs}
  \Fmat_{jl} = \frac{1}{N_k}\sumo{l}{N_k-1} ~ \e^{-\i k_jz_l} 
  ,
  \hspace{1cm}
  \Fmatinv_{ln} = \sumo{n}{N_k-1} ~ \e^{\i k_nz_l} .
\end{align}

\subsubsection{Discrete free energy equations}
\label{sec:DiscreteFreeEnergyEquations}
We now obtain evolution equations for the quadratic free energies $W_E$, $W_f$ \eqref{eq:FreeEnergyLinearized}.
These have neat expressions in terms of the \FH\ coefficients obtained by inserting the spectral representation \eqref{eq:SemiDiscreteFinal} into \eqref{eq:FreeEnergyLinearized},
\begin{align}
  \begin{split}
    W_f &= \int_{\Omega}\d z ~\intii \d v ~  \frac{f^2}{2f_0} 
    = \frac{1}{2}\sum_{j=-\Nt}^{\Nt}\sum_{m=0}^{N_m} |a_{jm}|^2,
  \\
  W_E &= \int_{\Omega} \d z ~  |E|^2 
   = \sum_{j=-\Nt}^{\Nt} |\E_j|^2.
  \end{split}
\end{align}
Evolution equations for these are obtained by manipulating the moment equations \dvp.
Multiplying the $m=0$ moment equation by $a^*_{j0}/k_j^2$, using \eqref{eq:DiscreteElectricField} and \eqref{eq:DiscretePoissonEquation} to insert the electric field, adding the result to its complex conjugate and summing over $j$ we obtain 
\begin{align}
  \fd{W_E}{t}
  + {\cal F} = 0,
  \label{eq:WEeqn}
\end{align}
where 
\begin{align}
{\cal F} = \sqrt{2}\ \Real\lp \sum_{j=-\Nt}^{\Nt} \frac{\i a_{j0}^*a_{j1}}{k_j}\rp,
\end{align}
is a flux between the first two Hermite moments.
Similarly, multiplying \eqref{eq:DiscreteHermiteMomentKineticEquation} by $a_{jm}^*$, adding the result to its complex conjugate, 
and summing over all $m$ and $j$, we obtain 
\begin{align}
  \fd{W_f}{t}
  - {\cal F} 
  + {\cal T}
  = 
  {\cal C},
  \label{eq:Wfeqn}
\end{align}
where
\begin{align}
{\cal T}
= \Real \lp\sum_{j=-\Nt}^{\Nt}\sum_{m=0}^{N_m} a_{jm}^* \Nmat_{jm} \rp,
\\
{\cal C}
= \Real \lp \sum_{j=-\Nt}^{\Nt}\sum_{m=0}^{N_m} a_{jm}^* C_{jm} \rp,
\end{align}
are the nonlinear term viewed as a free energy source and the collisional sink of free energy.

Combining \eqref{eq:WEeqn} and \eqref{eq:Wfeqn} we have the global budget equation
\begin{align}
  \fd{}{t}\lp W_f + W_E \rp
  +
  {\cal T}
  = 
  {\cal C}
  \label{eq:GlobalBudgetEquation}
\end{align}
In the linear case 
without collisions
(${\cal T}={\cal C}=0$),
this expresses the exact discrete conservation of the truncated free energy $W=W_f+W_E$.
Nonlinearly, 
${\cal T}$
accounts for the terms omitted in the linearization of free energy $W_{\mathrm{exact}}$ \eqref{eq:FreeEnergy}. 

Finally, equation \eqref{eq:WEeqn} shows that $W_E$ only changes through the flux term ${\cal F}$.
We show later (\sec~\ref{sec:HermiteFlux}) that this term represents net the flux of free energy through Hermite space.
Therefore the electric field only decays or grows as the result of net forwards or backwards Hermite flux respectively.

\subsection{Algorithm description}

We now discuss details of the algorithm 
to solve \dvp.
These equations may be combined and written schematically as
\begin{align}
  \fd{\a}{t} = \tsop\left[\a\right] , 
  \label{eq:dadt=Aa}
\end{align}
where 
$\a$ denotes the 
coefficients $a_{jm}$. 
The algorithm has two main steps: 
forming the \rhs\ $\tsop$, 
and numerically integrating $\tsop$ in time to find the coefficients $\a$. 
Time integration is performed by third-order Adams--Bashforth as discussed in \sec\ \ref{sec:TimeIntegration}.
To form $\tsop$, 
we must 
determine the electric field,
calculate the nonlinear term,
and properly treat the fine scales that appear in space and velocity space due to the nonlinear term and particle streaming respectively. 
These are discussed in \sec s \ref{sec:FieldSolve}--\ref{sec:Recurrence}.
We consider the parallelization and communication patterns in the code in \sec~\ref{sec:sgk}.

As noted in \sec~\ref{sec:VlasovPoissonSystem}, the \vps\ is very similar to the long-wavelength limit of gyrokinetics.
Therefore we solve using \sgk, a full 5D gyrokinetics code which also features a mode for solving in the (1+1)D long wavelength limit.
To solve the \vps\ rather than gyrokinetics, we make two modifications to \sgk\ 
which are described in \sec\ \ref{sec:sgk}
along with details of the parallelization scheme.
These changes 
do not affect the key algorithms or structure of the code
and so  
the test problems in \sec\ \ref{sec:NumericalResults} act to validate \sgk.

\subsubsection{Time integration}
\label{sec:TimeIntegration}
The solution of \eqref{eq:dadt=Aa}
is approximated using the explicit third-order Adams--Bashforth scheme,
\begin{align}
  \label{eq:AB3}
  \a^{i+1} &= \a^{i} + \Delta t\left( \frac{23}{12} \tsop\left[ \a^i \right] 
  - \frac43 \tsop\left[\a^{i-1}\right] 
  + \frac{5}{12} \tsop\left[ \a^{i-2}\right] \right) ,
\end{align} 
where 
$\a^i$ denotes the 
coefficients $a_{jm}$ at the $i$th time level, 
and $\Delta t$ is the timestep.
\sgk\ also implements a variable time-spacing version of this formula to allow changing the timestep during execution. 

The advantages of third-order Adams--Bashforth are given in \cite{Durran91,Durran99}.
It is stable and accurate for non-dissipative wave phenomena, with fourth-order errors in amplitude and wave speed.
It is also appropriate for problems like ours where the calculation of $\tsop$ dominates the computation work.
\cite{Durran91} defines the ``efficiency factor'' of an integration scheme, the maximum stable timestep in an oscillatory test problem divided by the number of evaluations of $\tsop$ per timestep.
By this measure Adams--Bashforth is the most efficient third-order scheme, as while it has a smaller stable timestep than other schemes such as Runge--Kutta, it only requires one $\tsop$ evaluation per timestep.

The main disadvantage of \toab\ is that two previous timestep operators must be stored.
This is potentially a problem as the operators are each the same size as the total problem size;
however this is not limiting in the (1+1)D \vps.

One must also ensure that the scheme is third order accurate: 
as \eqref{eq:AB3} uses three past values at each timestep,
we must amend the scheme for the first and second timesteps where fewer past values are available. 
For these two timesteps we use explicit Euler and second-order Adams--Bashforth timestep respectively.
In principle, the use of explicit Euler makes the global time integration error second order (the local truncation error is first order, but as explicit Euler is only used once, the error does not accumulate over $\sim1/\Delta t$ timesteps).
In practice however, the error in explicit Euler time integration is insignificant relative to other errors.

\subsubsection{Field solve}
\label{sec:FieldSolve}

The electric field for use in $\tsop$ is readily obtained from \eqref{eq:DiscreteElectricField} and \eqref{eq:DiscretePoissonEquation}:
if $k_j=0$ then $E_j=0$,
otherwise
$E_j = \i a_{j0}/k_j$.
The truncated Hermite expansion is equivalent to a discretization in $v$-space on a grid the roots of $H_{N_m}$,
and so $a_{j0}$ is equal to the zeroth moment of the distribution function obtained via $(2N_m-1)$th order Gauss--Hermite quadrature.
However unlike $v$-space discretizations which require a sum over all grid points, 
the evaluation of the field in Hermite space requires only the coefficient $a_{j0}$.
This has communication benefits discussed in \sec~\ref{sec:sgk}.

\subsubsection{Nonlinear term}

We calculate the nonlinear term \eqref{eq:DiscreteNonlinearTerm}
using the product of \DFT s.
As derived in \sec\ \ref{sec:DiscretizedSystem}, 
the discrete wavenumbers and $z$-grid are
\begin{align}
  k_j=2\upi j/L,
  \hspace{1cm}
  z_l = lL / N_k.
\end{align}
We use the FFTW library \citep{FFTW3}, which implements unnormalized \DFT s, \ie\ \eqref{eq:DFTs} but without the factor $1/N_k$ in the first transform.
Note that with the normalization in \eqref{eq:DFTs} the forward transform of $\e^{\i k_jz_l}$ has magnitude one.

With this \ps\ approach, the problem of Fourier aliasing occurs.
The product of inverse transforms in \eqref{eq:DiscreteNonlinearTerm} is a sum over Fourier modes $\e^{\i(k_n+k_{n'})z_l}$.
When $k_n+k_{n'}>k_{\Nt}$ the largest wavenumber in the simulation, its contribution to the nonlinear term should be neglected.
However as the \DFT\ is periodic, this mode contributes to the Fourier transform at the wavenumber $k_n+k_{n'}-k_{\Nt}$.
This spurious appearance of high wavenumber contributions in the low wavenumbers is called aliasing.
Dealiasing is often performed by the two-thirds rule, a Fourier filter where the Fourier coefficients for the highest third of wavenumbers are set to zero before the nonlinearity is calculated.
For quadratic nonlinearities 
such as the Fourier convolution \eqref{eq:ModifiedNonlinearTerm}, 
this removes all spurious modes as all modes with $|k_n+k_{n'}|>k_{\Nt}$ remap onto wavenumbers that are neglected from the simulation \citep{Orszag71,Boyd01}.

The two-thirds rule works well, but costs one-third of the resolution.
In addition, the sharp transition from unmodified coefficients to zeroed coefficients acts like a reflecting boundary condition in wavenumber space.
This causes error in the highest resolved wavenumbers unless the Fourier coefficients are negligible at the point the filter is imposed.  
To ensure this is the case we multiply the distribution by the Hou--Li filter \citep{Hou07}
\begin{align}
  \label{eq:HouLiK}
  \exp\lp -36(|k|/\max(k))^{36}\rp ,
\end{align}
before the calculation of the nonlinear term.
In fact, this damps the highest modes so strongly
that both smoothing and dealiasing is effected.
Indeed using the Hou--Li filter instead of two-thirds filtering allows one to keep 12--15\% more Fourier modes \citep{Hou07}.

\subsubsection{Recurrence}
\label{sec:Recurrence}
The particle streaming term $v\tpd{f}{z}$ is a phase space shear that causes infinitesimally small scale structure to form in velocity space.  
For any discretization, these structures become finer than grid-scale after some finite, resolution-dependent time.  
The discretization fails to capture this structure and is invalid after this time.

In Hermite space, particle streaming corresponds to nearest-neighbour mode coupling due to the recurrence relation \eqref{eq:rr},
where each mode represents the velocity space scale $\sim\upi (2/m)^{1/2}$.
The moment hierarchy is not closed as the $m$th equation depends on the $(m+1)$th Hermite mode.
To truncate the hierarchy we set $a_{jm}=0$ for all $m\geq N_m$, which determines the finest resolved velocity scale, $\sim\upi (2/N_m)^{1/2}$. 
In the highest moment equation, truncation forces $a_{jN_m}=0$, \ie\ forces streaming to finer scales to vanish.
Thus $a_{N_m}=0$ is like a hard-wall boundary condition for quantities like $|a_{jm}|^2$, the contribution of \FH\ mode $(j,m)$ to the free energy 
\eqref{eq:FreeEnergyLinearized}.
In the linear system one may observe an initial forward flux of free energy in Hermite space from large to small scales, a reflection at the point $m=N_m$, and a subsequence backwards flow of free energy. 
Recurrence occurs when this spuriously reflected free energy reappears in the low moments that represent the physical quantities.
Recurrence is so called because this returning free energy causes a sudden increase in the magnitude of a previously decaying quantity, such as the Landau-damped electrostatic potential.   

\cite{Schekochihin14} showed that the Hermite coefficients decompose as the sum of forward and backward propagating modes.
Linearly these modes decouple apart from at the boundary $m=N_m$, where incoming forward modes excite the backwards propagating modes which cause recurrence.
Recurrence is therefore prevented by damping the distribution function with a filter or collision operator so that $a_{jN_m}=0$. 
For low resolution ($N_m\sim 10$), the damping must be smooth across Hermite space (\ie\ algebraic in $m$), 
and hypercollisional operators such as the iterated L\'enard--Bernstein operator \citep{Lenard58}
\begin{align}
  - \nu (m/N_m)^{\alpha} a_{jm} , 
\end{align}
are effective.
In linear simulations, the critical collision frequency $\nu^*$ decreases with resolution as $\nu^*\sim 1/N_m^{\alpha}$.
Thus the factor $1/N_m^{\alpha}$ allows a constant $\nu$ to be used for different resolutions. 
We have also used this operator for higher resolution and nonlinear simulations of the \VPS; 
however the collision frequency $\nu$ must be tuned with resolution making convergence studies awkward.
Instead we smooth velocity space with the Hou--Li-type filter 
\begin{align}
  \label{eq:HouLiM}
  \exp\lp -36(m/(N_m-1))^{36}\rp . 
\end{align}
While this is too sharp for low resolutions (small $N_m$),
it is sufficiently smooth for high resolutions,
and yields exponential convergence as demonstrated in \sec~\ref{sec:Convergence}.

\subsubsection{\sgk}
\label{sec:sgk}

The system \dvp\ is implemented using a reduced-dimension version of the gyrokinetics code \sgk.
For \sgk\ to solve the \vps, we make two modifications.
Firstly we replace the gyrokinetic quasineutrality condition \eqref{eq:Quasineutrality} with Gauss' law \eqref{eq:PoissonEquation}. 
As the code is spectral in Fourier space, this change is trivial.
Secondly we insert the nonlinear term \eqref{eq:DiscreteNonlinearTerm} which is absent in gyrokinetics (at the order solved by \sgk).

The \sgk\ parallelization scheme is described in detail in \cite{SGK}.
The basic idea is to divide the five-dimensional distribution function evenly among processors, while forcing all the parallel wavenumbers for a given phase space point to be local to a processor.
Each processor also has a copy of the smaller, three-dimensional electromagnetic field.
For the (1+1)D \vps\ we therefore parallelize over $m$ while keeping $k$ local.
This is optimal.
The main potential source of communication is in the nonlinear term \eqref{eq:DiscreteNonlinearTerm} where the Fourier transforms are sums over all $k$ at a fixed $m$. 
By keeping all $k$ on processor, communication is entirely eliminated from this term.

Besides the nonlinear term, the equations are largely local in phase space, and only two parts require communication.
Firstly the mode coupling and nonlinear term in \eqref{eq:DiscreteHermiteMomentKineticEquation} require communication of neighbouring Hermite modes when these fall on different processors.
This is small point-to-point communication and is entirely vectorized in $k$.
Secondly, in the discrete Poisson equation \eqref{eq:DiscretePoissonEquation} the electric field is calculated on the processor which holds $a_{j0}$ and is then sent to all other processors.
Here the Hermite spectral method is preferable to velocity grid discretizations, where contributions to a sum approximating the integral in \eqref{eq:PoissonEquation} must be sent and received by each processor before the broadcast of the electric field.

\section{Numerical results}
\label{sec:NumericalResults}

We present the solution of the linearized and nonlinear \vps\ with \sgk. 
In the absence of exact nonlinear solutions, 
we benchmark \sgk\ 
against other codes
under grid refinement,
for two standard nonlinear test problems,
nonlinear Landau damping
and
the two stream instability.

Following convention,
the initial conditions unless otherwise stated are
\begin{align}
  \label{eq:InitialConditions}
  \oneddist(v) = f_0(v) + A\cos(kz)f_0(v)  , 
\end{align}
where $A=0.5$, $k=0.5$,
and we use a box length $L=4\upi$ so that wavenumbers are half-integers.


\subsection{Linear Landau damping}
\label{sec:LinearLandauDamping}

\begin{figure}
  \centering
  \subfigure[Linear\label{fig:LinearLandauDamping}]{
  \includegraphics[trim=3.5cm 9.5cm 4.0cm 9.5cm,width=0.48\textwidth,clip]{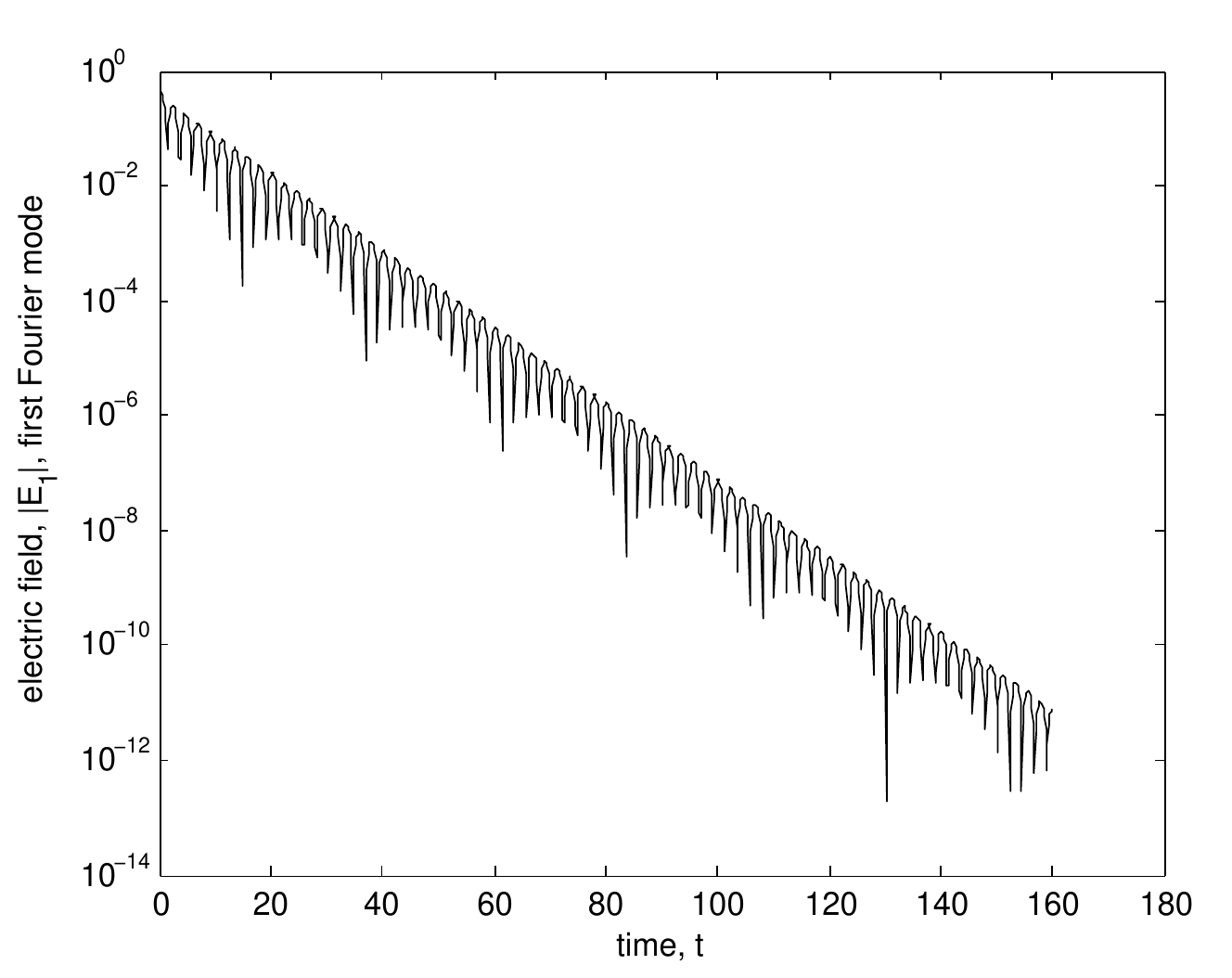}
  }
  \subfigure[Nonlinear\label{fig:NonlinearLandauDamping}]{
  \includegraphics[trim=3.5cm 9.5cm 4.0cm 9.5cm,width=0.48\textwidth,clip]{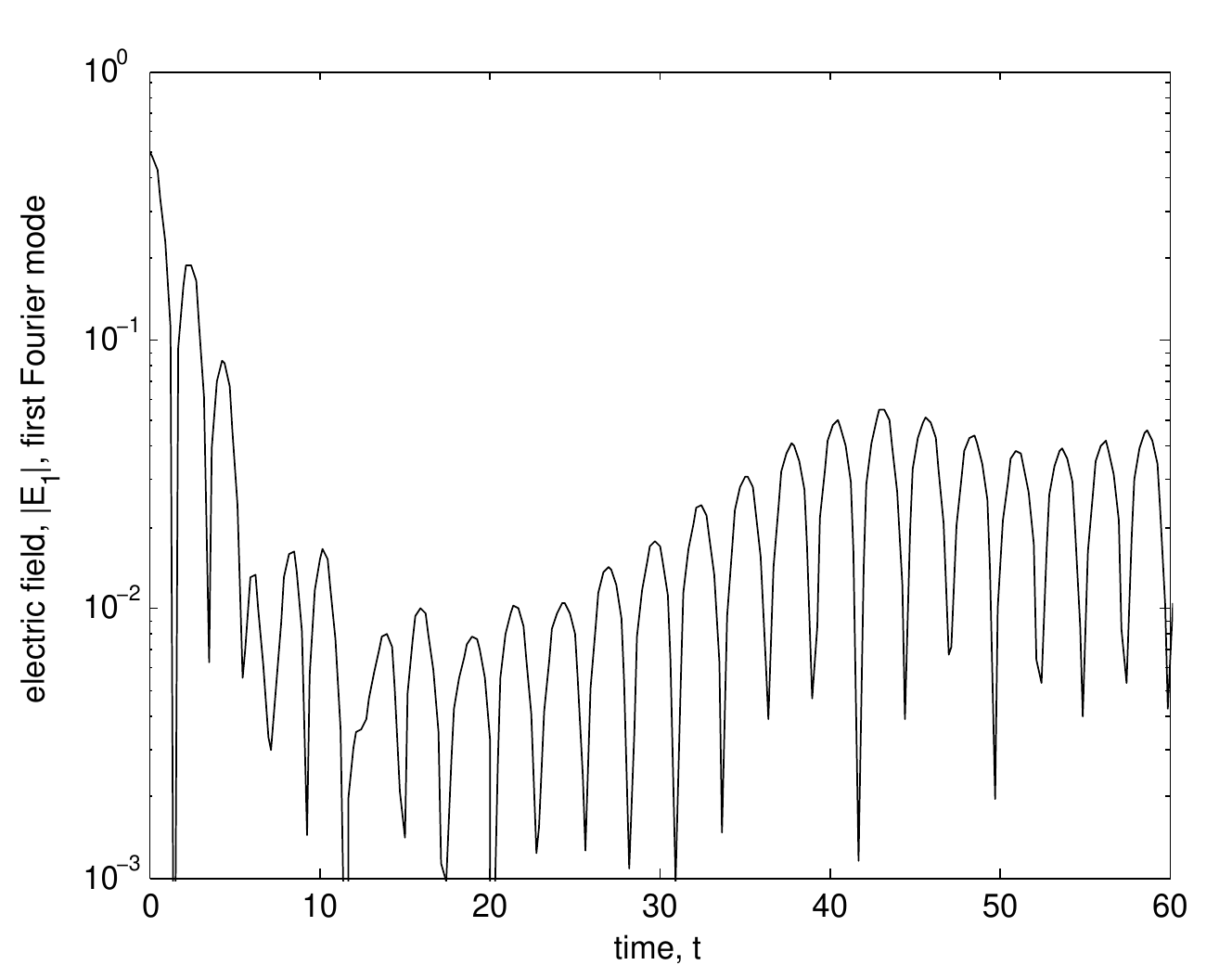}
  }
  \caption{The first electric field Fourier mode versus time. 
  \label{fig:ElectricFieldFit}}
\end{figure}

The linearized system obtained by neglecting the term $-E\tpd{f}{v}$ in \eqref{eq:KineticEquation} 
exhibits Landau damping,
according to the dispersion relation
    \begin{align}
  D(\omega) = ik^3 + 2ik + 2i\omega  Z(\omega/k) = 0,
    \end{align}
    where $Z(\zeta)=\pi^{-1/2}\int e^{-v^2}/(v-\zeta)\,\d v$ is the plasma dispersion function \citep{Fried61}.
    This has the property that roots $\omega$ appear in frequency pairs $\pm\omega_R+i\gamma$ corresponding to left and right travelling waves. 
    Thus there are two dominant modes with equal growth rate and opposite frequency.

    The discretized system is equivalent to 
    the matrix initial value problem
\begin{align}
  \pd{f}{t} = Mf,
\end{align}
with timestep operator $M$. 
The exact solution is obtained in terms of eigenvalues $i\omega$ and eigenvectors $x$ of $M$,
\begin{align}
  f(z,v,t) = \sum_{l=1}^{N_m} \alpha_l x_{l} e^{i(k_jz-\omega_lt)}.
\end{align}
The coefficients $\alpha_l=y_l^*\beta/(y_l^*x_l)$ where $y_l$ is the $l$th eigenvector of the adjoint matrix $M^*$, and $\beta$ is the initial value of the distribution function in velocity space.  
The dominant eigenvalues of $M$ also occur in the frequency pair $\pm\omega_R+i\gamma$.
Generic initial conditions excite both dominant eigenmodes and after sufficient long time leads to an oscillation with frequency $2\omega_R$ via the interference pattern of the two modes
\begin{align}
  |\varphi|^2 \sim |\varphi_1e^{-i\omega_Rt + \gamma t} +\varphi_2 e^{i\omega_Rt + \gamma t}|^2
  = \left( |\varphi_1|^2 + |\varphi_2|^2 + 2\Real(\varphi^*_1\varphi_2e^{2i\omega_Rt}) \right) e^{2\gamma t} .
  \label{eq:TwoModeInterference}
\end{align}
Thus we may also determine the frequency of the dominant mode from the initial value problem.

\begin{figure}
  \centering
  \subfigure[All modes excited.\label{fig:LinearFreeEnergy}]{
  \includegraphics[trim=3.5cm 9.5cm 4.0cm 9.5cm,width=0.48\textwidth,clip]{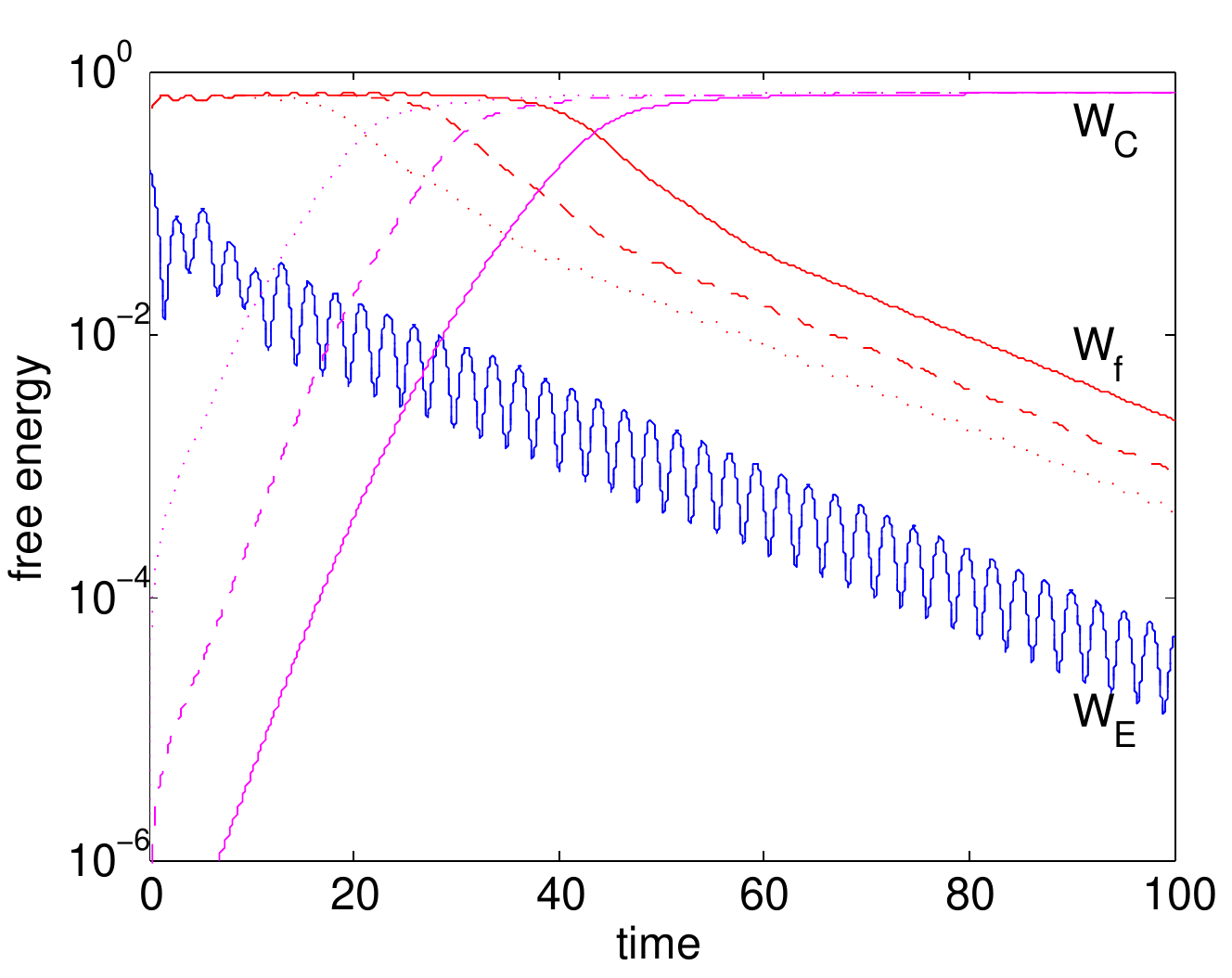}
  }
  \subfigure[One dominant mode removed.\label{fig:LinearFreeEnergyModeRemoved}]{
  \includegraphics[trim=3.5cm 9.5cm 4.0cm 9.5cm,width=0.48\textwidth,clip]{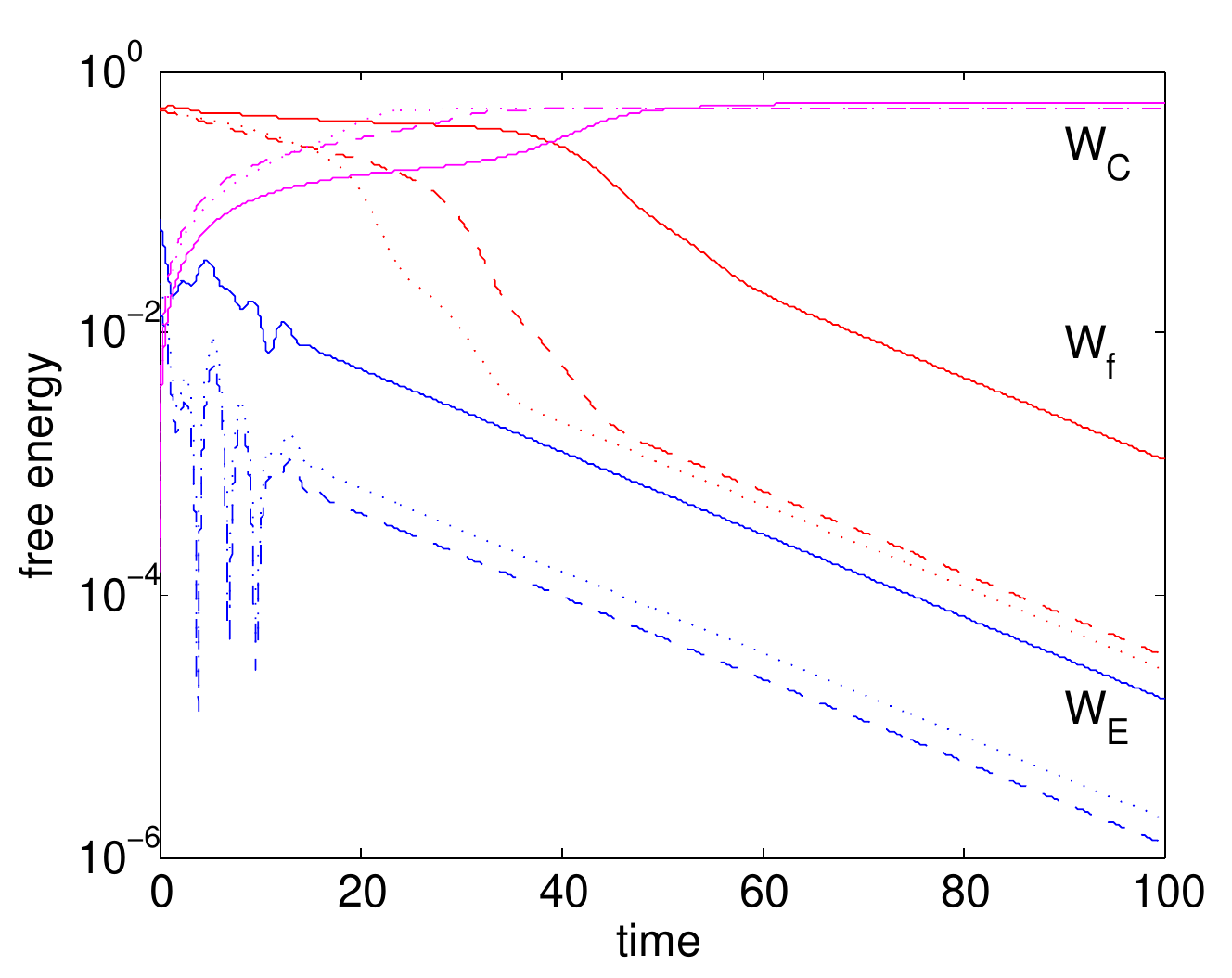}
  }
  \caption{Free energy contributions in the linear case. Line style corresponds to resolution: $N_m=128$ (dotted), $N_m=256$ (dashed), $N=512$ (unbroken).  
  \label{fig:LinearFreeEnergies}}
\end{figure}

In \fig~\ref{fig:LinearLandauDamping} we plot the $k=0.5$ mode for a linear simulation with the Hou--Li filter \eqref{eq:HouLiM} applied in Hermite space.
The frequency $\omega_R=1.415$ and damping rate $\gamma=-0.153$ are in agreement with \cite{Cheng76}. 
In \fig~\ref{fig:LinearFreeEnergy} we plot the corresponding free energy time trace: the free energies of the electric field $W_E$ and distribution function $W_f$, and the time-integrated collisional sink ${\cal S}$.
After an initial transient, by $t=20$ the system enters the collisionless regime in which
$W_E$ decays at the Landau rate with a superimposed oscillation due to two mode interference, 
while $W_f$ oscillates in antiphase to $W_E$ without decaying.
This is reminiscent of Landau's Laplace transform solution \citep{Landau46}, and in contrast to the regime one might expect where $W_E$ would not decay linearly until $W_f$ was also linearly decaying.

The collisionless regime lasts until the free energy in the distribution function reaches collisional scales and is damped.
The system enters the asymptotic regime where both $W_E$ and $W_f$ decay at the Landau rate.
This behaviour is the same as that described by \cite{Ng99} for systems with weak \fp\ collisions. 
The time for free energy to reach collisional scales increases with resolution so that the onset of the eigenmode regime may be delayed by increasing the number of Hermite modes, as shown in \fig~\ref{fig:LinearFreeEnergy}.

Both $W_E$ and $W_f$ oscillate as they decay due to two-mode interference as in \eqref{eq:TwoModeInterference}.
We obtain smooth decay by choosing initial conditions which do not project onto one of the dominant modes. 
Now the free energy traces are smooth (\fig~\ref{fig:LinearFreeEnergyModeRemoved}) with the dominant mode a single travelling wave.


\subsection{Nonlinear Landau damping}
\label{sec:NonlinearLandauDamping}
\label{sec:NonlinearLandauDampingSubsection}

We now present simulations of nonlinear Landau damping, 
which has been treated extensively in the literature 
\citep[\eg][]{Grant67,Cheng76,Zaki88,Nakamura99,Filbet01,Zhou01,Heath12}.
We benchmark \sgk\ by reproducing known results and demonstrating convergence, before giving a description of the system via its Hermite space behaviour.

\begin{figure}
  \centering
  \subfigure[$k=0.5$]{\includegraphics[trim=3.5cm 9.5cm 4.0cm 9.5cm,width=0.49\textwidth,clip]{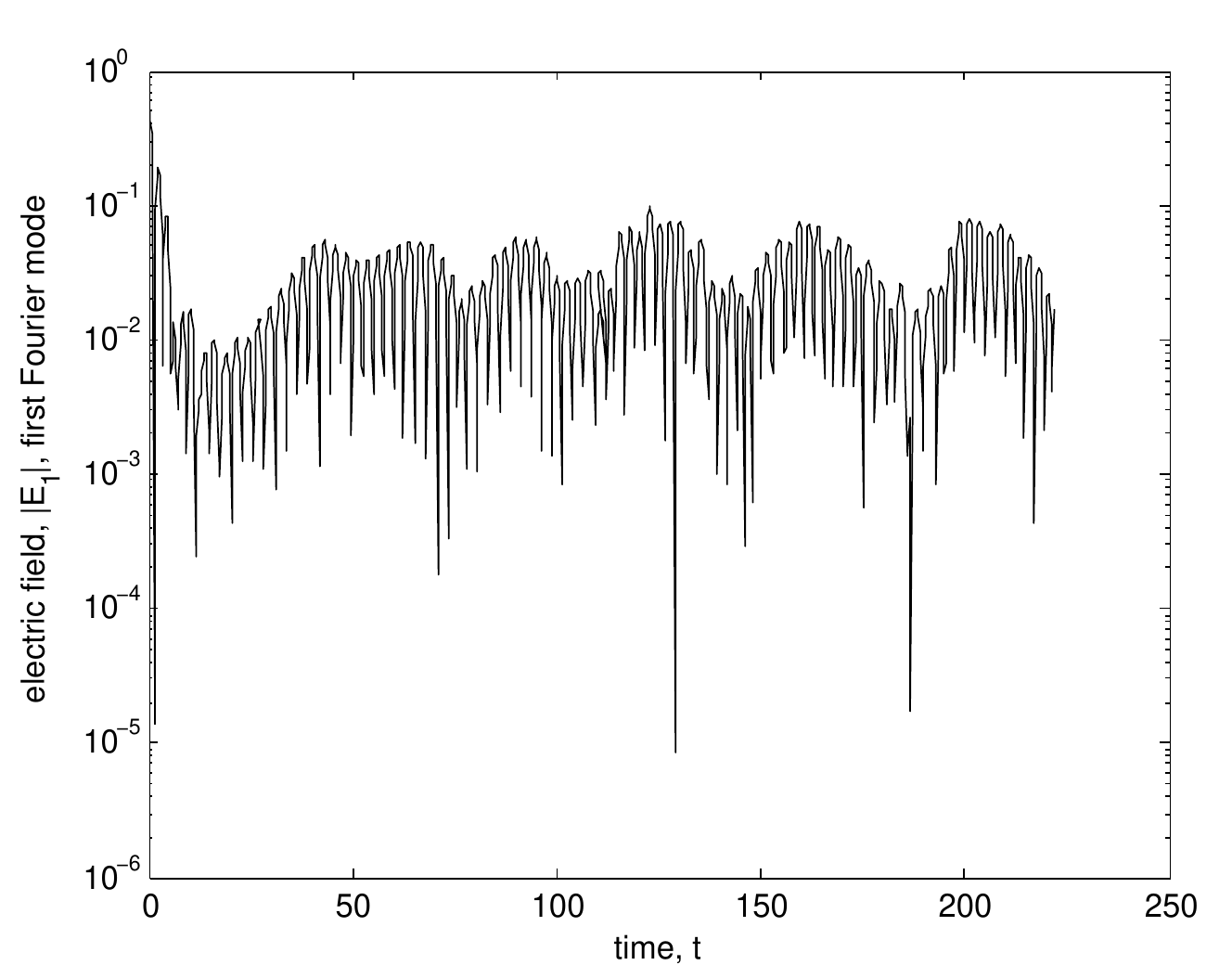}}
  \subfigure[$k=1.0$]{\includegraphics[trim=3.5cm 9.5cm 4.0cm 9.5cm,width=0.49\textwidth,clip]{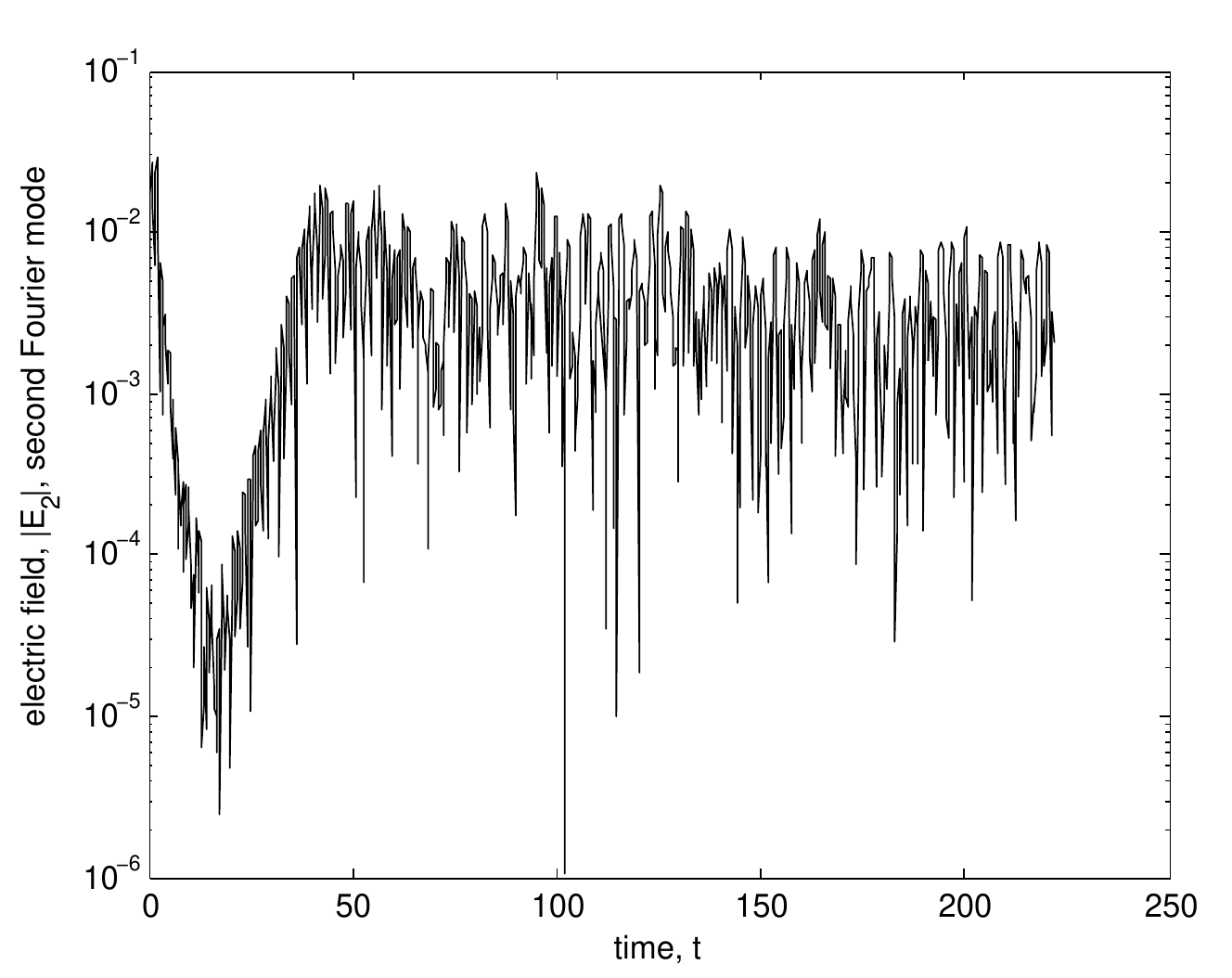}}
  \subfigure[$k=1.5$]{\includegraphics[trim=3.5cm 9.5cm 4.0cm 9.5cm,width=0.49\textwidth,clip]{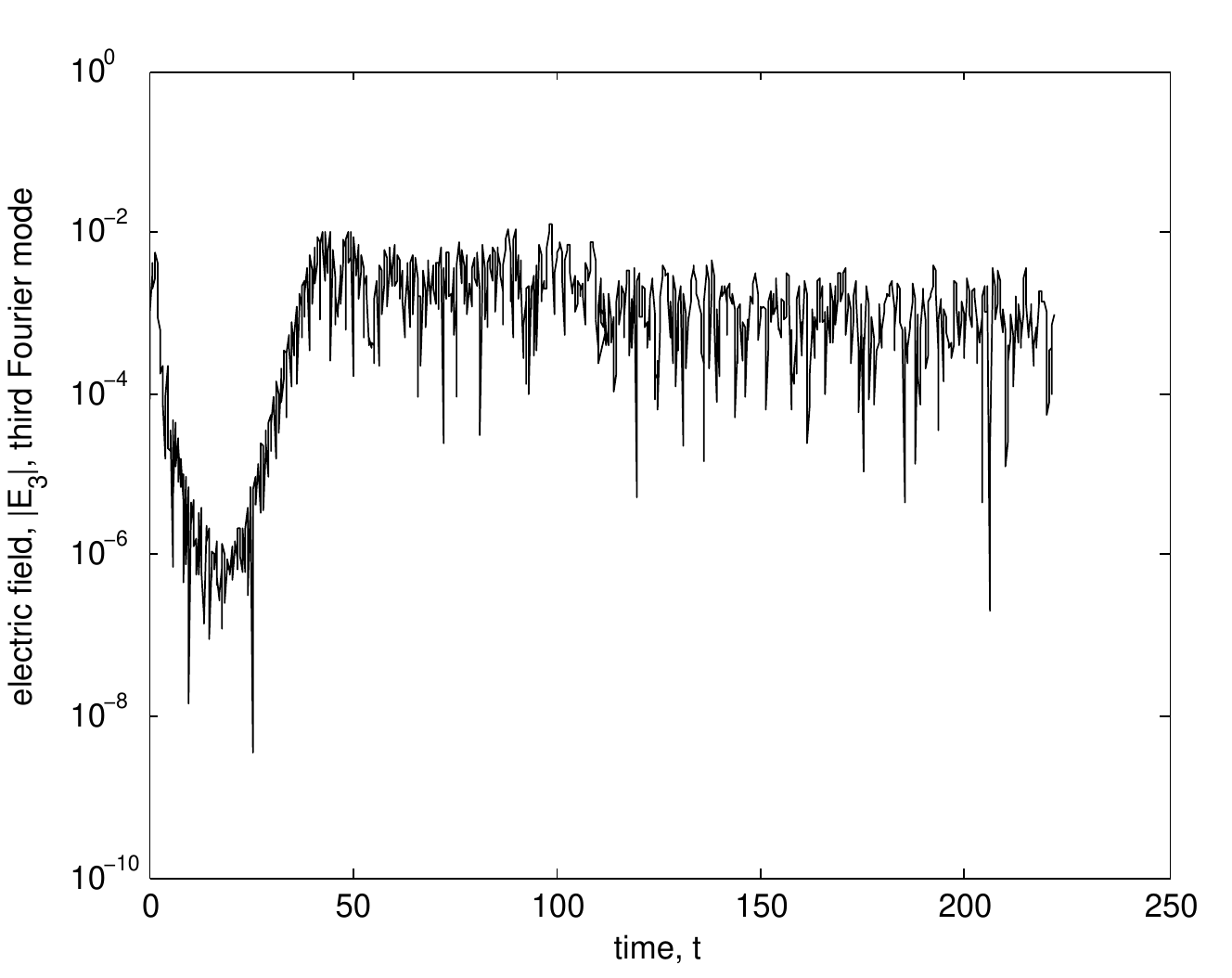}}
  \subfigure[$k=2.0$]{\includegraphics[trim=3.5cm 9.5cm 4.0cm 9.5cm,width=0.49\textwidth,clip]{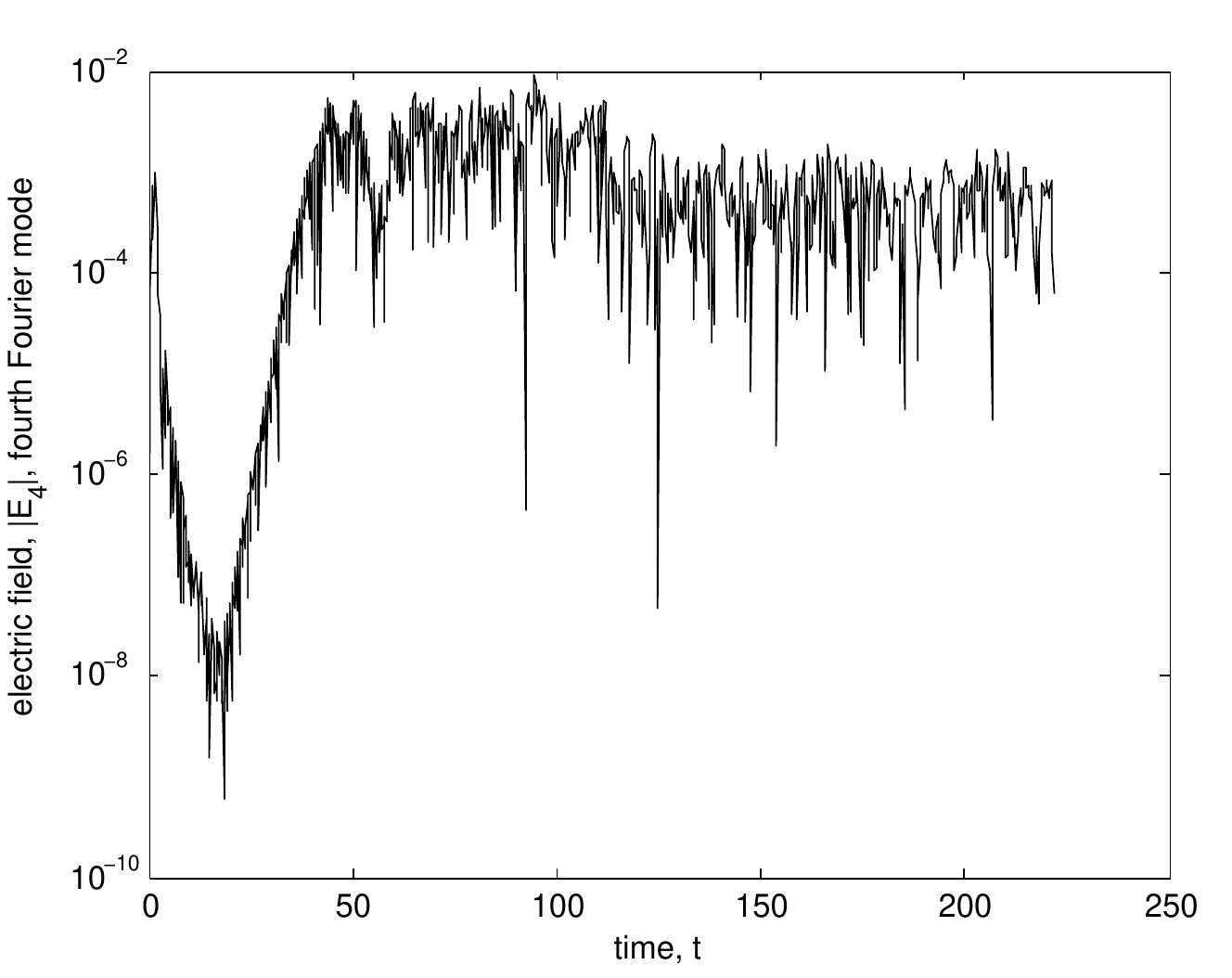}}
  \caption{Electric field versus time for the first four Fourier modes.
  \label{fig:ElectricFieldVsTime}}
\end{figure}

The electric field for the dominant Fourier mode at early times is given in \fig~\ref{fig:NonlinearLandauDamping},
and longer time traces of the four lowest modes are plotted in \fig~\ref{fig:ElectricFieldVsTime}. 
These are in agreement with previous simulations \citep[\eg][]{Heath12}.

\begin{figure}
  \centering
  \subfigure[$t=1$]{\includegraphics[trim=1.5cm 0.5cm 1.5cm 1.0cm,width=0.49\textwidth,clip]{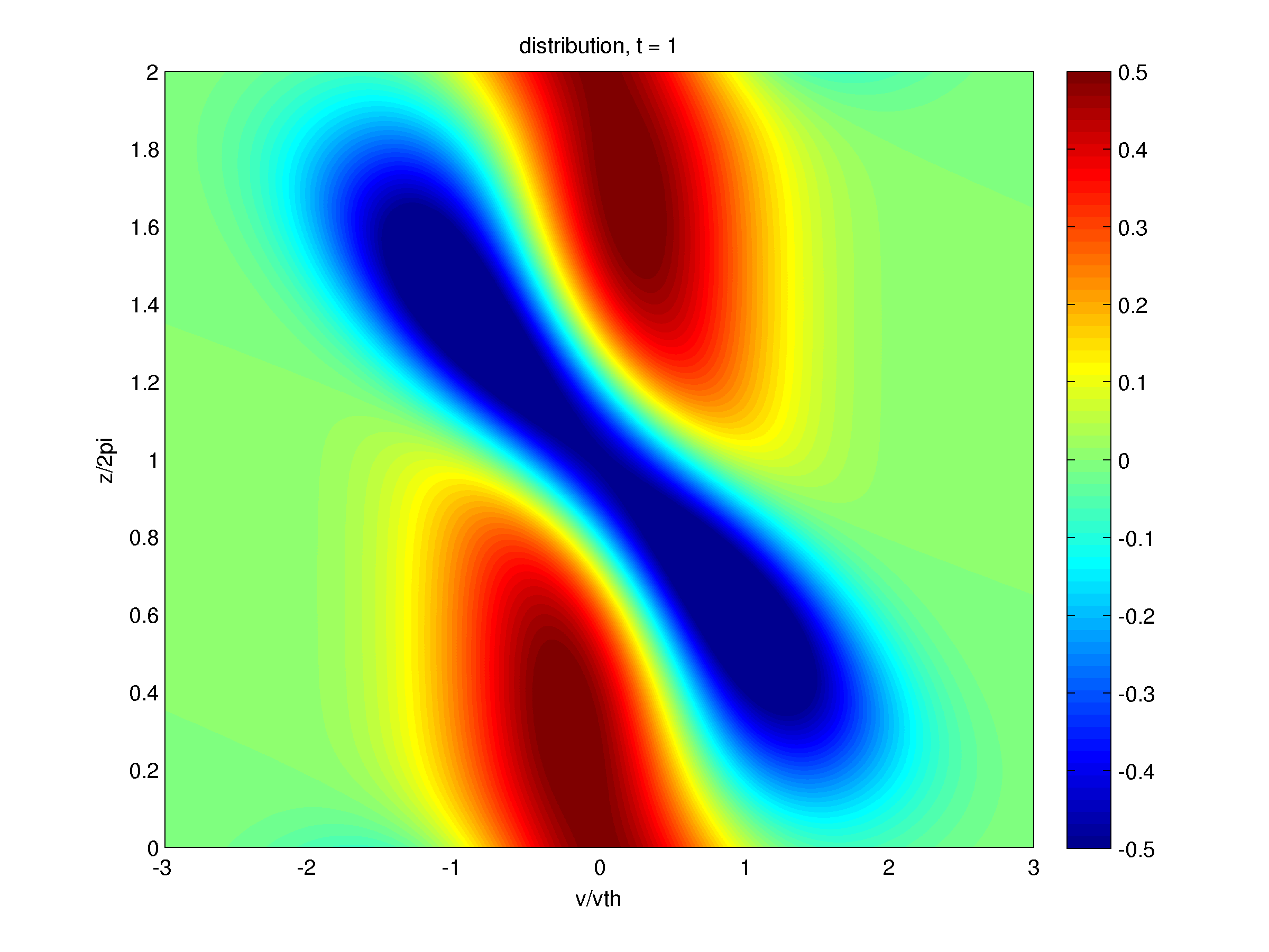}}
  \subfigure[$t=5$]{\includegraphics[trim=1.5cm 0.5cm 1.5cm 1.0cm,width=0.49\textwidth,clip]{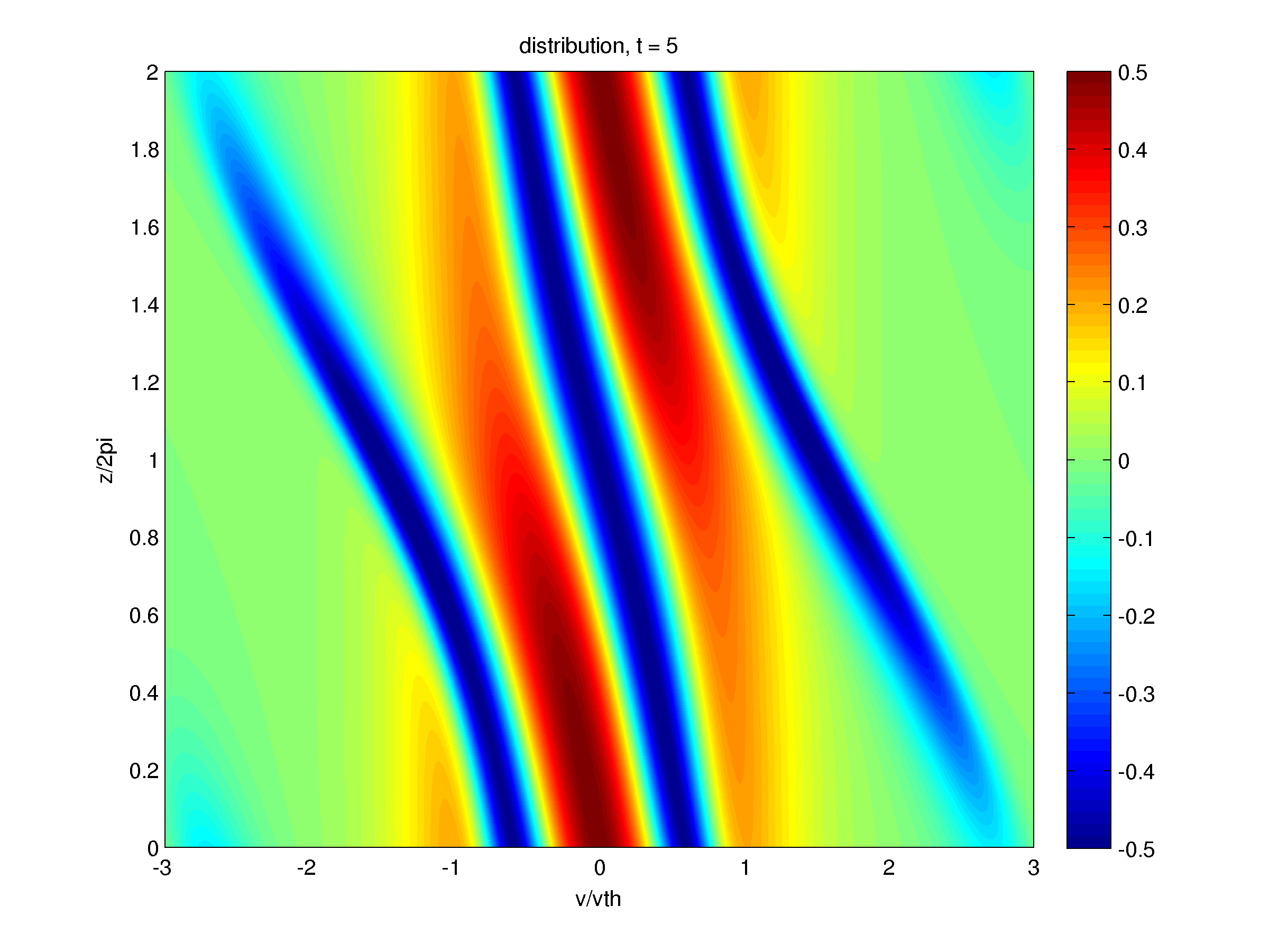}}
  \subfigure[$t=10$\label{fig:t10}]{\includegraphics[trim=1.5cm 0.5cm 1.5cm 1.0cm,width=0.49\textwidth,clip]{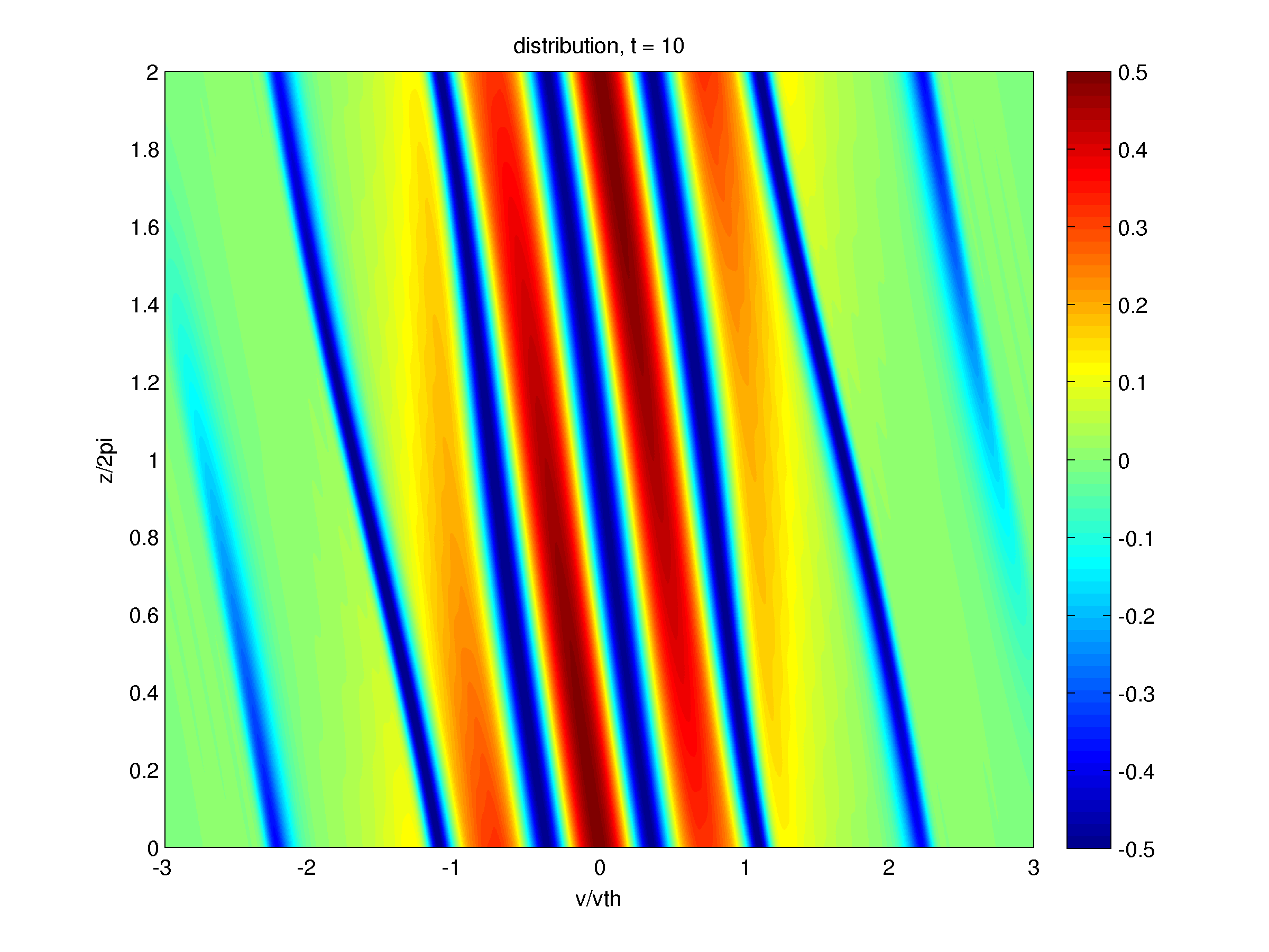}}
  \subfigure[$t=20$]{\includegraphics[trim=1.5cm 0.5cm 1.5cm 1.0cm,width=0.49\textwidth,clip]{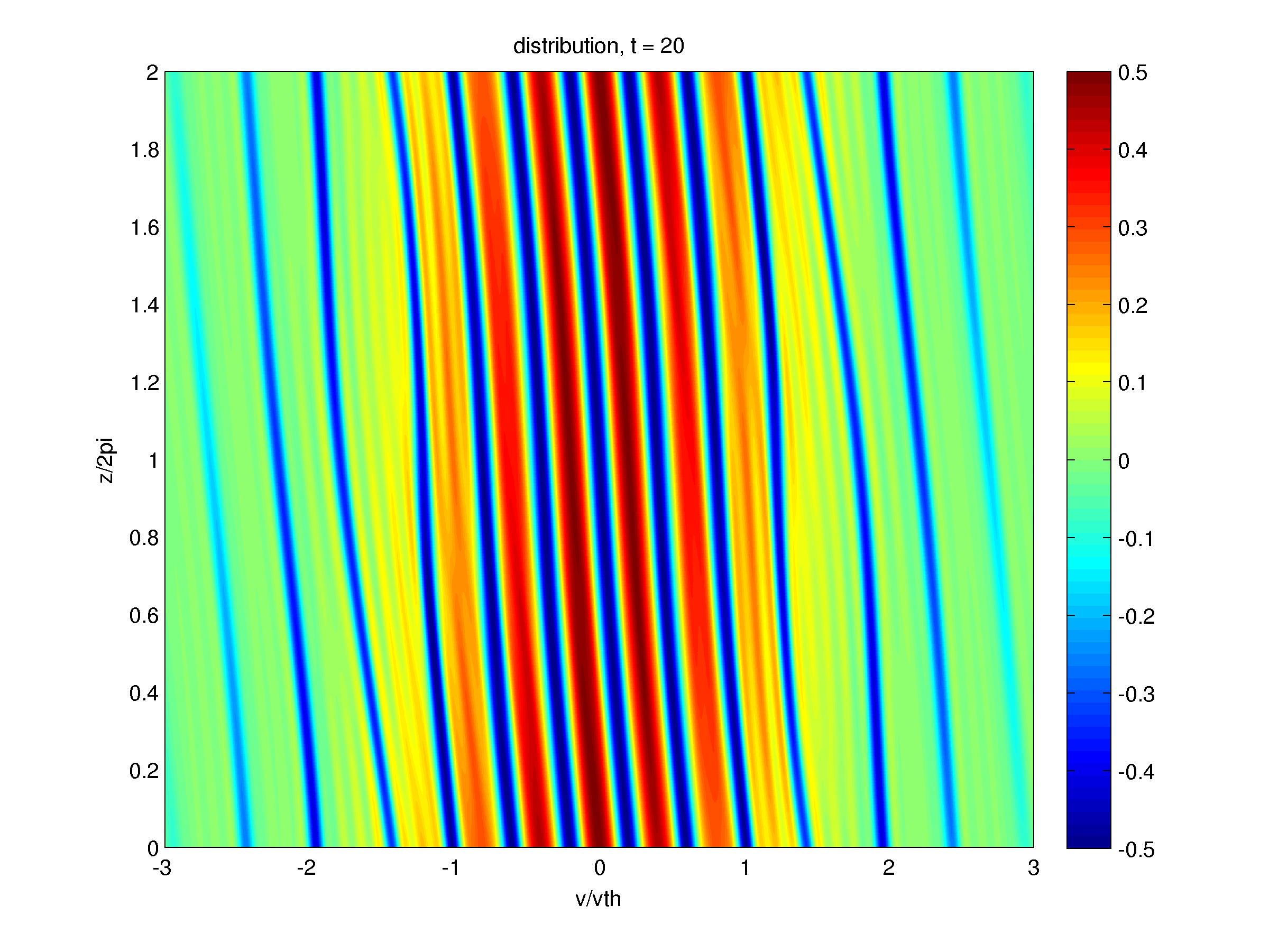}}
  \subfigure[$t=40$\label{fig:t40}]{\includegraphics[trim=1.5cm 0.5cm 1.5cm 1.0cm,width=0.49\textwidth,clip]{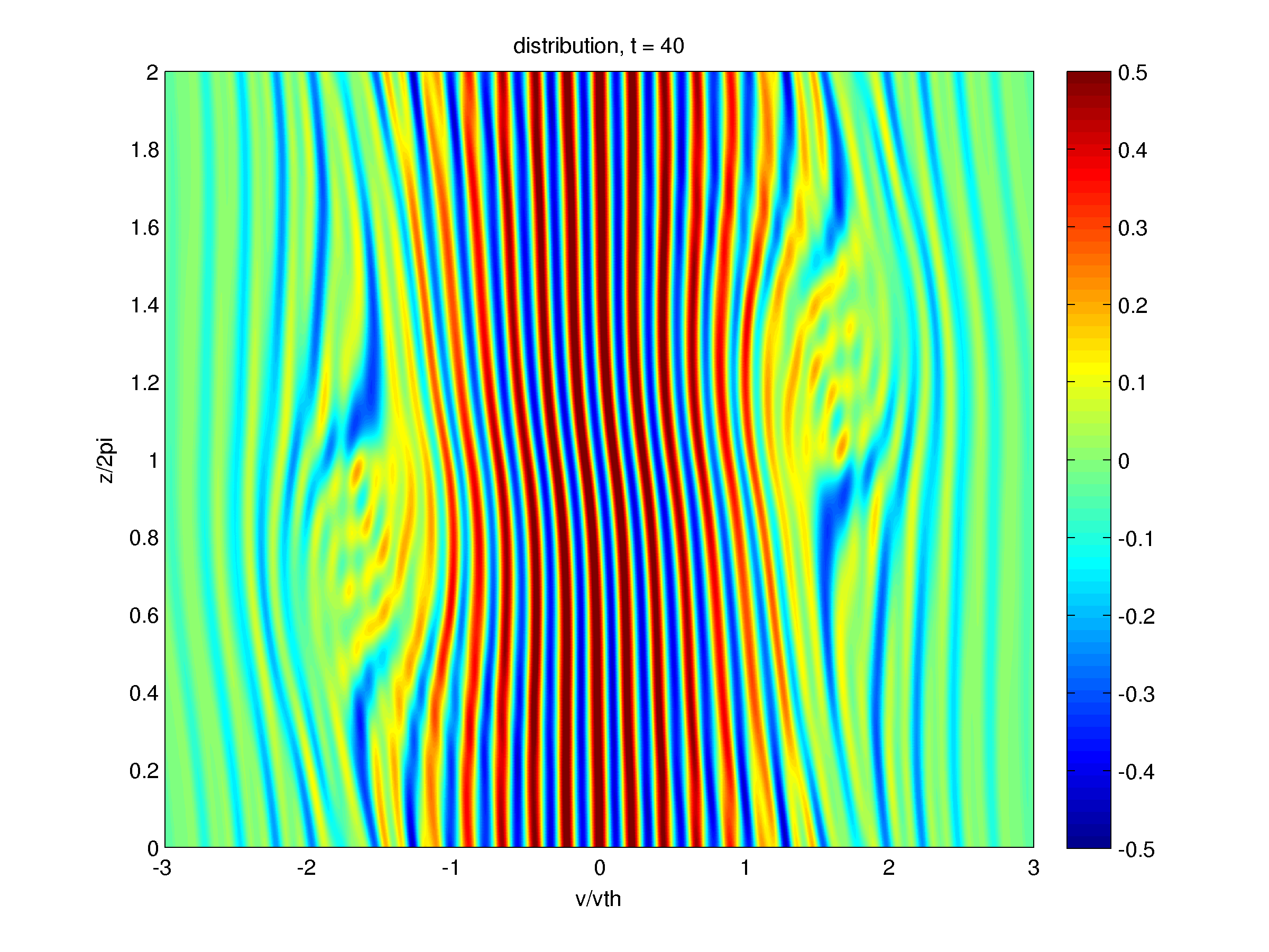}}
  \subfigure[$t=60$\label{fig:t60}]{\includegraphics[trim=1.5cm 0.5cm 1.5cm 1.0cm,width=0.49\textwidth,clip]{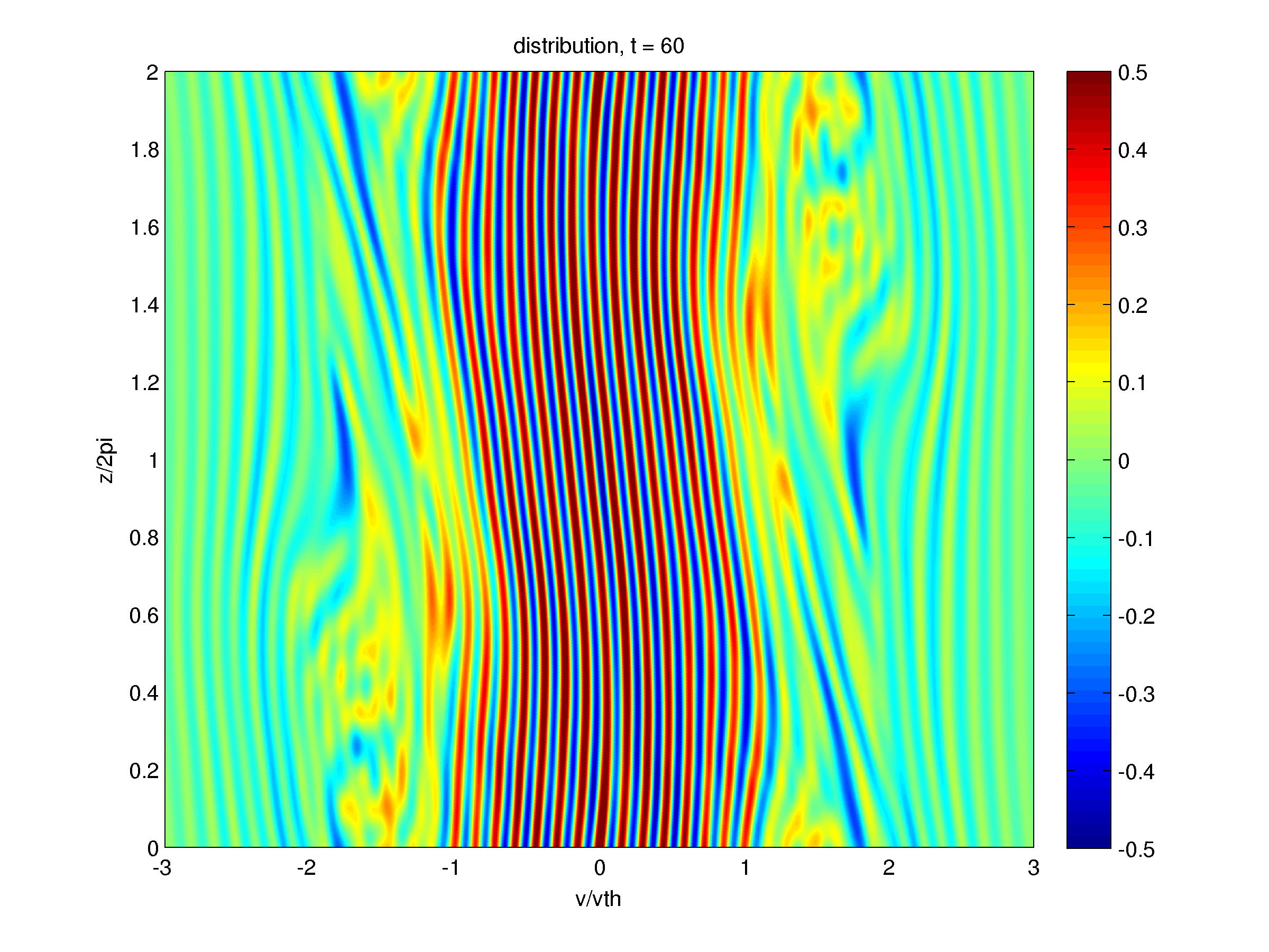}}
  \caption{Time slices of the perturbed distribution function \eqref{eq:SemiDiscreteFinalGridPoints} in $(z,v)$ space with resolution $(N_k,N_m)=(257,2048)$.
    The distribution is plotted on $z$-collocation points using the \DFT\ of the \FH\ coefficients.  
    As the Hermite transform is spectral, any velocity grid is permitted, and we plot on a grid of 2048 points with $v\in[-3,3]$.
  \label{fig:DistributionTimeSlices}
}
\end{figure}

We also plot time slices of the perturbed distribution $f$ in $(z,v)$-space in \fig~\ref{fig:DistributionTimeSlices}.
These show phase space shearing at early-times leading to a striped, highly-oscillatory pattern in velocity space that is characteristic of phase mixing \citep[and is indeed similar to the linear Landau damping case plotted in][\fig~3]{Heath12}. 
Here however nonlinear effects are visible as the stripes are not straight lines, but are wave-like with contours of the distribution function oscillating in $z$ with wavenumber $k=0.5$.
The waves at larger velocities oscillate with larger amplitude in $v$.
At about $t=30$, the oscillations in the region $|v|\in(1,2)$ roll up, forming vortex-like structures which propagate in the direction of the shearing (see \fig~\ref{fig:t40}).
By $t=60$ (\fig~\ref{fig:t60}), shearing has elongated these structures to be on the box scale, and they persist, flowing in the shearing direction in the region $|v|\in(1,2)$.
The region $v\in(-1,1)$ retains the striped phase mixed pattern, but also has a clear oscillation with wavenumber $k=0.5$.

\begin{figure}
  \centering
  {\includegraphics[trim=3.5cm 9.5cm 4.0cm 9.5cm,width=0.49\textwidth,clip]{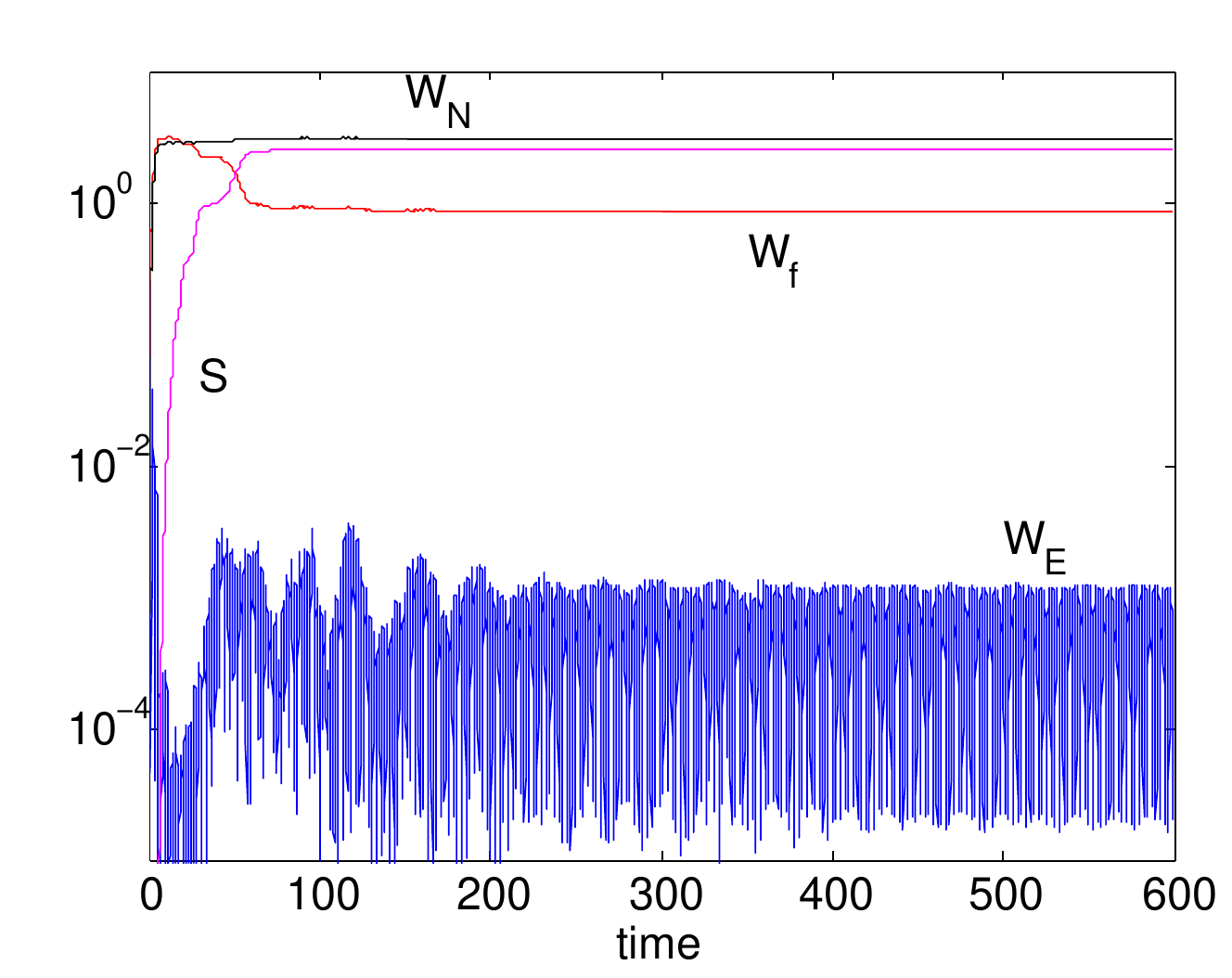}}
  {\includegraphics[trim=3.5cm 9.5cm 4.0cm 9.5cm,width=0.49\textwidth,clip]{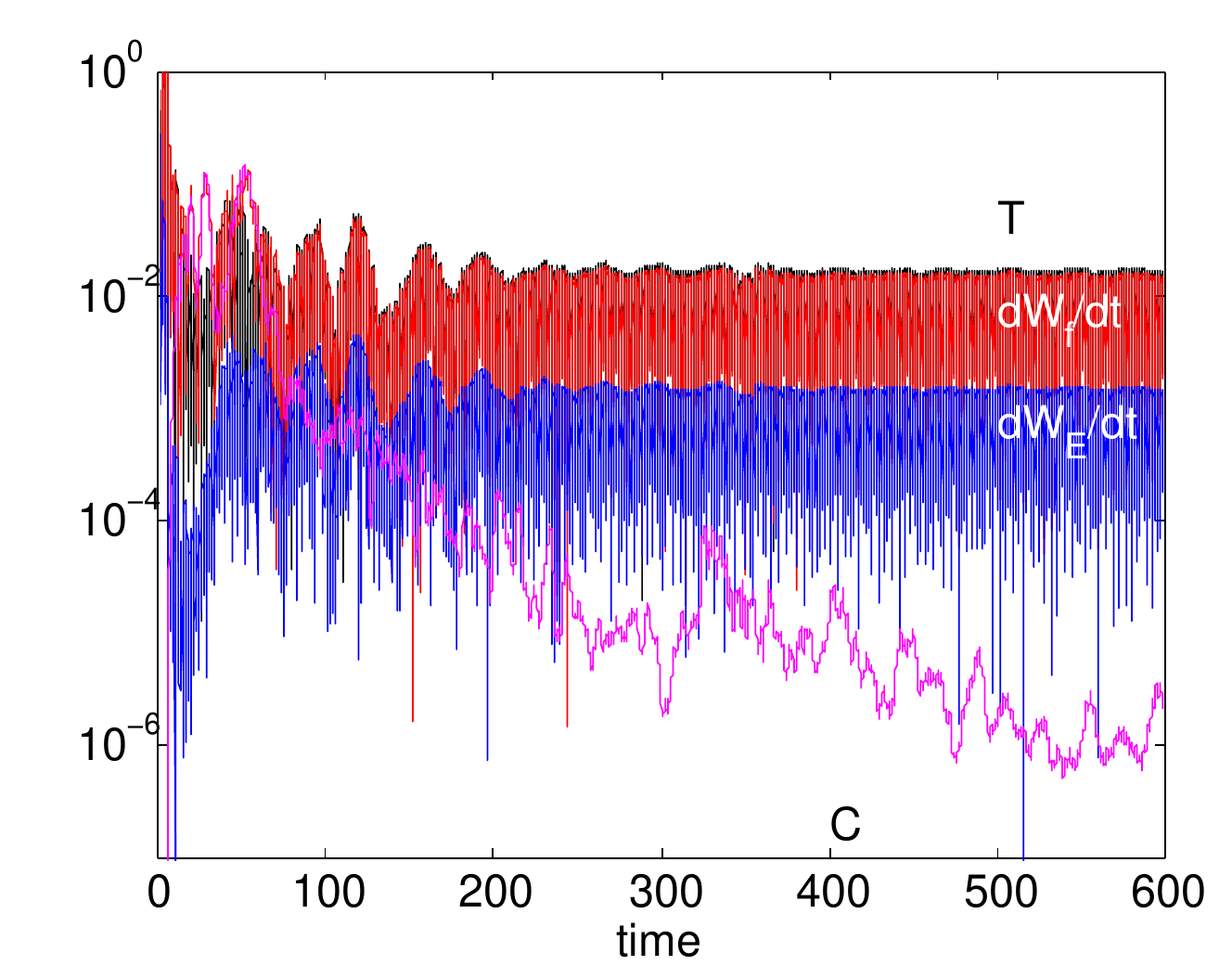}}
  \caption{Free energy time traces for the nonlinear system.}
  \label{fig:NonlinearFreeEnergyTrace}
\end{figure}

In \fig~\ref{fig:NonlinearFreeEnergyTrace}(a) we  plot the free energy contributions $W_E$, $W_f$, $W_N$ and ${\cal S}$
and in \fig~\ref{fig:NonlinearFreeEnergyTrace}(b) plot their respective time derivatives $\dot{W}_E$, $\dot{W_f}$, ${\cal T}$ and ${\cal C}$.
At long times, the free energies reach a steady state.
In particular the collisional sink ${\cal C}$ tends to zero so that no free energy is removed from the system.
Thus at long times free energy is exchanged between $W_f$, $W_N$ and $W_E$.
Moreover we see from \fig~\ref{fig:NonlinearFreeEnergyTrace}(b) that $\dot{W}_f$ and ${\cal T}$ are approximately equal, and thus there are only small changes in the electric field free energy as $\dot{W}_E\approx\dot{W}_f-{\cal T}$.

\subsubsection{Convergence}
\label{sec:Convergence}

\begin{figure}
  \centering
  \subfigure[]{\includegraphics[trim=3.5cm 9.5cm 4.0cm 9.5cm,width=0.49\textwidth,clip]{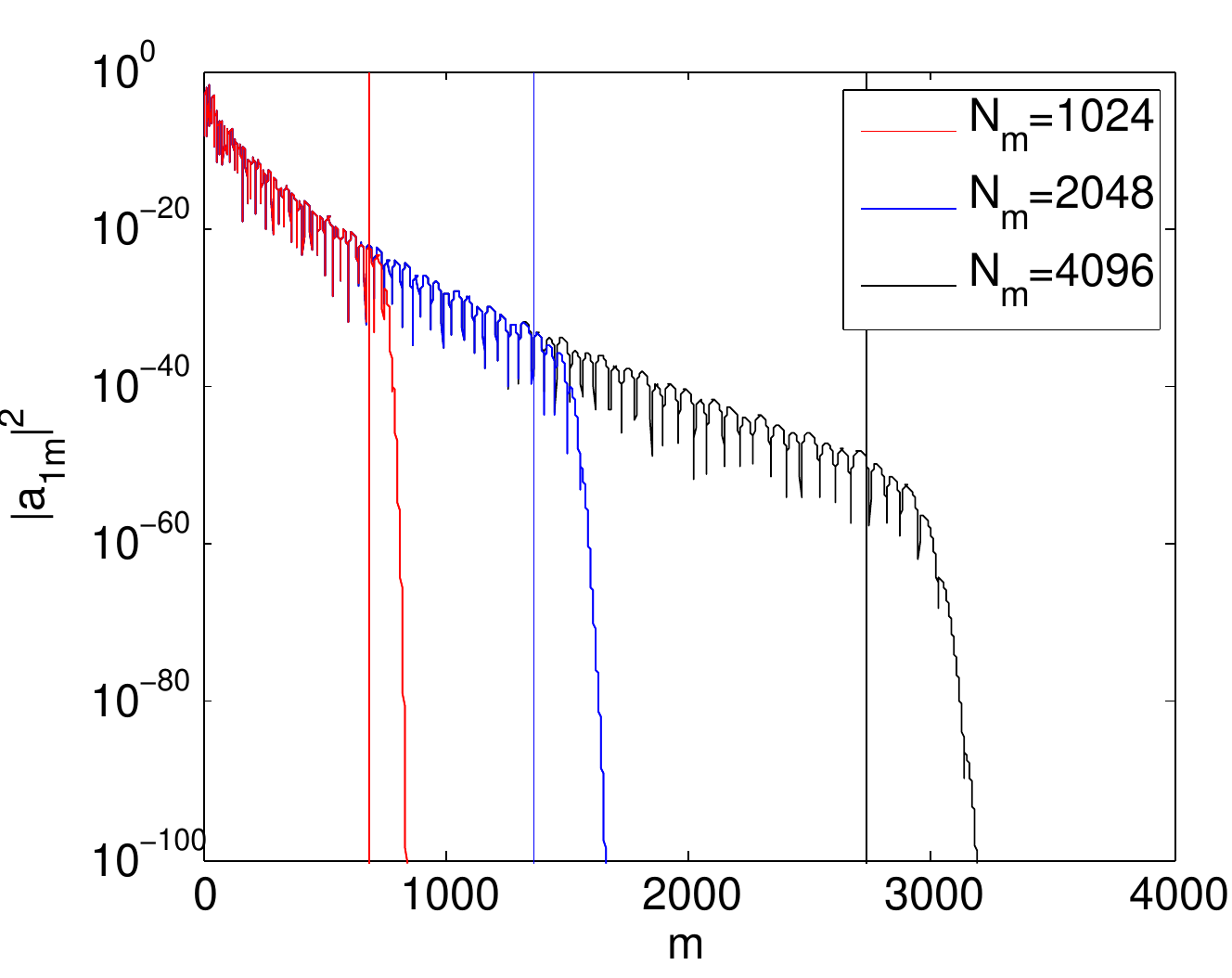}}
  \subfigure[]{\includegraphics[trim=3.5cm 9.5cm 4.0cm 9.5cm,width=0.49\textwidth,clip]{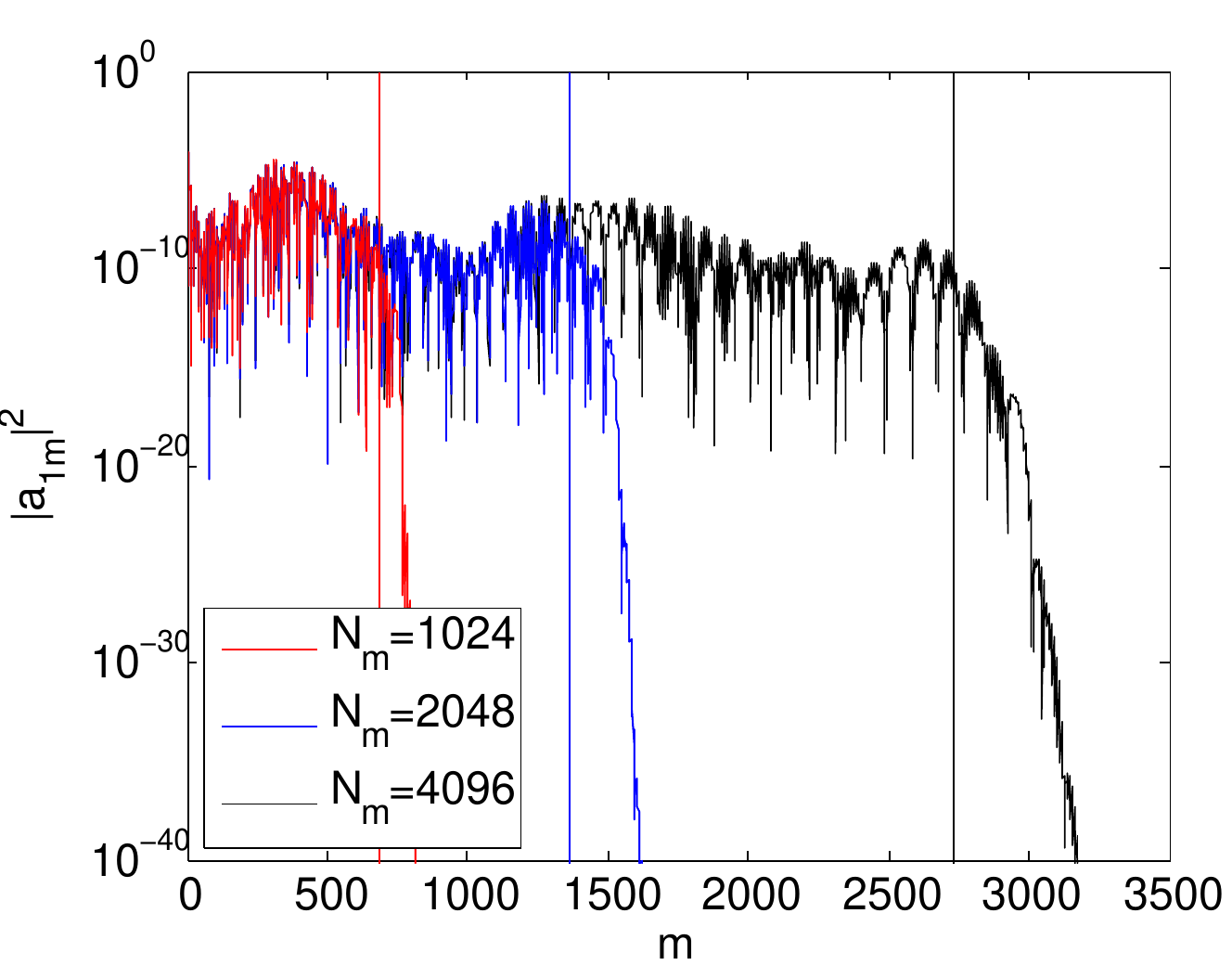}}
  \caption{Hermite spectra with Hou--Li filtering at $t=10$ (left) and $t=40$ (right). The vertical lines are at $m=2N_m/3$ for each resolution. All spectra are for wavenumber $k_1$ and have $N_k=257$.}
  \label{fig:Spectra}
\end{figure}

We show convergence behaviour by making a series of runs, repeatedly doubling resolution: $N_m$ from 32 to 4096, and $(N_k-1)$ from 16 to 256.
At a fixed time, we compare the \FH\ coefficients of a run to those of the best resolved run $(N_k,N_m)=(257,4096)$.
That is if $a_{jm}$ and $\bar{a}_{jm}$ are the \FH\ modes of the two runs, we define the error 
\begin{align}
  \sum_{j=-\Nt^*}^{\Nt^*}\sum_{m=0}^{N_m^*} | a_{jm} - \bar{a}_{jm}|^2,
  \label{eq:ErrorDefinition}
\end{align}
where $\Nt^*$, $N_m^*$ define which modes are included in the comparison.
Since both sets of coefficients are subject to resolution-dependent Fourier filtering, we consider only the modes which are unaffected by filtering in both runs.
We therefore take 
$\Nt^*= \left\lfloor 2\Nt/3\right\rfloor$
and
$N_m^*= \left\lfloor 2N_m/3\right\rfloor$,
for $(\Nt,N_m)$ of the lesser resolved run.
In \fig~\ref{fig:Spectra} we plot Hermite spectra for different resolutions to illustrate which modes achieve convergence.

The difference in spectral coefficients (without mode selection)
is related to the squared difference of the distribution functions
via Parseval's theorem
\begin{align}
  \sum_{j=-\Nt}^{\Nt}\sum_{m=0}^{N_m} | a_{jm} - \bar{a}_{jm}|^2
  =
  \frac{1}{N_k}\sum_{l=0}^{N_k-1}\int \d v ~ \frac{|f(z_l,v)-\bar{f}(z_l,v)|^2}{f_0},
\end{align}
where $f$ and $\bar{f}$ are the distribution functions corresponding to $a_{jm}$, $\bar{a}_{jm}$.
Thus the error \eqref{eq:ErrorDefinition} is similar to the error in the distribution function on collocation points in $(z,v)$-space, but with an extra factor of $1/f_0$ which exaggerates errors in the velocity tail of the distribution.

\begin{figure}
  \centering
  \subfigure[Convergence with $N_k$, $t=10$]{\includegraphics[trim=3.5cm 9.5cm 4.0cm 9.5cm,width=0.49\textwidth,clip]{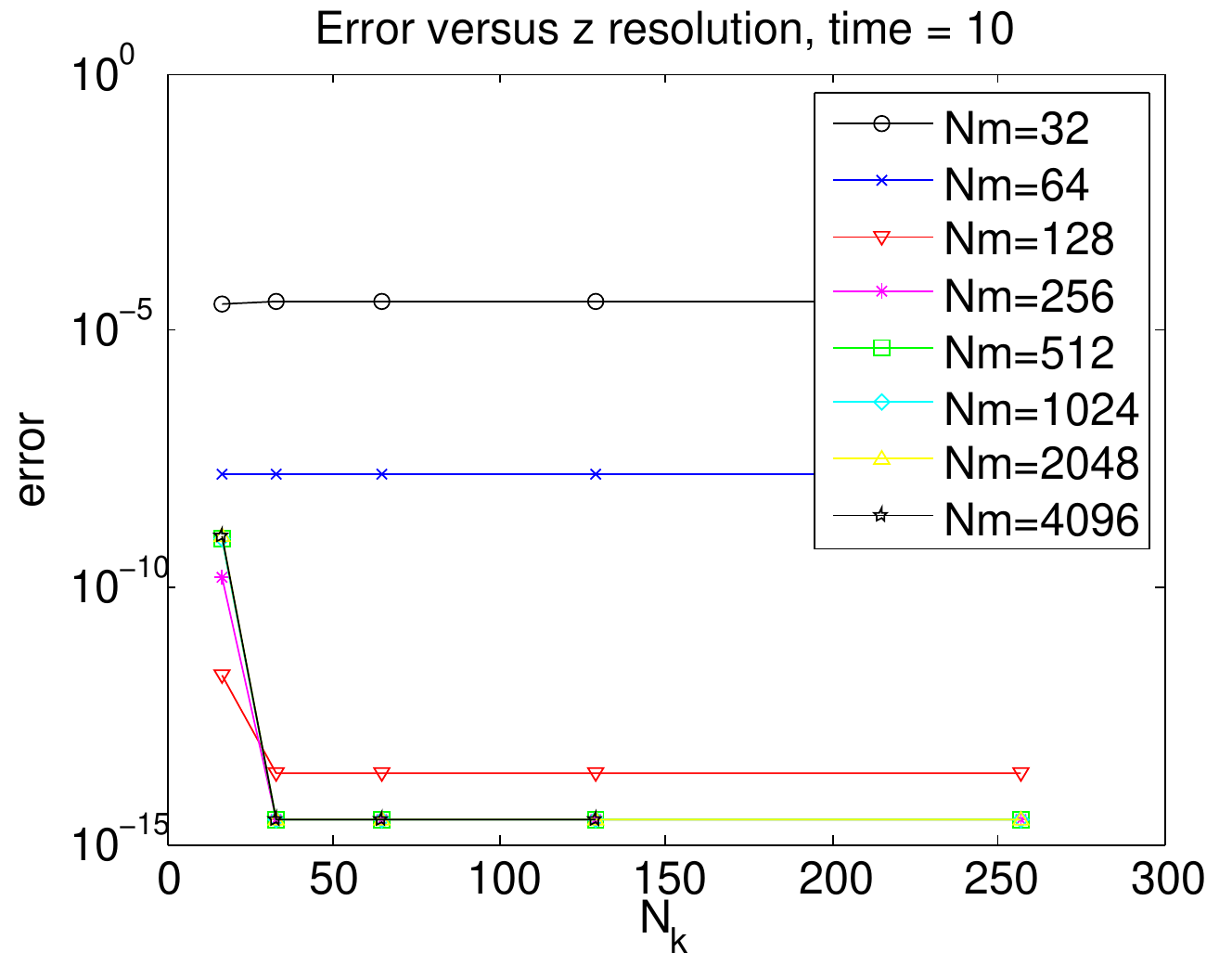}}
  \subfigure[Convergence with $N_m$, $t=10$]{\includegraphics[trim=3.5cm 9.5cm 4.0cm 9.5cm,width=0.49\textwidth,clip]{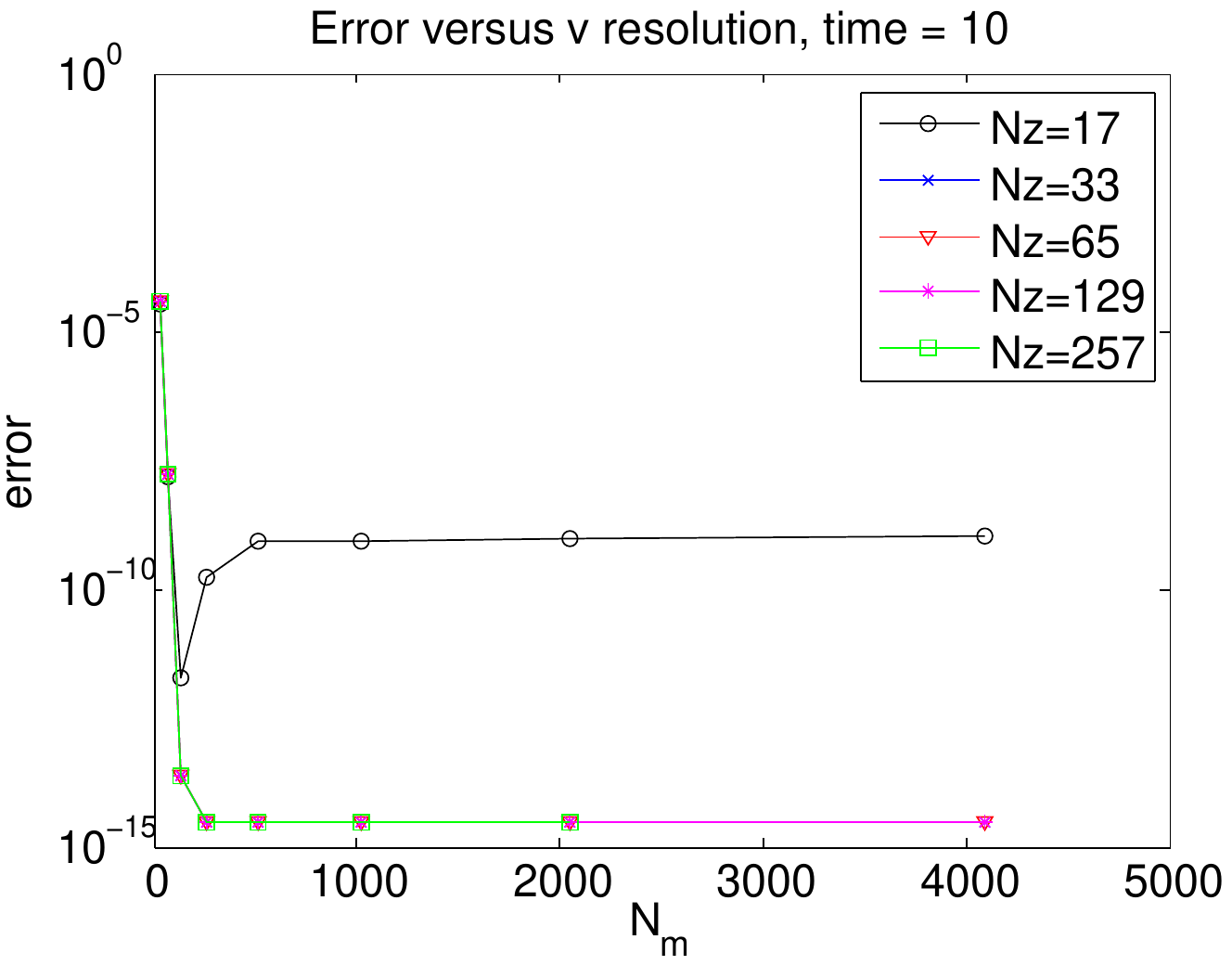}}
  \subfigure[Convergence with $N_k$, $t=40$]{\includegraphics[trim=3.5cm 9.5cm 4.0cm 9.5cm,width=0.49\textwidth,clip]{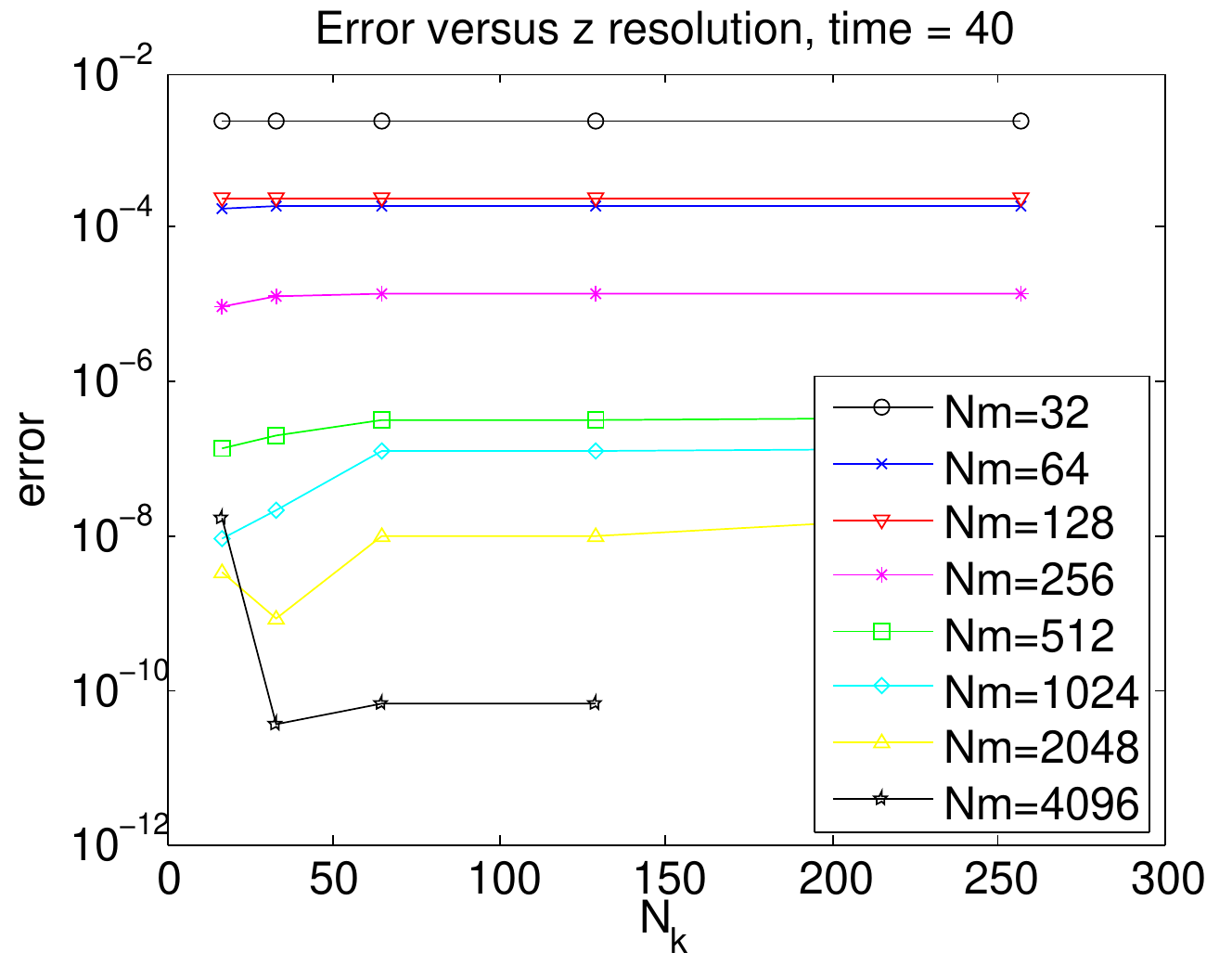}}
  \subfigure[Convergence with $N_m$, $t=40$\label{fig:errv40}]{\includegraphics[trim=3.5cm 9.5cm 4.0cm 9.5cm,width=0.49\textwidth,clip]{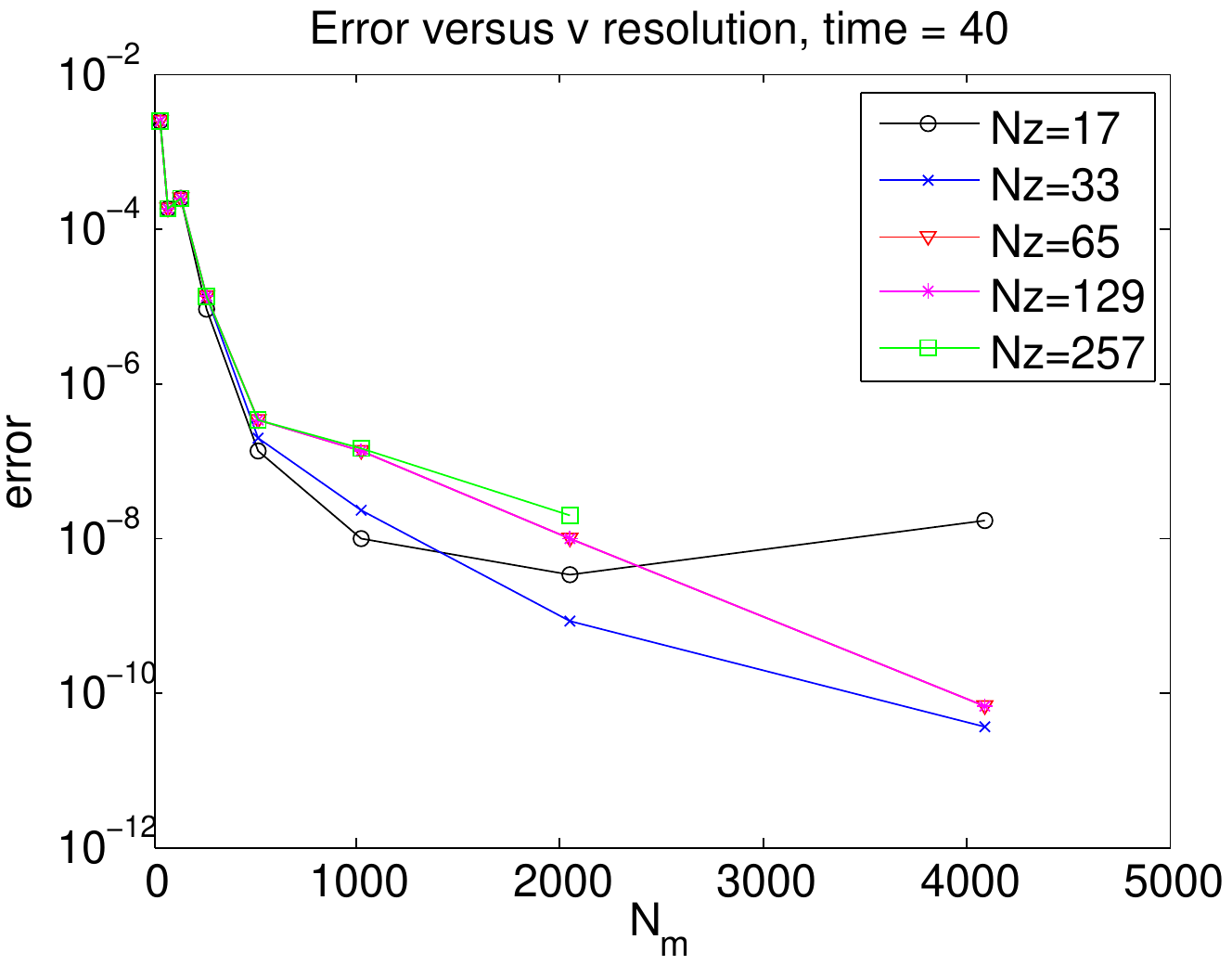}}
  \caption{Convergence.}
  \label{fig:ConvergencePlots}
\end{figure}

In \fig~\ref{fig:ConvergencePlots} we plot the error for two times, $t=10$ and $t=40$, which correspond to phase space diagrams \fig s~\ref{fig:t10} and \ref{fig:t40}.
At the earlier time $t=10$, the behaviour is similar to the linear case.
There is little structure in $z$ and once we have enough Fourier modes to capture this ($N_k=33$), the error does not decrease with $N_k$.
While there is finer structure in $v$, this is captured by $256$ Hermite modes, and as with $N_k$, increasing $N_m$ beyond this point does not reduce the error.

The later time $t=40$ corresponds to the top of the roll-over in the electric field (see \fig~\ref{fig:NonlinearLandauDamping}).
The physical space structure is still captured by a small number of Fourier modes after which there is no improvement in convergence.
In Hermite space, the scheme converges exponentially in $N_m$ once at least 33 Fourier modes are used.

\subsubsection{Hermite flux}
\label{sec:HermiteFlux}

We now describe the behaviour of the system in \FH\ phase space 
with  a view to explaining two nonlinear effects:
firstly that after its initial decay, the electric field grows in the absence of linear instability; 
and secondly that the electric field does not decay at long times.

The magnitude of the coefficients $|a_{jm}|^2$ are contributions to the leading order relative entropy \eqref{eq:RelativeEntropyLinearized} from each \FH\ mode
\begin{align}
  \begin{split}
  \RE[\oneddist|f_0] 
  = \intii\d z ~\intii \d v ~  \frac{f^2}{2f_0} 
  = \sum_{m=0}^{N_m-1}\sum_{j=-\Nt}^{\Nt} |a_{jm}|^2.
  \end{split}
  \label{eq:RelativeEntropyFourierHermite}
\end{align}
Moreover the coefficients may be used to describe the flow of relative entropy in \FH\ space.
\cite{Zocco11} studied the flow of free energy in a reduced gyrokinetic model,
and showed that by introducing $\at_{jm}=(\i\ \sgn\ k_j)^ma_{jm}$, particle streaming is a Hermite space flux.
Applying this transform for $m>1$ in the linearization of \eqref{eq:DiscreteHermiteMomentKineticEquation} 
we obtain
\begin{align}
  \frac{1}{2}\pd{\at^2_{jm}}{t} + \lp \Gamma_{j,m+1/2} - \Gamma_{j,m-1/2}\rp = 0,
  \label{eq:FluxEqn}
\end{align}
where the Hermite flux is
\begin{align}
  \Gamma_{j,m-1/2} 
  = |k_j|\sqrt{m/2}\ \at_{jm}\at_{j,m-1}
  = k_j\sqrt{m/2}\ \Imag\lp a^*_{jm}a_{j,m-1}\rp,
  \label{eq:HermiteFlux}
\end{align}
and the coefficients $\at_{jm}$ are real.
The flux equation \eqref{eq:FluxEqn} may be approximated by
\begin{align}
  \frac{1}{2}\pd{\at^2_{jm}}{t} + \pd{\Gamma^{SV}_{jm}}{m} = 0,
  \label{eq:FluxEqnContinuous}
\end{align}
with the flux 
defined as
\begin{align}
  \Gamma^{SV}_{jm} = |k_j|\sqrt{m/2}\ \at^2_{jm} = |k_j|\sqrt{m/2}\ |a_{jm}|^2.
  \label{eq:LinearFlux}
\end{align}
Equation \eqref{eq:FluxEqnContinuous} may be written
\begin{align}
  \lp \frac{1}{2}\pd{}{t} + |k_j| \pd{}{\sqrt{2m}}\rp \lp \sqrt{2m}\at^2_{jm}\rp = 0,
  \label{eq:CharacteristicsPDE}
\end{align}
so that free energy propagates along characteristics $m=2|k_j^2|(t-t_0)^2$.
For eigenfunctions in time $\tfd{|a_{jm}|^2}{t}=2\gamma_j|a_{jm}|^2$, \eqref{eq:FluxEqnContinuous} gives the spectrum
\begin{align}
  |a_{jm}|^2 = \frac{C_{j}}{\sqrt{2m}} \exp\lp -\frac{2\sqrt{2}\gamma_j m^{1/2}}{|k|} \rp,
\end{align}
for constants $C_{j}$, which is in excellent agreement with numerically-calculated linear spectra \citep{Hypercollisions}.

The approximation $\Gamma=\Gamma^{SV}$ holds provided that $\at_{jm}$ is slowly varying in $m$ in the sense that $\at_{jm}\approx\at_{j,m+1}$;
equation \eqref{eq:FluxEqn} also supports alternating solutions with $\at_{jm}\approx -\at_{j,m+1}$.
Therefore \citet[see also  \citet{Kanekar14}]{Schekochihin14} introduced the decomposition
$\at_{jm} = \at^+_{jm} + (-1)^m\at^-_{jm}$, where 
\begin{align}
  \at^+_{jm} = \frac{\at_{jm}+\at_{j,m+1}}{2},
  \hspace{1cm}
  \at^-_{jm} = (-1)^m\frac{\at_{jm}-\at_{j,m+1}}{2},
\end{align}
are both continuous in $m$.
Substituting these into \eqref{eq:FluxEqn} we obtain
\begin{align}
  \begin{split}
  \frac{1}{2}\pd{\lp\atpm_{jm}\rp^2}{t} 
  & \pm \frac{|k_j|}{2} \lp (s_{m+2}+s_{m+1})\atpm_{j,m+1}\atpm_{jm}
  - (s_{m+1}+s_m)\atpm_{j,m-1}\atpm_{jm}\rp
  \\ &
  \pm \frac{|k_j|(-1)^m}{2} \lp (s_{m+2}-s_{m+1})\atmp_{j,m+1}\atpm_{jm}
  - (s_{m+1}-s_m)\atpm_{jm}\atmp_{j,m-1}\rp
  = 0,
  \end{split}
  \label{eq:atpmEvolution}
\end{align}
where $s_m=\sqrt{m/2}$ and the differences $s_{m+2}-s_{m+1}$ and $s_{m}-s_{m-1}$ are both $O(1/\sqrt{m})$.
Thus for large $m$, the particle streaming is always a flux, but in a different direction for $\at^+$ and $\at^-$.
The ``phase-mixing'' mode $\at^+$ propagates from low to high $m$, while the ``un-phase-mixing'' mode $\at^-$ propagates from high to low $m$. 

Comparing the true Hermite flux $\Gamma$ to the approximation 
$\Gamma^{SV}$,
by defining the normalized flux
\begin{align}
  \hat{\Gamma}_{jm} = \frac{\Gamma_{jm}}{\Gamma^{SV}_{jm}} 
  = \frac{\sgn\ k_j\ \Imag \lp a^*_{j,m+1}a_{jm}\rp}{|a_{jm}|^2}
  = \frac{\at_{j,m+1}\at_{jm}}{\at_{jm}^2}
  = \frac{(\at^+_{jm})^2-(\at^-_{jm})^2}{(\at^+_{jm}+(-1)^m\at^-_{jm})^2} ,
  \label{eq:NormalizedFlux}
\end{align}
we see that the approximation $\Gamma\approx\Gamma^{SV}$ is valid when the $\at^+$ mode is dominant;
otherwise significant amounts of $\at^-$ modifies the streaming.
We therefore use the normalized Hermite flux to describe the transfer of free energy in phase space.
This quantity is of particular interest in determining the behaviour of the electric field, as we recall from \sec~\ref{sec:DiscreteFreeEnergyEquations} that the electric field only grows or decays as the result of net Hermite flux.

The nonlinear \vps\ \dvp\ is written in terms of $\atpm$ as
\begin{align}
  \begin{split}
  \pd{\atpm_{jm}}{t} 
  + S^{\pm}_{jm}
  + B^{\pm}_{jm}
  + N^{\pm}_{jm}
  = 0,
  \end{split}
  \label{eq:atpmEntropy}
\end{align}
where the streaming term is
\begin{align}
  \begin{split}
  S^{\pm}_{jm} = 
  & \pm \frac{|k_j|}{2} \lp (s_{m+2}+s_{m+1})\atpm_{j,m+1}
  - (s_{m+1}+s_m)\atpm_{j,m-1}\rp
  \\ &
  \pm \frac{|k_j|(-1)^m}{2} \lp (s_{m+2}-s_{m+1})\atmp_{j,m+1}
  - (s_{m+1}-s_m)\atmp_{j,m-1}\rp
  = 0,
  \end{split}
\end{align}
the Boltzmann response is
\begin{align}
  B^{\pm}_{jm} =
  \mp \frac{\delta_{m0}+\delta_{m1}}{|k_j|\sqrt{2}} \lp \atp_{j0}+\atm_{j0}\rp ,
\end{align}
and the nonlinear term is
\begin{align}
  N^{\pm}_{jm} =
  \pm  \sum_{j'=-\Nt}^{\Nt} i\E_{j-j'}  \left[ (D^{m+1}_{jj'}+D^m_{jj'})  \atpm_{j',m-1} + (-1)^m(D^{m+1}_{jj'}-D^m_{jj'}) \atmp_{j',m-1} \right],
  \label{eq:pmNonlinearTerm}
\end{align}
where the electric field may be written in terms of $\atpm$ as
\begin{align}
  i\E_j =
  \begin{cases}
  - \frac{\atp_{j0}+\atm_{j0}}{k_j},
  \hspace{1cm} &  j\neq 0,\\
  0, 
  \hspace{1cm}
   & j=0,
  \end{cases}
\end{align}
and where $D^m_{jj'}=\sqrt{m/2}(\sgn\ k_j)^m(\sgn\ k_{j'})^{m-1}$.
Again the difference $D^{m+1}_{jj'}-D^m_{jj'}$ is $O(1/\sqrt{m})$.

\begin{figure}
  \centering
  \subfigure[$\log(|\atp_{1m}|^2)$, Hou--Li filtering]
  {\includegraphics[trim=1.0cm 0.0cm 2.1cm 1.0cm,width=0.49\textwidth,clip]{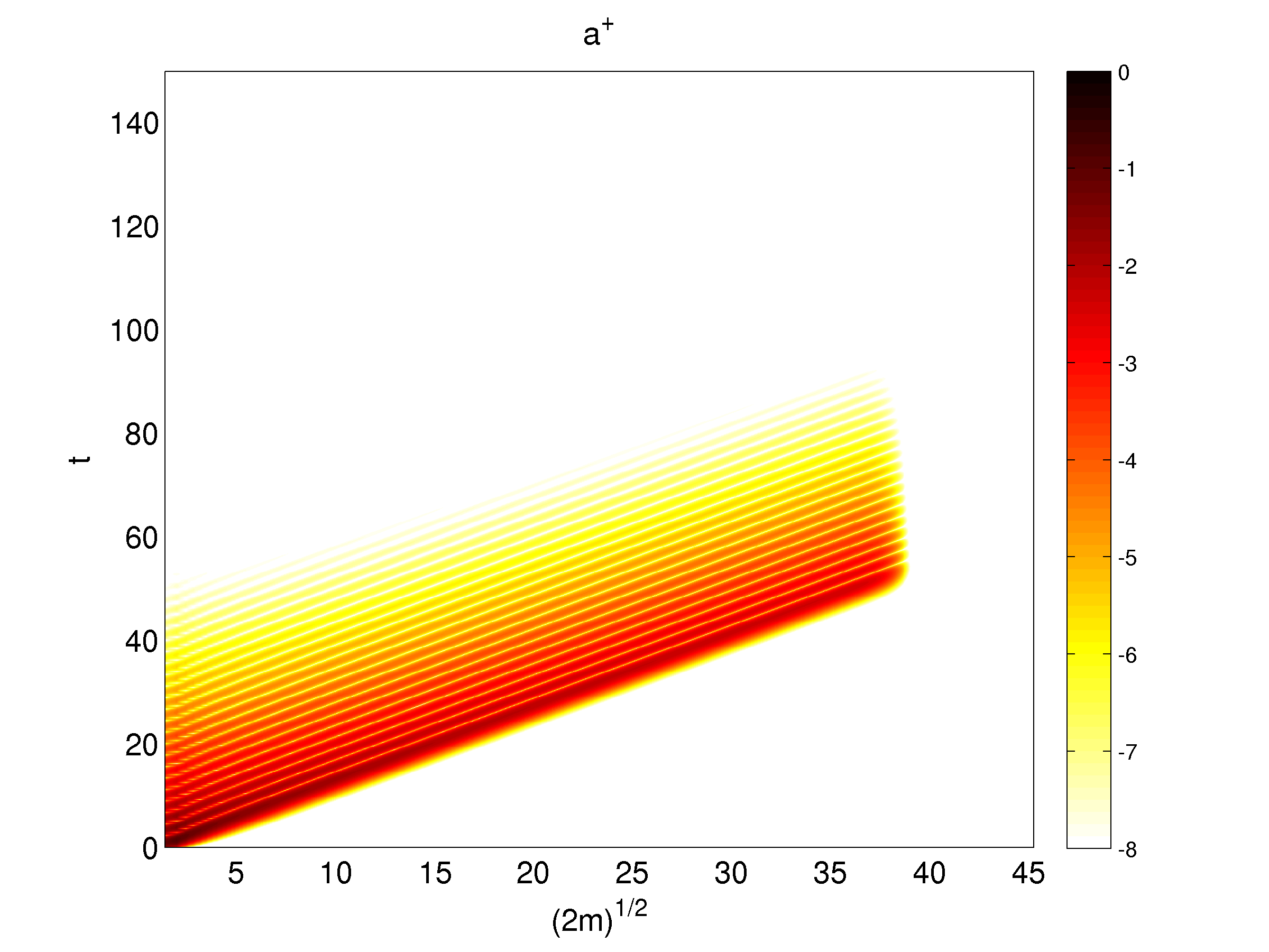}}
  \subfigure[$\log(|\atm_{1m}|^2)$, Hou--Li filtering]
  {\includegraphics[trim=1.0cm 0.0cm 2.1cm 1.0cm,width=0.49\textwidth,clip]{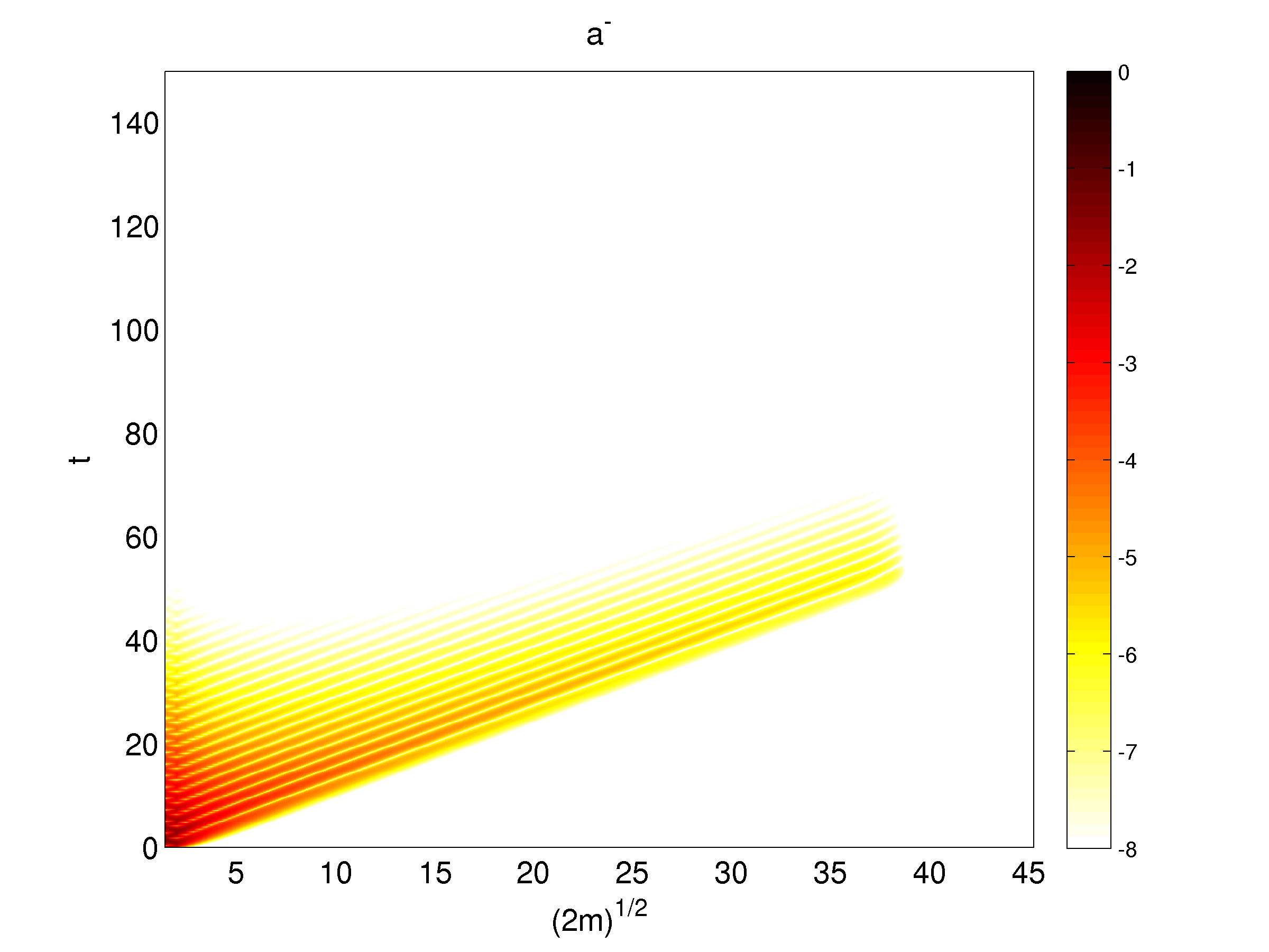}}
  \subfigure[$\log(|\atp_{1m}|^2)$, no velocity space dissipation]
  {\includegraphics[trim=1.0cm 0.0cm 2.1cm 1.0cm,width=0.49\textwidth,clip]{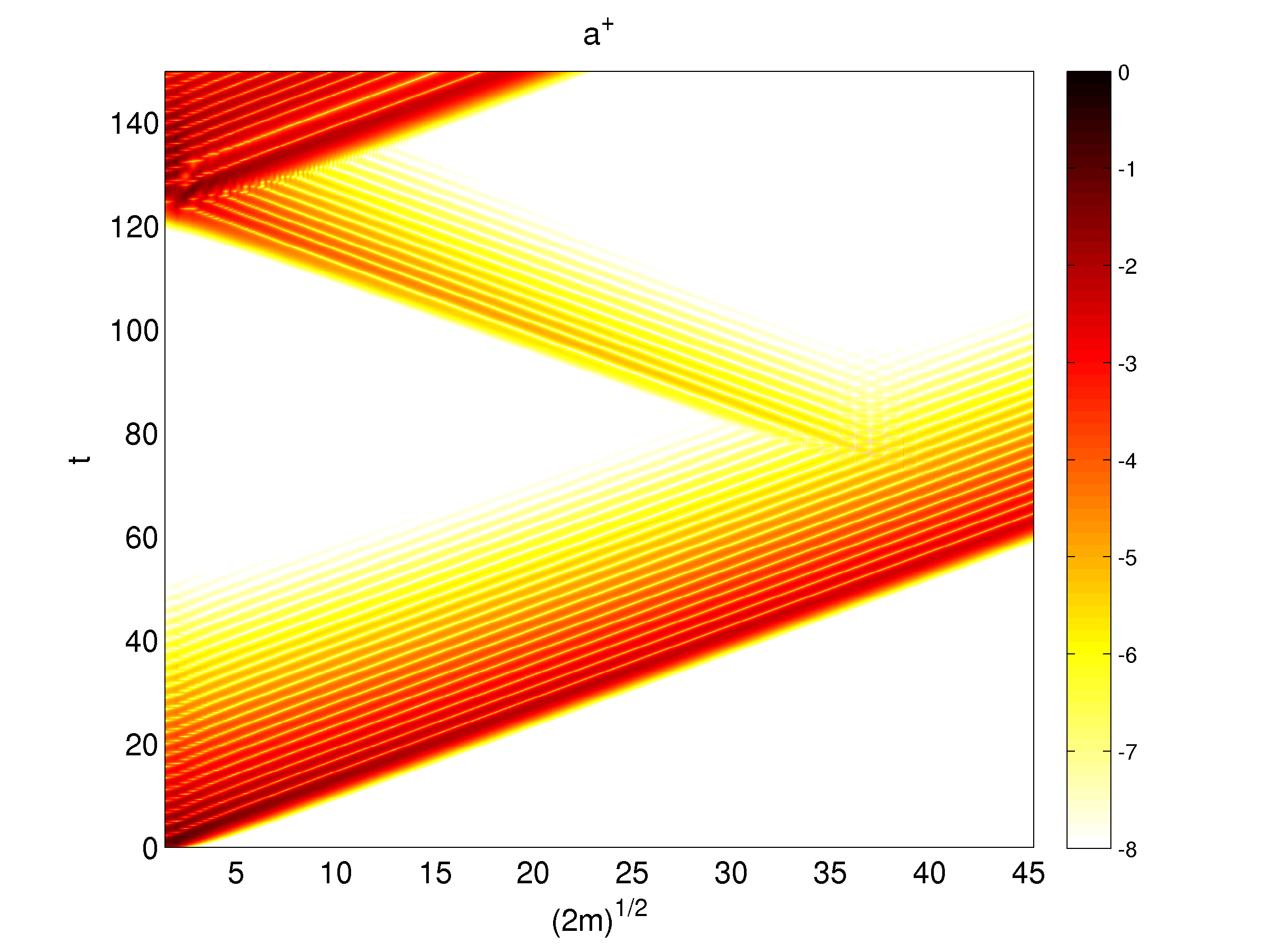}}
  \subfigure[$\log(|\atm_{1m}|^2)$, no velocity space dissipation]
  {\includegraphics[trim=1.0cm 0.0cm 2.1cm 1.0cm,width=0.49\textwidth,clip]{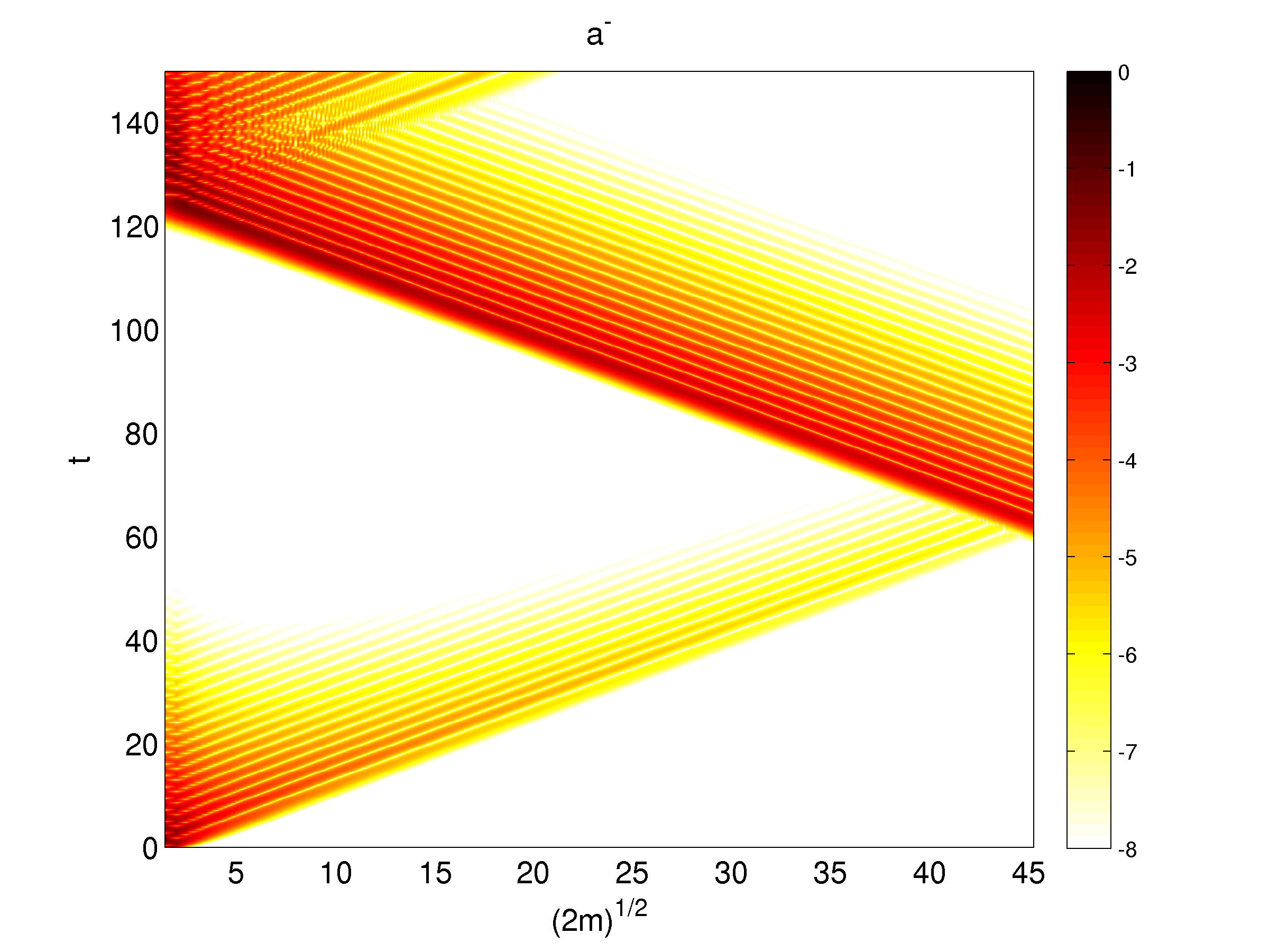}}
  \caption{The magnitude of forwards and backwards propagating modes for $k=0.5$ in the linearized system with and without velocity space dissipation.}
  \label{fig:at_contours_linear}
  \label{fig:atContoursLinear}
\end{figure}

\begin{figure}
  \centering
  \subfigure[$\log(|\atp_{1m}|^2)$]
  {\includegraphics[trim=1.0cm 0.0cm 2.1cm 1.0cm,width=0.49\textwidth,clip]{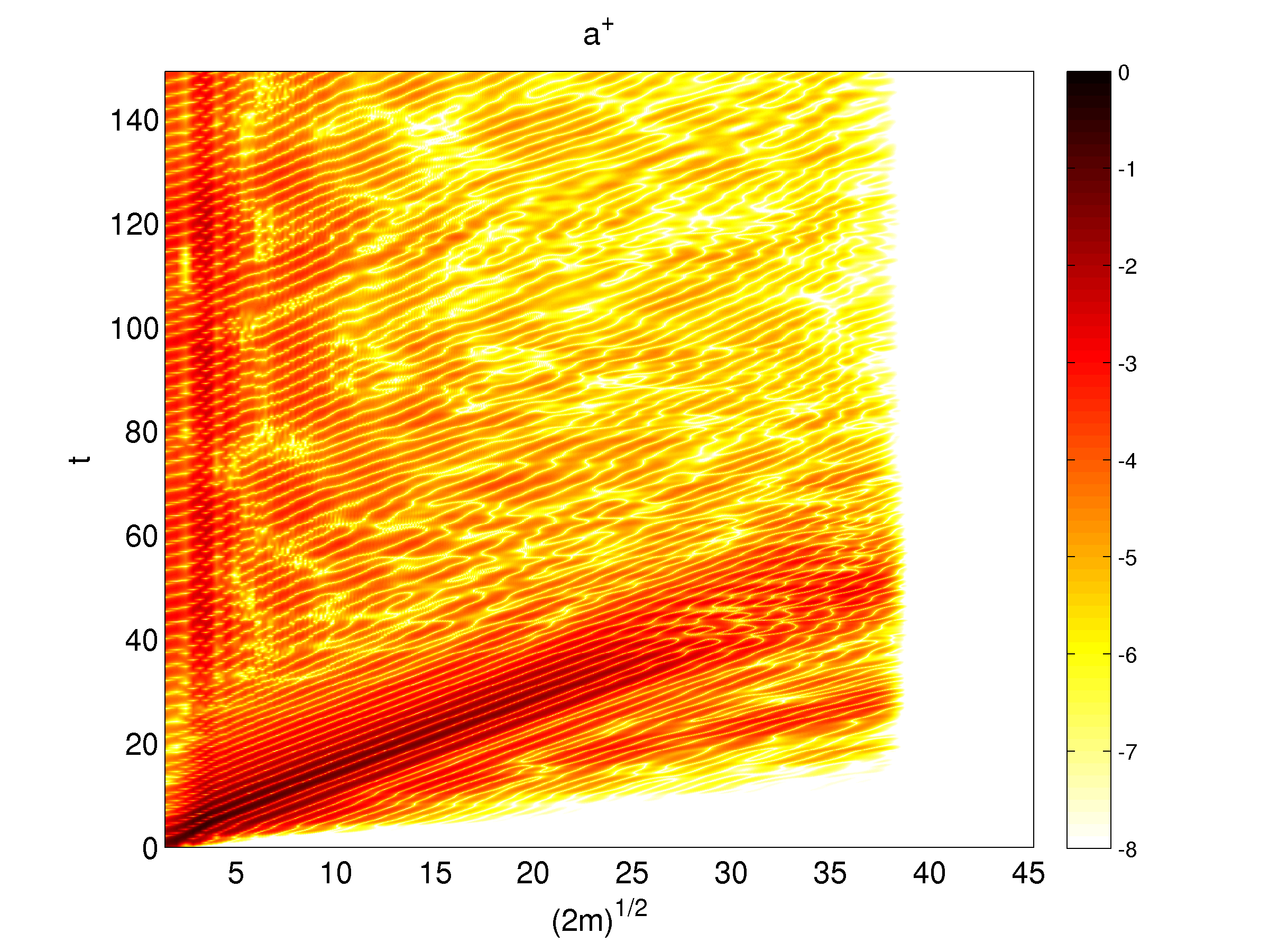}}
  \subfigure[$\log(|\atm_{1m}|^2)$]
  {\includegraphics[trim=1.0cm 0.0cm 2.1cm 1.0cm,width=0.49\textwidth,clip]{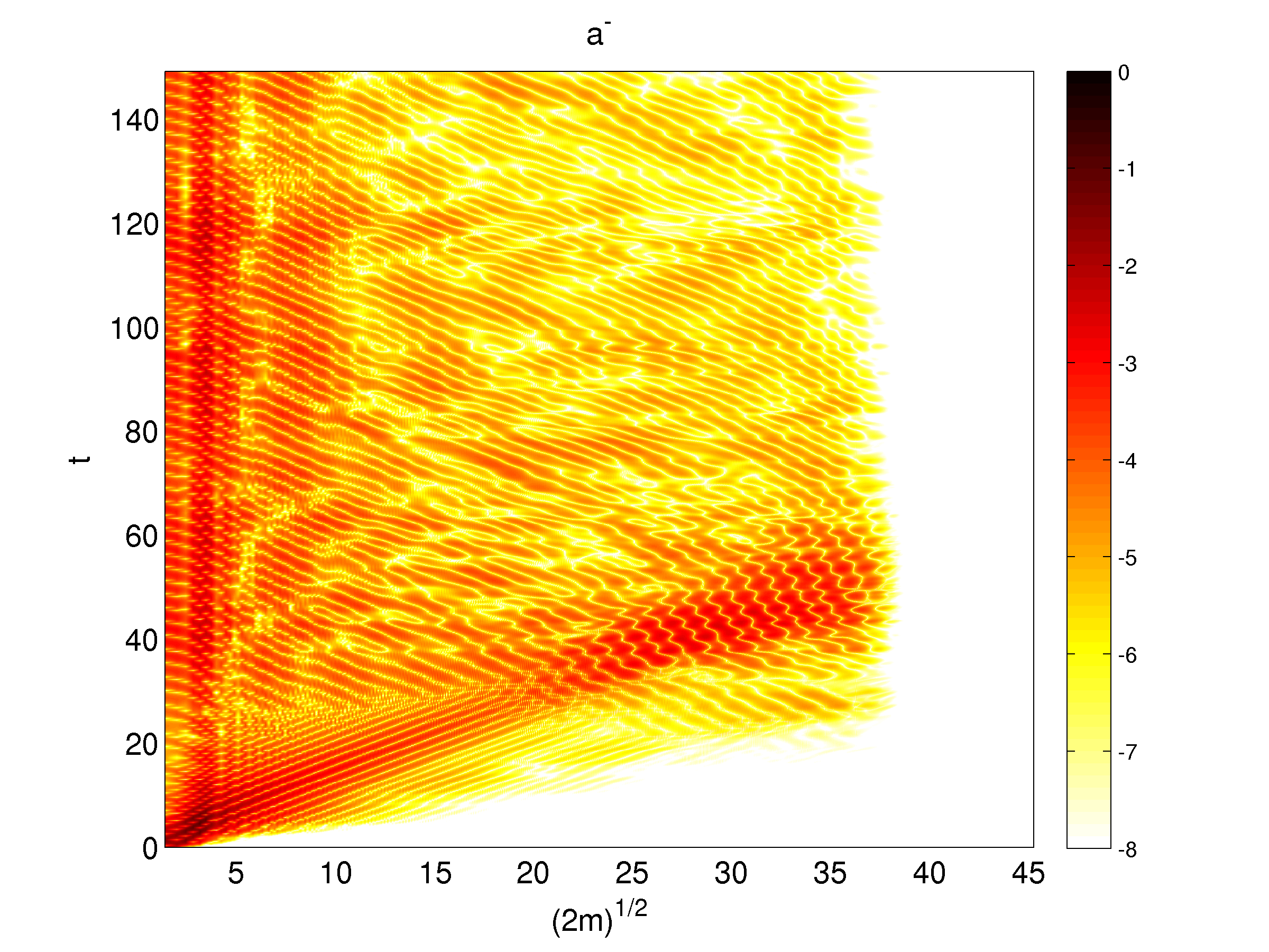}}
  \caption{The magnitude of forwards and backwards propagating modes for $k=0.5$ for the nonlinear system.}
  \label{fig:at_contours_nonlinear}
  \label{fig:atContoursNonlinear}
\end{figure}

In \fig s~\ref{fig:at_contours_linear} and \ref{fig:at_contours_nonlinear} we plot the Hermite spectra $(\atpm)^2$ against Hermite index and time for the first wavenumber $k=0.5$ in the linear and nonlinear systems. 
In the linear system the equations for $\atpm$ decouple, except for the Boltzmann response term at $m=0$ and $m=1$, and for the $O(1/\sqrt{m})$ cross-coupling term in the streaming.
Moreover the modes $\atpm$ propagate along characteristics $m=\pm 2k^2(t-t_0)^2$. 
We observe this behaviour in \fig s~\ref{fig:at_contours_linear}.
In the linear case with the Hou--Li filter (\fig s~\ref{fig:at_contours_linear}a,b)  
the free energy fluxes forward along very clear characteristics $m=2k^2(t-t_0)^2$ until reaching collisional scales where it is damped.
The decoupling is not perfect as there is some forward propagation observed in the $\atm$ mode in \fig~\ref{fig:at_contours_linear}.
The amplitude of the backward propagating mode increases as $m$ decreases suggesting it is due to the $O(1/\sqrt{m})$ cross-coupling term in the streaming,
however it is always significantly smaller than the $\atp$ mode at the corresponding $(\sqrt{2m},t)$ point.

In \fig s~\ref{fig:atContoursLinear}(c,d) we show the linear case with no velocity space dissipation, so that the reflection of free energy at the truncation point $a_{N_m}=0$ generates backwards flux. 
The backwards flux propagates along $m=-2k^2(t-t_0)^2$ characteristics in both the $\atp$ and $\atm$ modes, but has significantly larger magnitude in the $\atm$ plot. 
Overall we conclude the decomposition is generally accurate with forwards and backwards modes dominating the $\atp$ and $\atm$ plots respectively.

We now plot the corresponding $k=0.5$ spectra for the nonlinear case in \fig~\ref{fig:at_contours_nonlinear}.
The nonlinear term \eqref{eq:pmNonlinearTerm} introduces Fourier mode coupling where free energy in other wavenumbers excites both $\atp$ and $\atm$ in the $k=0.5$ wavenumber.
We observe this in \fig~\ref{fig:atContoursNonlinear} where the free energy propagates on the characteristics $m=\pm 2k^2(t-t_0)^2$ respectively.
These characteristics appear throughout phase space rather than only near the characteristics that correspond to the propagation of initial conditions, as in \fig~\ref{fig:atContoursLinear}.
\begin{figure}
  \centering
  \includegraphics[trim=3.5cm 9.5cm 4.0cm 10.3cm,width=0.49\textwidth,clip]{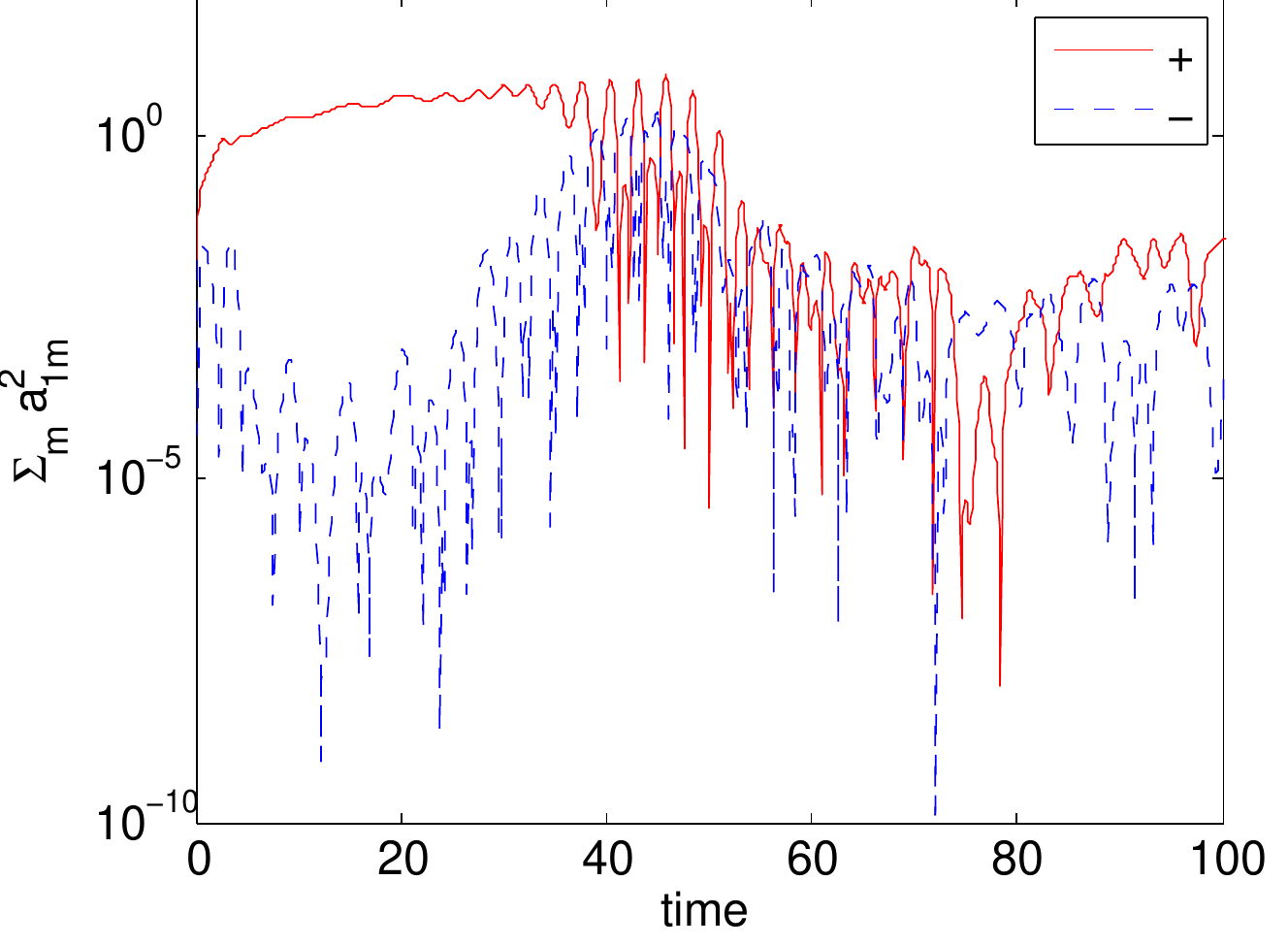}
  \includegraphics[trim=1.0cm 0.2cm 1.5cm 1.0cm,width=0.49\textwidth,clip]{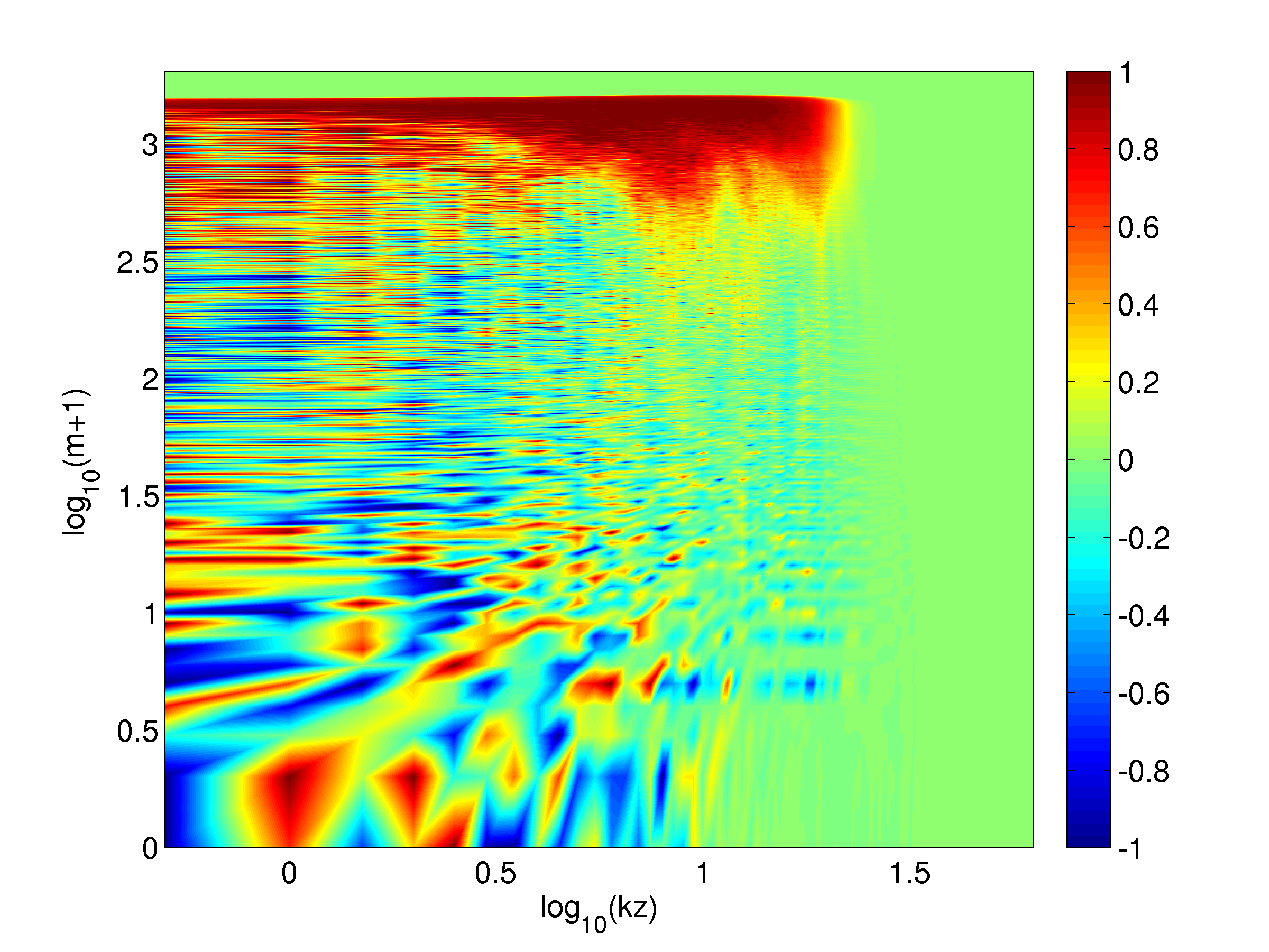}
  \caption{Left: contributions to the free energy from forwards and backwards modes, $\sum_m(\atpm_{1m})^2$. Right: time-averaged normalized Hermite flux for the interval $t\in[40,80]$.
  \label{fig:MFlux}
}
\end{figure}
This suggests that the $\atm$ modes that cause the increase in the electric field are excited by the nonlinear term.
However it remains possible that the back flux is generated by the Boltzmann response or the streaming correction through instabilities that are not excited linearly.
To determine which effect is responsible, we plot $\sum_{m}(\atpm_{1,m})^2$, the contribution to the free energy from the forwards and backwards modes in \fig~\ref{fig:MFlux}(a).
At $t=20$ when the electric field grows, the contribution to free energy from $\atm$, $\sum_m (\atm_{1m})^2$ grows exponentially.
This suggests the increase in free energy is due to a term like $\atm\atm\atp$, such as is found in the nonlinear term contribution to the free energy equation $\atm N^-$.

Turning to the absence of Landau damping at long times, we see from \fig~\ref{fig:NonlinearFreeEnergyTrace} that the free energy contributions reach a steady state where there is very little collisional damping.
Moreover from \fig~\ref{fig:MFlux}(a) we see that the free energy contributions from $\atpm$ balance, showing there is little net flux.
This is a statement only about $k=0.5$, so in \fig~\ref{fig:MFlux}(b) we plot the normalized Hermite flux \eqref{eq:NormalizedFlux} for all phase space, time-averaged over the interval $t\in[40,80]$.
This shows that indeed there is no systematic Hermite flux towards fine scales. 
However the growth or decay of the electric field over long timescales, requires a net Hermite flux to persist over long timescales; 
similarly collisional damping requires a systematic flux to fine scales. 
Therefore by generating a backward Hermite flux which on average balances with the forward flux, 
the nonlinearity has effectively suppressed Landau damping.

\subsection{Two stream instability}
\label{sec:TwoStreamInstability}

We now demonstrate the spectral method for a non-Maxwellian equilibrium $f_0$
by studying the two-stream instability.
This standard problem has been treated in great detail elsewhere
(see \eg\ \cite{Grant67,Denavit71,Cheng76,Zaki88,Klimas94,Nakamura99,Pohn05,Heath12})
and we wish only to illustrate that expected results are obtained with modified \sgk.

We use the new background distribution
\begin{align}
  f_0 = \frac{2v^2}{\sqrt{\upi}}\exp(-v^2),
  \label{eq:NewEquilibrium}
\end{align}
and the initial conditions \eqref{eq:InitialConditions} with 
$A=0.05$, $k=0.5$ and $L=4\upi$. 
The equilibrium is the sum of two Hermite functions
\begin{align}
  f_0 = \sqrt{2}\phi_2(v) + \phi_0(v),
\end{align}
from which we verify that \eqref{eq:EquilibriumIntegral} holds.
Thus the new equilibrium only enters the \vps\ in the kinetic equation \eqref{eq:KineticEquation} as a modification to the response term on the \rhs,
\begin{align}
  E\pd{f_0}{v} = -\lp 2\sqrt{3}\phi_3(v) + \sqrt{2}\phi_1(v)\rp E.
\end{align}
In Hermite space, this yields an extra source term in the moment equation \eqref{eq:DiscreteHermiteMomentKineticEquation},
\begin{align}
  \label{eq:TwoStreamDiscreteHermiteMomentKineticEquation}
  \pd{a_{jm}}{t} + \i k_j \lp \sqrt{\frac{m+1}{2}}a_{j,m+1} + \sqrt{\frac{m}{2}}a_{j,m-1}\rp 
  + \Nmat_{jm}
  = -\sqrt{2}\E_j\delta_{m1} - 2\sqrt{3}\E_j\delta_{m3}  ,
\end{align}
with the equations for the electric field \eqref{eq:DiscreteElectricField} and \eqref{eq:DiscretePoissonEquation} unchanged.

\begin{figure}
  \centering
  \subfigure[$t=0$\label{fig:TwoStreamt0}]{\includegraphics[trim=1.5cm 0.5cm 1.5cm 1.0cm,width=0.49\textwidth,clip]{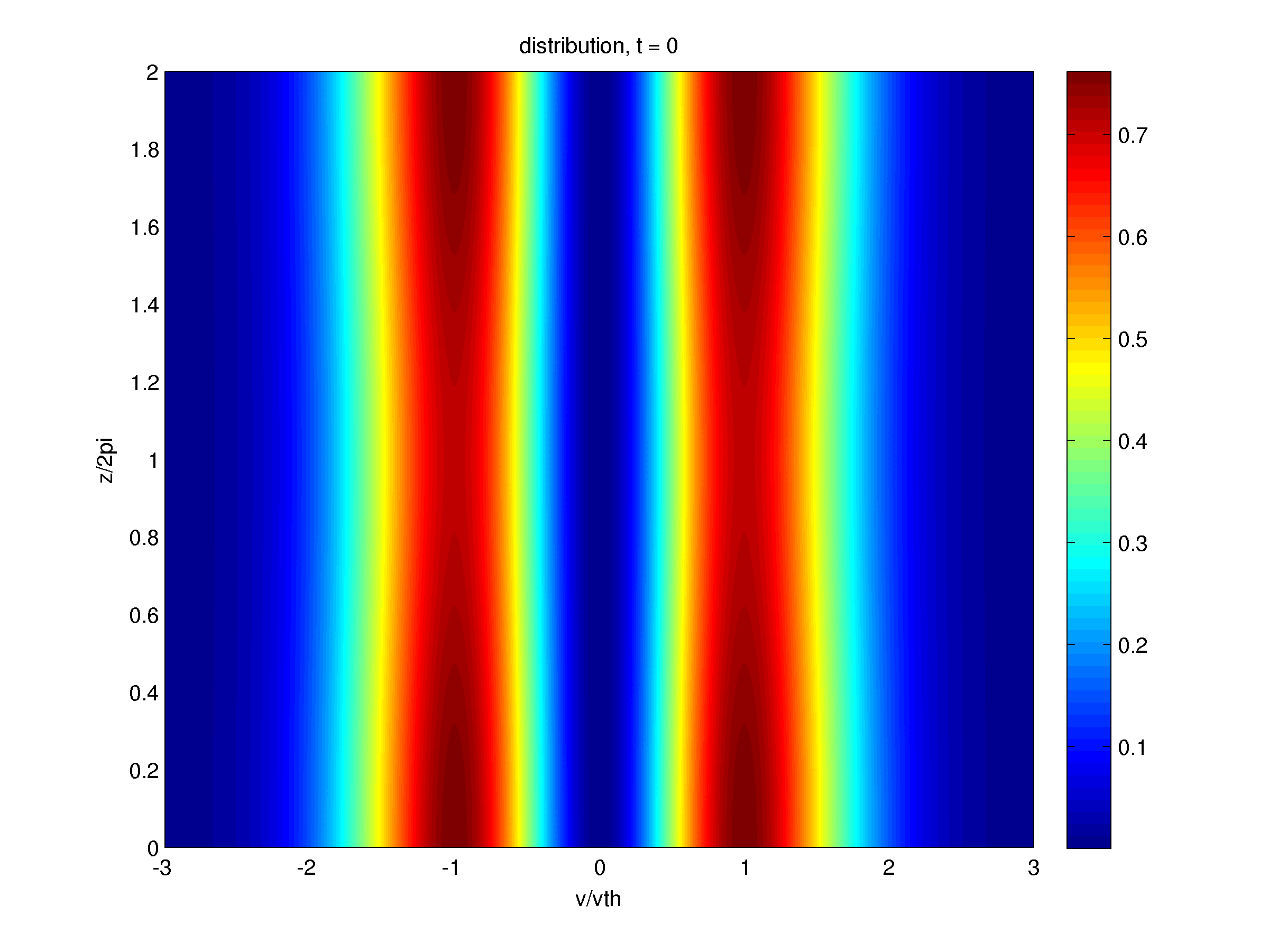}}
  \subfigure[$t=20$]{\includegraphics[trim=1.5cm 0.5cm 1.5cm 1.0cm,width=0.49\textwidth,clip]{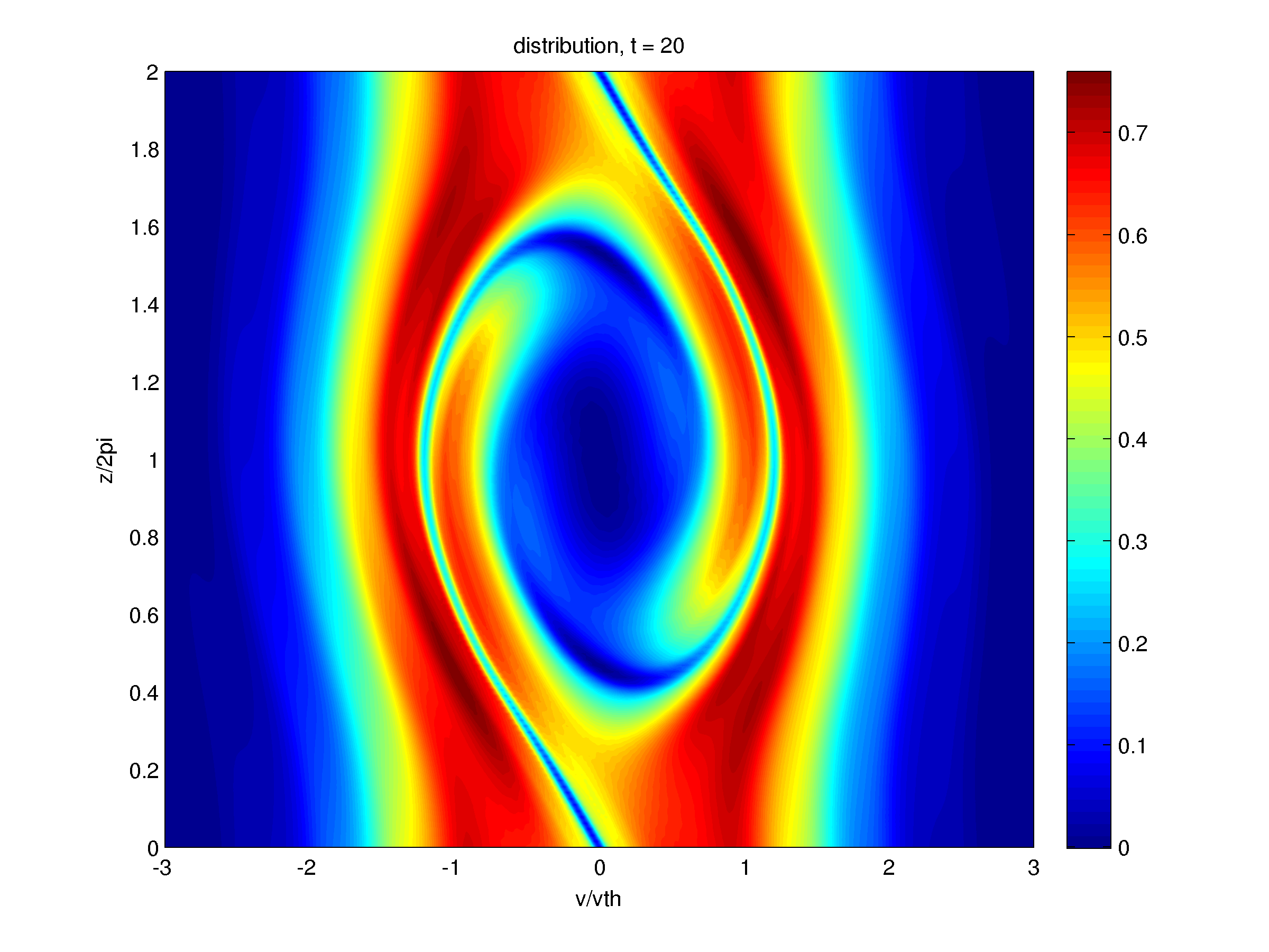}}
  \subfigure[$t=40$]{\includegraphics[trim=1.5cm 0.5cm 1.5cm 1.0cm,width=0.49\textwidth,clip]{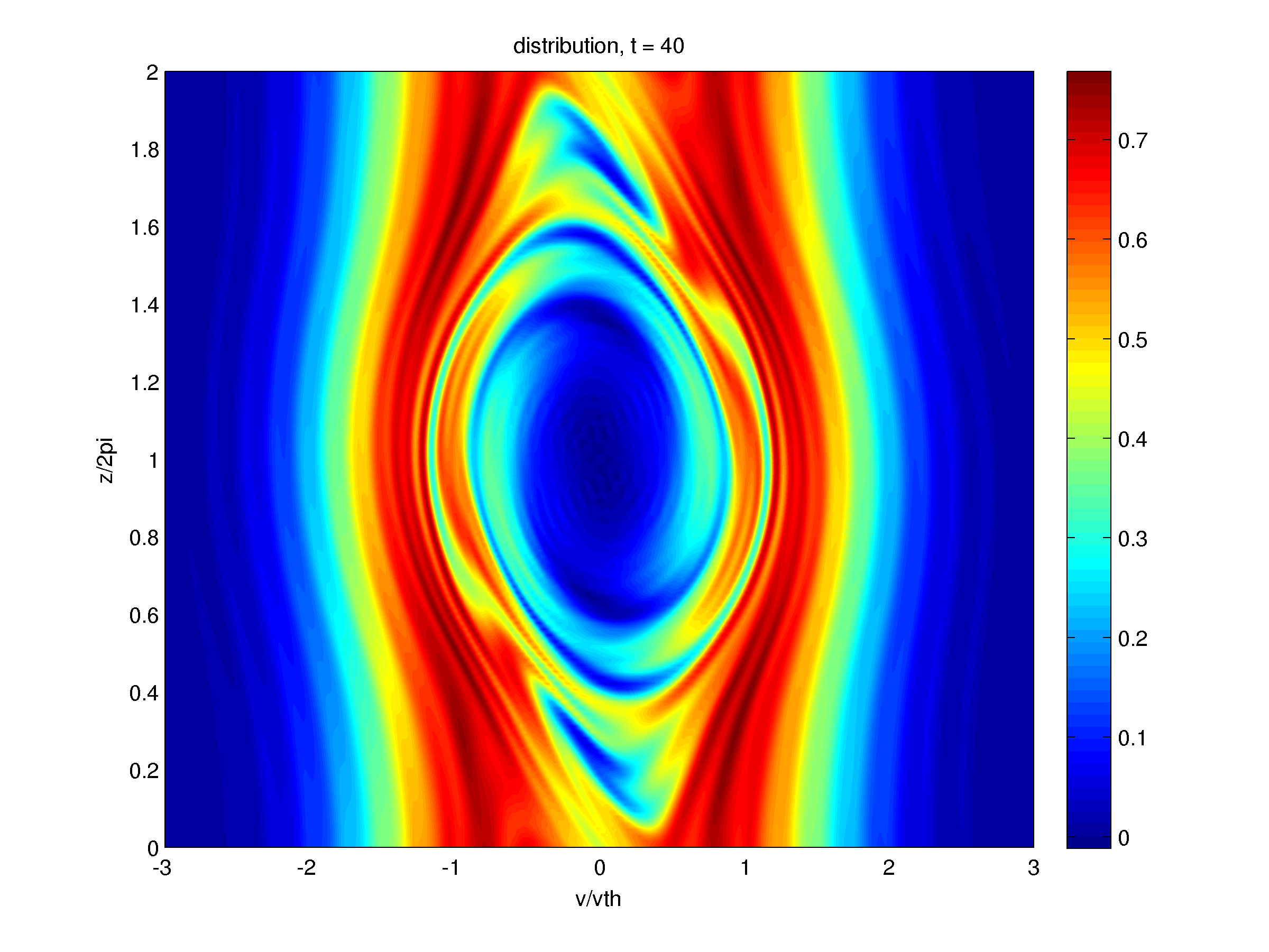}}
  \subfigure[$t=60$]{\includegraphics[trim=1.5cm 0.5cm 1.5cm 1.0cm,width=0.49\textwidth,clip]{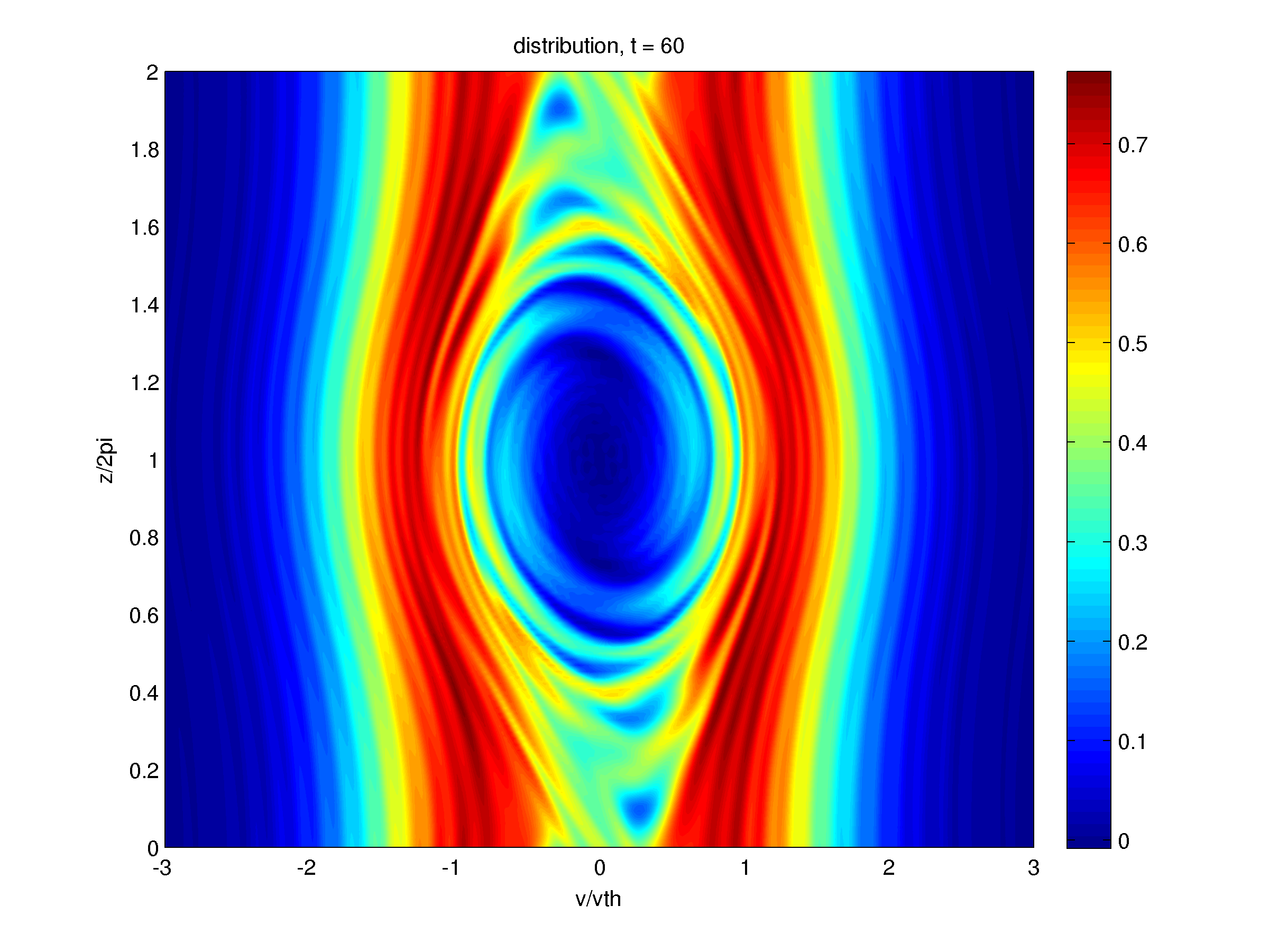}}
  \subfigure[$t=100$]{\includegraphics[trim=1.5cm 0.5cm 1.5cm 1.0cm,width=0.49\textwidth,clip]{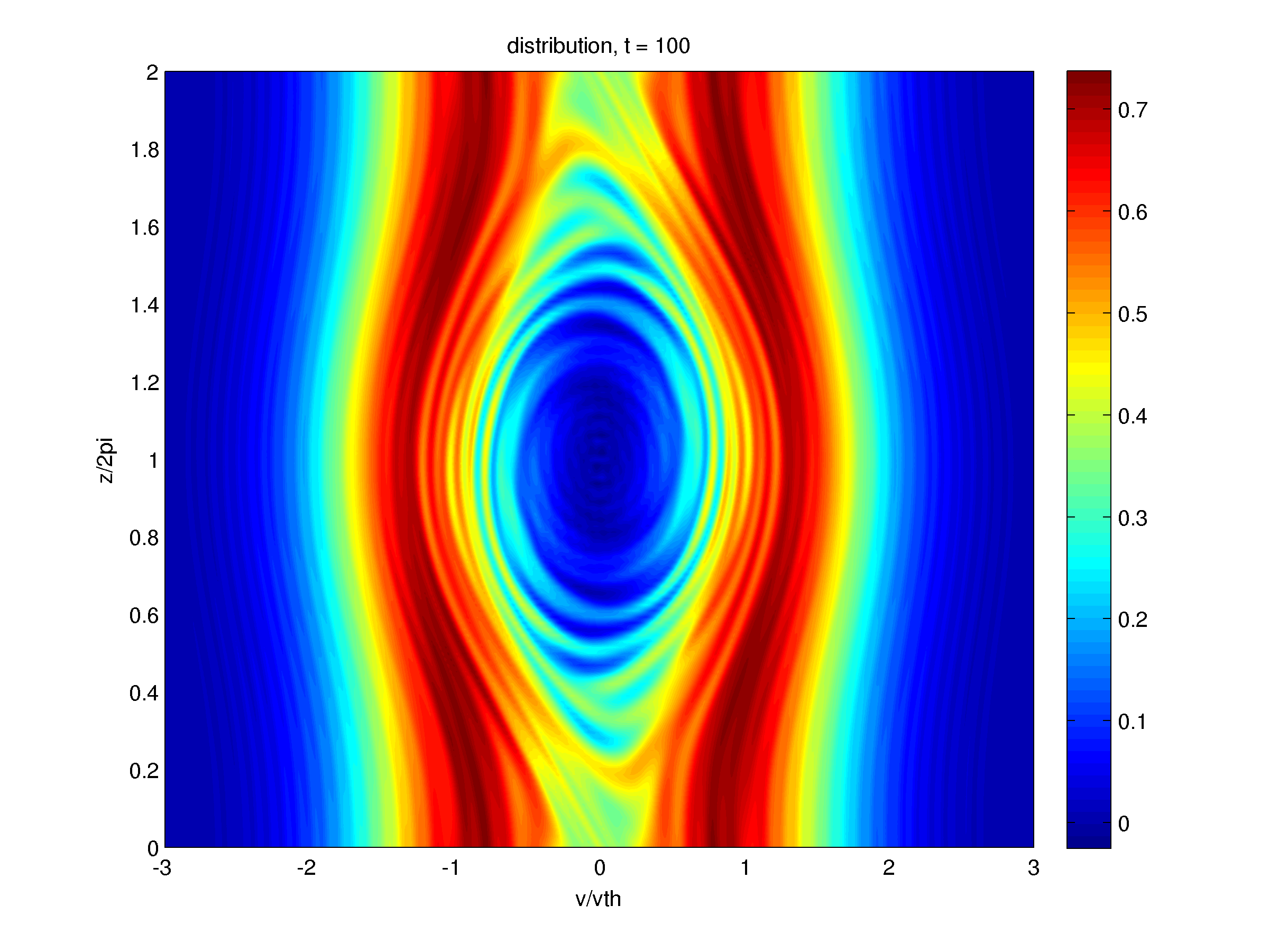}}
  \subfigure[$t=160$]{\includegraphics[trim=1.5cm 0.5cm 1.5cm 1.0cm,width=0.49\textwidth,clip]{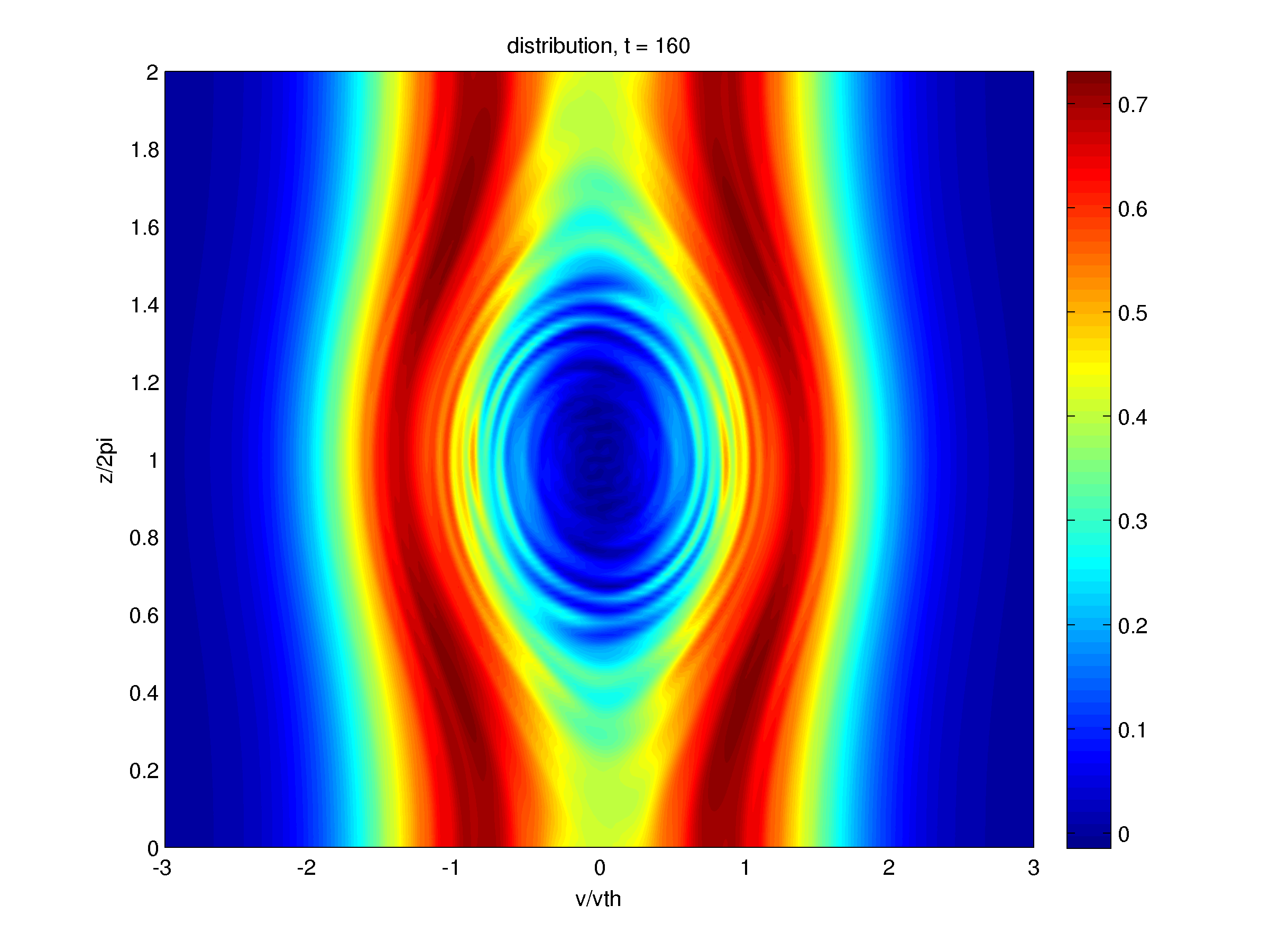}}
  \caption{Time slices of the full distribution function for the two stream instability in $(z,v)$ space, calculated with $(N_k,N_m)=(129,4096)$ and plotted on the Fourier collocation points and a uniform grid of 4096 point for $v\in[-3,3]$.
  \label{fig:TwoStreamDistributionTimeSlices}
}
\end{figure}

The bimodal equilibrium \eqref{eq:NewEquilibrium} describes two counter-streaming electron beams, which, with the small initial perturbation, is shown in \fig~\ref{fig:TwoStreamt0}.
The new source term in \eqref{eq:TwoStreamDiscreteHermiteMomentKineticEquation} introduces a linear instability for $k^2<2$ (see appendix~\ref{sec:LinearStability}). 
The perturbation is unstable and grows exponentially until the nonlinear term becomes important and saturates the linear growth.
In the long time limit, the distribution function approaches the Bernstein--Greene--Kruskal state \citep{Bernstein57}.
We see this in our solution, plotted in \fig~\ref{fig:TwoStreamDistributionTimeSlices}.

\section{Conclusion}

In this work we have illustrated the usefulness of \FH\ spectral method for treating the 1+1D \vps.
The \FH\ representation presented in \sec~\ref{sec:SpectralRepresentation} yields an attractive moment-based formulation,
which we implemented using a modified version of the \sgk\ gyrokinetics code.
The fine velocity space scales which arise due to particle streaming were smoothed using the Hou--Li spectral filter \citep{Hou07},
which had been successfully applied in physical space in fluid simulations, but had not previously been applied in velocity space.
This filtering eliminates recurrence, meaning the method is successful even when phase mixing and filamentation are dominant effects.
This is particularly important in regimes like nonlinear Landau damping where the nonlinearity generates structure at fine scales (see \fig~\ref{fig:Spectra}) which must be distinguishable from recurrence effects. 

In \sec~\ref{sec:NumericalResults} we replicated well-known results for nonlinear Landau damping and the two stream instability,
and demonstrated exponential convergence of \sgk\ in both space and velocity space.
This benchmarks \sgk\ against solutions obtained by early low-resolution \FH\ simulations \citep[\eg][]{Armstrong67,Grant67}, 
by PIC codes \citep[\eg][]{Denavit71}, 
by finite element methods \citep[\eg][]{Zaki88},
and by recent discrete Galerkin simulations \citep{Heath12}.

Finally, we studied the flow of free energy in \FH\ phase space using tools recently developed by \cite{Schekochihin14} for the gyrokinetic equations.
We expressed the distribution function as combination of forwards and backwards propagating modes in Hermite space, the difference of which represent the flux of free energy.
This net Hermite flux is associated with the change in the electric field via the free energy evolution equation \eqref{eq:WEeqn}.
We showed that the growth in the electric field at $t=20$ is associated with the generation of backwards propagating modes by the nonlinear term.
Both the electric field and the free energy in backwards propagating modes grow exponentially until the free energy content of forwards and backwards propagating modes roughly balance. 
Thereafter the magnitude of the electric field does not change significantly over time.
This is because there is no systematic net flux in Hermite space
and therefore no systematic change in the free energy of the electric field. 
Thus the electric field cannot grow or decay over long times, and so the nonlinearity effectively suppresses Landau damping.

The authors are grateful for fruitful conversations with 
I.~Abel, 
G.~Colyer, 
S.~Cowley,
W.~Dorland,
M.~Fox,
G.~Hammett,
E.~Highcock, 
A.~Kanekar,
G.~Plunk,
C.~Roach, 
A.~Schekochihin,
F.~van Wyk,
and A.~Zocco.
This work was supported by the UK Engineering and
Physical Sciences Research Council through a Doctoral Training Grant award to J.T.P. and an Advanced Research
Fellowship [grant number EP/E054625/1] to P.J.D., with additional support from Award No KUK-C1-013-04 made
by King Abdullah University of Science and Technology (KAUST).
The authors acknowledge the use of the IRIDIS HPC facility
through the e-Infrastructure South Centre for Innovation.
Some of the results of this research have been achieved using the
PRACE-3IP project (FP7 RI-312763) resource FIONN based in Ireland at 
the DJEI/DES/SFI/HEA Irish Centre for High-End Computing (ICHEC); 
though access to the HECToR HPC facility [grant
number EP/H002081/1];
and through the use of Hartree Centre resources in this work. The STFC Hartree Centre is a research collaboratory in association with IBM providing High Performance Computing platforms funded by the UK's investment in e-Infrastructure. The Centre aims to develop and demonstrate next generation software, optimised to take advantage of the move towards exa-scale computing.

\bibliographystyle{jpp}
\bibliography{vp,plasmas}

\appendix

\section{Linear dispersion relation}
\label{sec:LinearStability}

To illustrate the appearance of a linear instability for the non-Maxwellian equilibrium \eqref{eq:NewEquilibrium},
we derive the linear dispersion relation based on the first four Hermite moments in \eqref{eq:DiscreteHermiteMomentKineticEquation} and 
\eqref{eq:TwoStreamDiscreteHermiteMomentKineticEquation}. 
This may be viewed as a simple collisionless fluid model.
This does not capture decay rates, but is sufficient to determine regions of linear instability.

\begin{figure}
  \centering
  \includegraphics[trim=3.5cm 9.5cm 4.0cm 9.5cm,width=0.48\textwidth,clip]{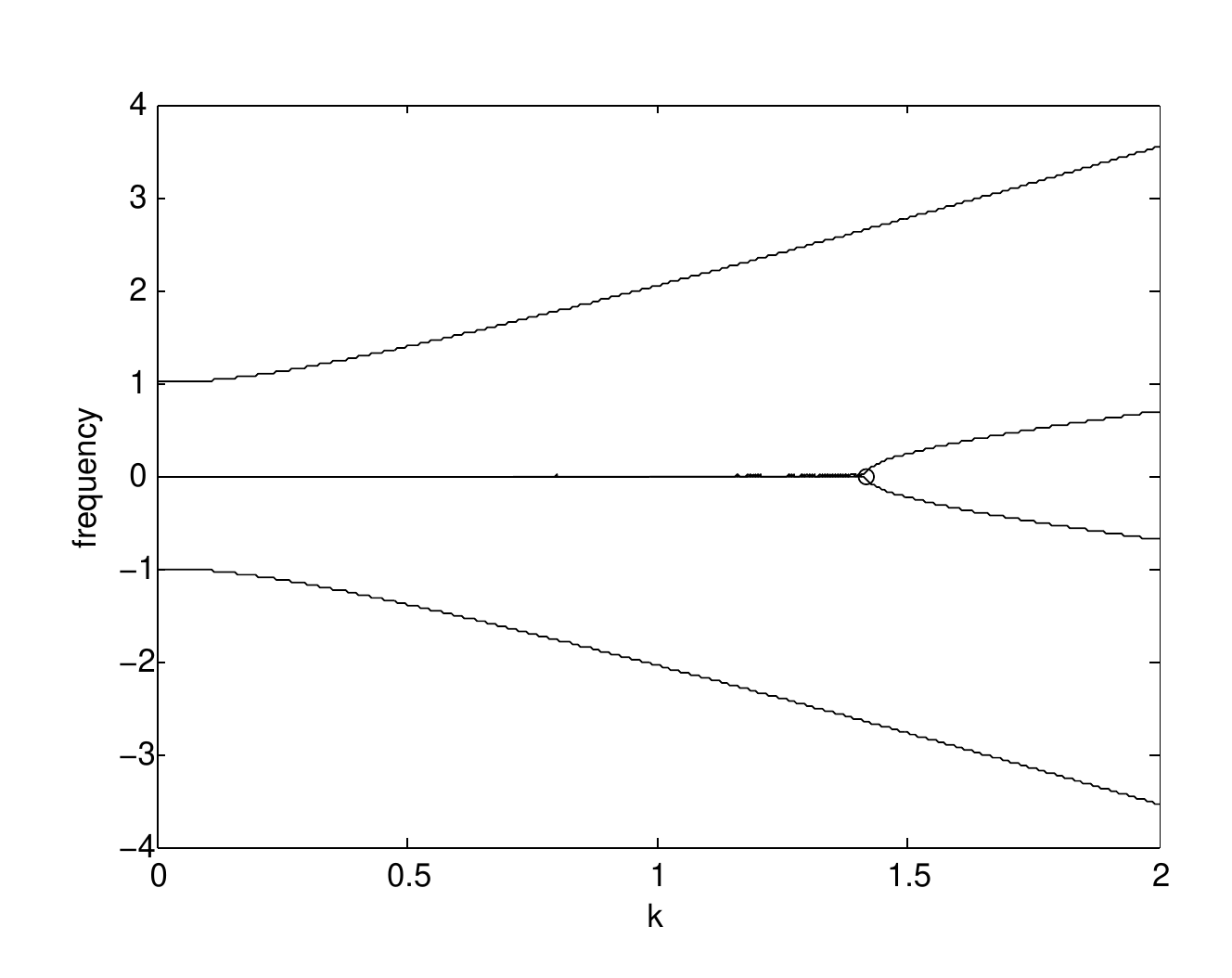}
  \includegraphics[trim=3.5cm 9.5cm 4.0cm 9.5cm,width=0.48\textwidth,clip]{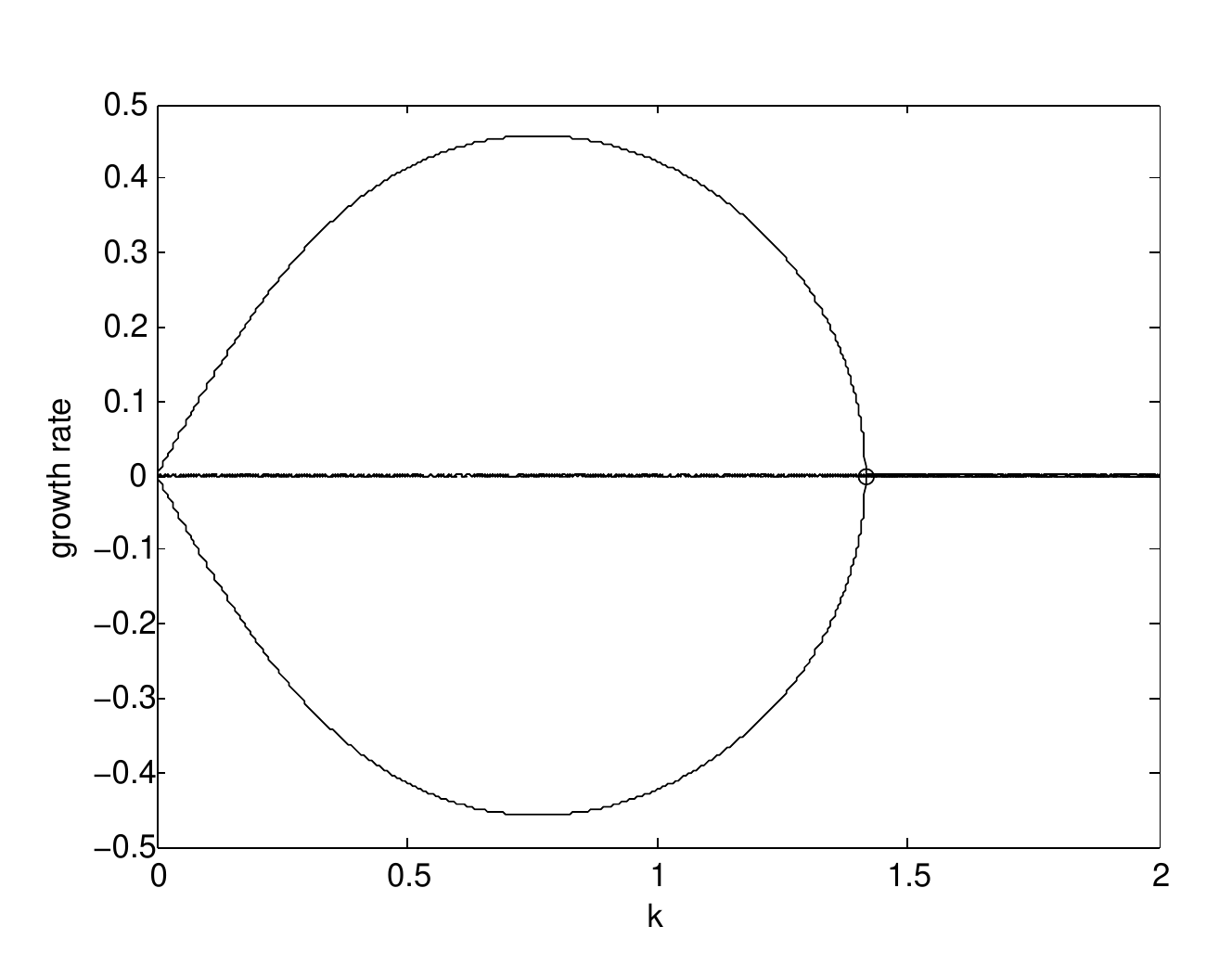}
  \caption{Dispersion relation for the two stream instability: frequency (left) and growth rate (right) against wavenumber $k$. 
    The small circle on the $k$-axis marks the wavenumber $k=\sqrt{2}$. 
  \label{fig:DispersionRelation}}
\end{figure}

The linear equation is parameterized by $k$, so we solve for time eigenfunctions of the form
$a_m(k,t) = \bar{a}_m(k) \e^{-i\omega t}$.
The first four moments are 
\renewcommand{\0}{0}
\begin{align}
  \left( \begin{array}{cccc}
      -\i\omega & \i k & \0 & \0 \\
      \i\lp \frac{k}{\sqrt{2}}+\frac{\sqrt{2}}{k} \rp & -\i\omega & \i k & \0 \\
      \0 & \i k & -\i\omega & \i k \sqrt{3/2} \\
      \i\beta 2\sqrt{3}/k & \0 & \i k\sqrt{3/2} & -\i\omega\end{array} \right)
\lp \begin{array}{c} a_0 \\ a_1 \\ a_2 \\ a_3 \end{array} \rp
      = 0 ,
\end{align}
where $\beta=0$ for the nonlinear Landau-damping problem, and $\beta=1$ for the two stream instability.
This yields the dispersion relation
\begin{align}
  \omega^4  - (3k^2+1)\omega^2 + \frac{3k^2}{4}\lp k^2 + 2 - 4\beta\rp = 0,
  \label{eq:DispersionRelation}
\end{align}
with solutions
\begin{align}
  \omega^2 = \frac{1}{2}\left[ (3k^2+1)\pm \sqrt{(3k^2+1)^2-3k^2(k^2 + 2 - 4\beta)} \right] .
  \label{eq:OmegaSquared}
\end{align}
The plasma is stable if $\omega^2$ is real and non-negative, that is if 
\begin{align}
  (3k^2+1)^2 \geq (3k^2+1)^2 - 3k^2\lp k^2 + 2 - 4\beta\rp .
\end{align}
For nonlinear Landau-damping ($\beta=0$) this is always true and all modes are linearly stable. 
For the two stream instability ($\beta=1$), 
modes corresponding to the $+$ sign in \eqref{eq:OmegaSquared} are stable,
while modes corresponding to the $-$ sign are unstable for wavenumbers $0<k<\sqrt{2}$.
The dispersion relation for the two stream instability is shown in \fig~\ref{fig:DispersionRelation}.

\end{document}